\documentclass[twocolumn]{aastex62} 
\usepackage{subfigure}
\usepackage{url}
\usepackage{graphicx,color}
\usepackage{gensymb}     

\usepackage{hyperref}
\newcommand{\CII}{\ion{C}{2}}

\newcommand{\MgII}{\ion{Mg}{2}}
\newcommand{\SiIV}{\ion{Si}{4}} 
\newcommand{\FeIX}{\ion{Fe}{9}}
\newcommand{\FeXII}{\ion{Fe}{12}}
\newcommand{\FeXIV}{\ion{Fe}{14}}
\newcommand{\FeXVI}{\ion{Fe}{16}}
\newcommand{\FeXVIII}{\ion{Fe}{18}}
\newcommand{\FeXXI}{\ion{Fe}{21}}
\newcommand{\HeII}{\ion{He}{2}}

\DeclareMathAlphabet{\mathitbf}{OML}{cmm}{b}{it}
\DeclareMathAlphabet{\mathf}{OML}{cmm}{c}{sl}

\newcommand{\eg}{e.g.,}

\newcommand{\kms}{km~s$^{-1}$}

\shorttitle{Hi-C 2.1 observations of fine-scale coronal energy release}
\shortauthors{Tiwari et al.}

\begin{document}

\title{Fine-scale explosive energy release at sites of prospective magnetic flux cancellation in the core of the solar active region observed by Hi-C 2.1, IRIS and SDO}



\correspondingauthor{Sanjiv K. Tiwari}
\email{tiwari@lmsal.com}

\author[0000-0001-7817-2978]{Sanjiv K. Tiwari}
\affiliation{Lockheed Martin Solar and Astrophysics Laboratory, 3251 Hanover Street, Bldg. 252, Palo Alto, CA 94304, USA}
\affiliation{Bay Area Environmental Research Institute, NASA Research Park, Moffett Field, CA 94035, USA}

\author[0000-0001-7620-362X]{Navdeep K. Panesar}
\affiliation{Lockheed Martin Solar and Astrophysics Laboratory, 3251 Hanover Street, Bldg. 252, Palo Alto, CA 94304, USA}
\affiliation{Bay Area Environmental Research Institute, NASA Research Park, Moffett Field, CA 94035, USA}

\author[0000-0002-5691-6152]{Ronald L. Moore}
\affiliation{NASA Marshall Space Flight Center, Mail Code ST 13, Huntsville, AL 35812, USA}
\affil{Center for Space and Aeronomic Research, The University of Alabama in Huntsville, Huntsville, AL 35805, USA}

\author[0000-0002-8370-952X]{Bart De Pontieu}
\affiliation{Lockheed Martin Solar and Astrophysics Laboratory, 3251 Hanover Street, Bldg. 252, Palo Alto, CA 94304, USA}
\affil{Rosseland Centre for Solar Physics, University of Oslo, P.O. Box 1029 Blindern, NO–0315 Oslo, Norway}
\affil{Institute of Theoretical Astrophysics, University of Oslo, P.O. Box 1029 Blindern, NO–0315 Oslo, Norway}

\author[0000-0002-5608-531X]{Amy R. Winebarger} 
\affiliation{NASA Marshall Space Flight Center, Mail Code ST 13, Huntsville, AL 35812, USA}

\author{Leon Golub} 
\affiliation{Harvard-Smithsonian Center for Astrophysics, 60 Garden Street, Cambridge, MA 02138, USA}

\author{Sabrina L. Savage} 
\affiliation{NASA Marshall Space Flight Center, Mail Code ST 13, Huntsville, AL 35812, USA}

\author{Laurel A. Rachmeler} 
\affiliation{NASA Marshall Space Flight Center, Mail Code ST 13, Huntsville, AL 35812, USA}

\author{Ken Kobayashi} 
\affiliation{NASA Marshall Space Flight Center, Mail Code ST 13, Huntsville, AL 35812, USA}

\author[0000-0002-0405-0668]{Paola Testa}
\affiliation{Smithsonian Astrophysical Observatory, 60 Garden Street, MS 58, Cambridge, MA 02138, USA}
\affil{Lockheed Martin Solar and Astrophysics Laboratory, 3251 Hanover Street, Bldg. 252, Palo Alto, CA 94304, USA}

\author{Harry P. Warren} 
\affiliation{Space Science Division, Naval Research Laboratory, Washington, DC 20375 USA}

\author{David H. Brooks} 
\affiliation{College of Science, George Mason University, 4400 University Drive, Fairfax, VA 22030 USA}

\author{Jonathan W. Cirtain}
\affiliation{BWX Technologies, Inc., 800 Main St \# 400, Lynchburg, VA 24504}

\author{David E. McKenzie} 
\affiliation{NASA Marshall Space Flight Center, Mail Code ST 13, Huntsville, AL 35812, USA}

\author{Richard J. Morton} 
\affiliation{Department of Mathematics, Physics and Electrical Engineering, Northumbria University, Newcastle Upon Tyne, NE1 8ST, UK}

\author[0000-0001-9921-0937]{Hardi Peter} 
\affiliation{Max Planck Institute for Solar System Research, 37077 Göttingen, Germany}

\author{Robert W. Walsh} 
\affiliation{University of Central Lancashire, Preston, PR1 2HE, UK}


\graphicspath{{Figures/}}     


\begin{abstract}
The second Hi-C flight (Hi-C2.1) provided unprecedentedly-high spatial and temporal resolution ($\sim$250km, 4.4s) coronal EUV images of Fe IX/X emission at 172 \AA, of AR 12712 on 29-May-2018, during 18:56:21-19:01:56 UT. Three morphologically-different types (I: dot-like, II: loop-like, III: surge/jet-like) of fine-scale sudden-brightening events (tiny microflares) are seen within and at the ends of an arch filament system in the core of the AR. Although type Is (not reported before) resemble IRIS-bombs (in size, and brightness wrt surroundings), our dot-like events are apparently much hotter, and shorter in span (70s). We complement the 5-minute-duration Hi-C2.1 data with SDO/HMI magnetograms, SDO/AIA EUV images, and IRIS UV spectra and slit-jaw images to examine, at the sites of these events, brightenings and flows in the transition-region and corona and evolution of magnetic flux in the photosphere. Most, if not all, of the events are seated at sites of opposite-polarity magnetic flux convergence (sometimes driven by adjacent flux emergence), implying likely flux cancellation at the microflare's polarity inversion line. In the IRIS spectra and images, we find confirming evidence of field-aligned outflow from brightenings at the ends of loops of the arch filament system. In types I and II the explosion is confined, while in type III the explosion is ejective and drives  jet-like outflow. The light-curves from Hi-C, AIA and IRIS peak nearly simultaneously for many of these events and none of the events display a systematic cooling sequence as seen in typical coronal flares, suggesting that these tiny brightening-events have chromospheric/transition-region origin.
\end{abstract}


\section{Introduction} \label{sec:intr}

The second sounding-rocket flight of the High-Resolution Coronal Imager \citep[Hi-C 2.1:][]{rach19} took coronal extreme ultra-violet (EUV) images of NOAA active region (AR) 12712 in 172 \AA\ (\FeIX/X emission) with unprecedented spatial and temporal resolutions ($\sim$250 km, 4.4 s). The data was collected for about five minutes, during the period of 18:56:21 -- 19:01:56 UT on May 29, 2018, near solar disk center (AR position: N15E10). The Hi-C 2.1 (hereafter `Hi-C') data have revealed multiple small-scale activities inside the AR core and in the AR's surroundings. These small-scale brightenings remained unnoticed in earlier EUV observations.       

Solar ARs contain the brightest and hottest coronal EUV loops \citep{golu80,real14} -- the core of an AR is typically the brightest structure inside the AR \citep{warr12}. In the chromosphere the AR core often contains a set of cool loops, known as an arch filament system \citep{bruz67}, long observed in H-alpha filtergrams. Usually emerging flux regions \citep[EFRs:][]{ziri72efr} in the cores of ARs are seen as cool arch filament systems \citep{bruz67,fraz72}. Because the field is arched and emerging, these arch filament systems are found to have blueshifts (of up to 10 \kms) in their central parts (apex) in the chromosphere and redshifts (of up to 40 \kms) at both ends \citep{geor90,tsir92,gonz18}. These flow patterns weaken as the field emergence ends -- hardly any significant flows are noticed after the emergence has stopped. The AR in the present study is near the end of global emergence of its overall bipolar field (but local flux emergence at multiple places, often recurrently, continues).  

Small-scale polarity inversion lines (PILs), also known as neutral lines, are often present in the cores of EFRs \citep{fraz72}. These emerging flux regions (with cool chromospheric but hot coronal environment) can have multiple explosive events such as Ellerman bombs \citep[EBs:][]{elle17,rutt13}, surges \citep{newt42,roy73} and IRIS bombs \citep[IBs:][]{pete14}. IBs and EBs both have mixed-polarity photospheric magnetic field, and often have common properties to each other but their plasma temperatures (of $<$10,000 K for EBs vs $\simeq$80,000 K for IBs) are apparently different. Both EBs and IBs might form in the photosphere \citep{tian16}. Recent magnetohydrodynamic simulations however support the idea that both form in the higher atmosphere, i.e., in the low chromosphere \citep{hans19}. 

Surges are more explosive (than EBs), can be hotter than chromospheric temperature, and have a rapid cool plasma outflow (from the source region), often followed by a weaker inflow (plasma flowing towards the source/base of the surge) \citep{newt42}. EBs are sometimes present at the base of surges \citep{roy73,mats08,youn18}, which have mixed-polarity photospheric magnetic flux similar to EBs, and are a consequence of flux emergence and/or flux cancellation \citep{roy73,yu04,jian07,lope18}. 
 
In the present work we report on three types of fine-scale transient brightening events in the core of the AR 12712 observed in 172 \AA\ by Hi-C: type I, dot-like; type II, loop-like; and type III, surge/jet-like events (described in Section \ref{resu}). 
Type I events were not identified earlier in AIA 171 \AA\ probably due to their small size, but possibly partially due to AIA's somewhat narrower bandwidth than that for Hi-C.

\section{Data and Methods} \label{data}
The five minutes of Hi-C observations \citep[obtained at a cadence of 4.4 s and a spatial resolution of $\sim$250 km:][]{rach19} were complemented by the Interface Region Imaging Spectrograph \cite[IRIS:][]{depo14IRIS}, the Solar Optical Telescope \citep[SOT:][]{tsun08,ichi08,suem08,shim08,lites13SP} onboard Hinode \citep{kosu07}, the Helioseismic Magnetic Imager \citep[HMI:][]{scho12} and Atmospheric Imaging Assembly \citep[AIA:][]{leme12} onboard SDO, and several other instruments. We mainly analyze the data from Hi-C, IRIS and SDO (AIA+HMI) in the present work.  

IRIS captured slit-jaw (SJ) movies in \MgII\ 2796, Mg continuum 2832, \SiIV\ 1400 and \CII\ 1330 \AA\ at a cadence of 13 s with a pixel size of 0.33 arcsec. These SJ images sample plasma from 6000 K to $\sim$100,000 K.

\begin{figure*}
	\centering
	\includegraphics[trim=6.65cm 0.15cm 5.5cm 0.04cm,clip,width=0.8\textwidth]{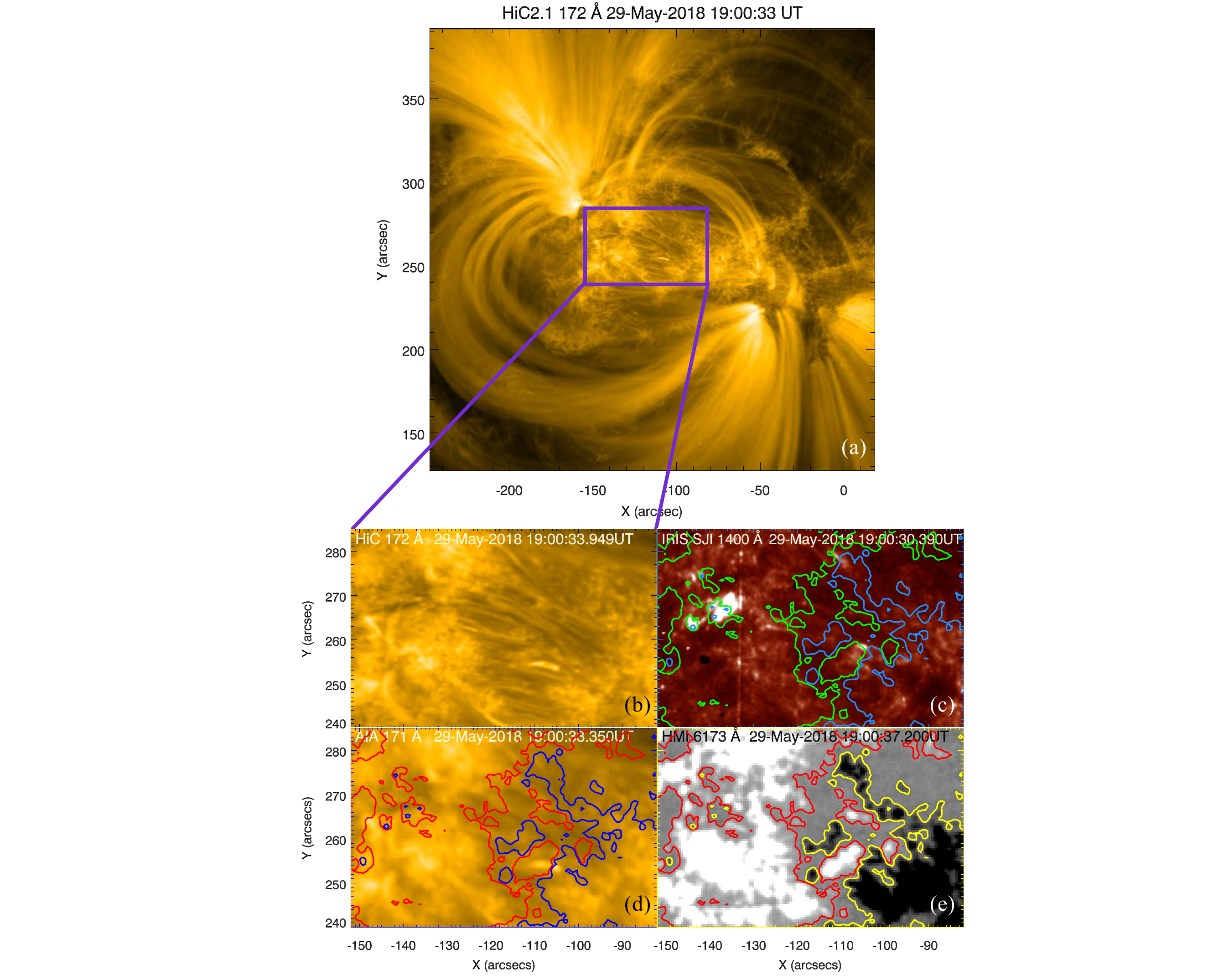}
	\caption{Context image of the Hi-C 2.1 observations of NOAA AR 12712 (at disk position: N15E10). (a) The full field of view (FOV) observed by Hi-C in 172 \AA. Within the core of the active region the region of interest for the present research is outlined by the purple box, a zoomed in view of which is displayed in panel (b). In panels (c), and (d) images of the same FOV as (b) observed with IRIS SJI 1400 \AA, and SDO/AIA 171 \AA\ are displayed. In panel (e) a line-of-sight (LOS) magnetogram (saturated at $\pm$400 G) obtained with SDO/HMI is shown. The red and yellow (green and blue for IRIS SJI 1400 \AA, red and blue for AIA 171 \AA) contours are respectively for positive and negative LOS magnetic field at a level of $\pm$25 G. }
	\label{fcontext1}
\end{figure*}

The IRIS slit scanned an east part of the region of our interest with an 8-step raster, at a step size of 1 arcsec and a step cadence of 3.2 s, thus resulting into a raster cadence of 25 s (OBS ID 3600104031). A total of 256 rasters were obtained for about 1 h and 50 minutes, including the five minutes of Hi-C observations. The exposure time for each slit position of each wavelength is 2 s. The slit width is 0.33 arcsec, and each pixel of the spectrum image spans 0.33 arcsec along the slit and a wavelength increment of 0.02 \AA\ (or a Doppler shift increment of 3 \kms) along the dispersion axis. The Hi-C field of view (FOV) and the FOV of our interest are shown in Figure \ref{fcontext1}. 

Similar to that of IRIS, the SDO AIA (12 s cadence for EUV images and 24 s cadence for UV images, 0.6 arcsec pixel size) and HMI (45 s cadence for line of sight (LOS) magnetograms, a pixel size of 0.5 arcsec) data are used to follow the brightness and magnetic field in the cool arch filament system. The random per-pixel photon noise for 45 s cadence HMI LOS magnetograms is $\approx$7 G \citep{couv16}. Small-scale dynamic events are followed for two and a half hours, centered at the Hi-C observations. 

AIA covers a broad range of temperatures. We use all EUV channels: AIA 304, 171, 193, 211, 335, 94, and 131 \AA, which show emission from plasma at $\sim$50,000 K (\HeII), 700,000 (\FeIX/X), 1.5 MK (\FeXII), 2 MK (\FeXIV), 2.5 MK (\FeXVI), 6 MK (\FeXVIII), and 10 MK (\FeXXI), respectively. Note that AIA 94 and 131 channels also see cooler components at about 1 MK and 0.5 MK, respectively, and the AIA 193 channel sees a hotter component at 20 MK. The AIA 193 and 211 \AA\ channels also see some cooler plasma, see \cite{leme12} for details. \cite{warr12} describe a method to remove the warm component from the AIA 94 channel. We have used this method to calculate ``hot 94" images.   

The Hi-C, IRIS, and SDO data are well aligned using SSW routines. Note that all corrections, including roll angle, as well as by manual fine tuning whenever required, were performed for alignment of Hi-C, IRIS, AIA images, and HMI magnetograms to about the spatial resolution of these images. HMI LOS magnetograms at a 45 s cadence are used to follow flux cancellation, emergence and/or the presence of mixed-polarity magnetic field. 

We have created three videos from Hi-C, IRIS SJ, and AIA images to track features over time and compare them in multiple wavelengths. To allow better tracking we have over plotted HMI LOS magnetogram contours of $\pm$25 G on each frame in these movies.
 
 We have created movies of the Dopplergrams from the spectral data of \MgII\ 2796, \SiIV\ 1400 and \CII\ 1330 \AA\ lines obtained with IRIS. Dopplergrams are intensity differences at fixed wavelength offsets (fixed Doppler-shift offsets) from line center in the blue and red wings of the line. For best visibility of redshift and blueshift our selected offsets are $\pm$50 \kms\ for \MgII\ 2796 \AA\ \citep[as in][]{depo14Science,tiw18}, and $\pm$25 \kms\ for \CII\ 1330 and \SiIV\ 1400 \AA\ lines (based on different trials in this work). These Dopplergrams show signatures of blueshift and/or redshift along the slit at the location where the slit cuts across the surge or other brightening event.  To suppress local fluctuations in the Dopplergrams, we have averaged Dopplergrams created by integrating the signal over a range of 10 \kms\ centered at around $\pm$50 \kms\ for \MgII, and around $\pm$25 \kms\ for \SiIV\ and \CII\ lines. The Dopplergrams for the \MgII\ line provide structure and dynamics (redshift and blueshift) of chromospheric plasma, whereas those for \SiIV\ and \CII\ lines provide structure and dynamics of transition region plasma \citep{depo14IRIS,depo14Science,tiw18}. 


\section{Results} \label{resu}

We identified 15 brightening events in the core of the AR observed by Hi-C, by combining Hi-C data with IRIS and SDO/AIA data. Based on different observed characteristics we assigned each event to one of three categories: type I -- dot-like transient brightening in Hi-C 172 \AA\ and AIA 171 \AA; type II -- transient elongated brightenings along small magnetic loops; and type III -- surge/jet -like transient eruptions with outflows often followed by inflows. Most of the observed properties i.e., lifetimes, visibility in AIA 94 \AA\ (or hot 94), the presence of mixed-polarity flux, flux convergence, measurable flux cancellation, flux emergence, field-guided flows (assuming all plasma flows and elongations in the UV and EUV images are along the magnetic field), Doppler flows in \MgII\ 2796 \AA\ (when IRIS slit covers at least a part of the event), and the presence/absence of underlying neutral line are listed in Table \ref{t1}.

\begin{deluxetable*}{cccccccccccc}
	\setlength{\tabcolsep}{1pt} 
	\tablenum{1}
	\tablecaption{List of 15 brightening events in the core of the NOAA AR 12712 caught by IRIS SJI, a few also by the IRIS slit, and/or by Hi-C 2.1. All these events are complemented by SDO (AIA and HMI) data in the present work. \label{t1}}
	\tablewidth{0pt}
	\tablehead{\colhead{Event no.,} & \colhead{Hi-C/AIA peak} & \colhead{Neutral} & \colhead{Hi-C} & \colhead{IRIS SJ/} & \colhead{Field-aligned} & Dopplergram &Flux& Meas. Flux\tablenotemark{j} & \colhead{Flux} & Lifetime\tablenotemark{b}  & Visibility in\\
		\colhead{type \& name\tablenotemark{a}} & \colhead{ time ($\pm$1 frame)} & \colhead{line} & \colhead{data}& \colhead{Spectra} & \colhead{flows\tablenotemark{g}} & feature & convergence & cancellation & \colhead{emergence} & (s) & AIA94/hot94}
	\startdata
	1. I (Dot 1)           & 18:58:32 & Yes             & Yes & Yes/No & No             & N/A                           & Yes & No       & No       & 83($\pm$10)  & Yes/Yes       \\ 
    2. I (Dot 2)  		   & 18:58:58 & Yes             & Yes & Yes/No & No             & N/A                          & Yes & Yes       & Yes      & 61($\pm$10)  & Yes/Yes       \\ 
	3. II (Loop 1)         & 18:56:47 & Yes             & Yes & Yes/No & Unidirectional & N/A                     & Yes & No       & No       & 117($\pm$10) & No/No        \\ 
	4. II (Loop 2\tablenotemark{h})&18:59:02&No         & Yes & Yes/No & Unidirectional & N/A          & No  & No       & No       &  35($\pm$10) &  Yes/Yes     \\  
    5. II (Loop 3)	       & 19:00:42 & Yes\tablenotemark{d}&Yes&Yes/No& Unidirectional & N/A          & Yes & No       & Not clear& 108($\pm$17) &  Yes/No      \\  
	6. II (Loop 4)         & 19:35:45 & Yes             & No  & Yes/No & Unidirectional & N/A                    & Yes & No       & No       & 60($\pm$24)  &  Yes/Yes     \\
    7. II (Loop 5\tablenotemark{c})&19:37:45&Yes        & No  & Yes/No & Not clear      & N/A            & Yes & No      & No       & 192($\pm$24) & Yes/No       \\ 
	8. II (Loop 6)         & 20:03:33 & Yes             & No  & Yes/No & Unidirectional & N/A                   & Yes & No       & Yes      & 192($\pm$24) & Yes/Yes      \\ 
	9. II (Loop 7)         & 20:25:21 & Yes             & No  & Yes/No & Unidirectional & N/A                    & Yes & No       & Yes      & 204($\pm$24) &  Yes/No      \\
	10. III (Surge 1)      & 18:38:33 & Yes             & No  & Yes/Yes& Unidirectional & blue\tablenotemark{i}&Yes& Yes & Yes      & 156($\pm$24) & Yes/Yes       \\ 
    11. III (Surge 2)      & 18:42:21 & Yes             & No  & Yes/Yes& Bidirectional  & red/blue               & Yes & Yes      & Yes      & 156($\pm$24) & Yes/No        \\
    12. III (Surge 3)      & 18:49:57 & Yes             & No  & Yes/Yes& Bidirectional  & red/blue              & Yes & Yes       & Yes      & 168($\pm$24) & Yes/Yes       \\ 
	13. III\tablenotemark{f} (Surge 4) & 19:01:56 & Yes & Yes & Yes/No & Bidirectional  & red/blue    & Yes  & Yes      & Yes      & 69($\pm$17)  & No/No         \\
	14. III (Surge 5)      & 19:07:33 & Yes             & No  & Yes/Yes& Unidirectional & red/blue            & Yes  & Yes      & Yes      & 132($\pm$24) & Yes/Yes\tablenotemark{e} \\ 
	15. III (Surge 6)      & 19:33:21 & Yes             & No  & Yes/Yes& Bidirectional  & red/blue              & Yes  & Yes      & Yes      & 120($\pm$24) & Yes/Yes       \\ 	
	\enddata
    \tablenotetext{a}{Event types I, II and III are described in Section \ref{resu}.}
	\tablenotetext{b}{The lifetime is calculated based on the appearance of the event in Hi-C 172 \AA, or when outside Hi-C duration, in AIA 171 \AA.}
   	\tablenotetext{c}{a 3-step event}
	\tablenotetext{d}{at the bright end}
	\tablenotetext{e}{there is a delay of 12-40 s in its appearance in AIA 94 \AA.}
	\tablenotetext{f}{This event looked very similar to type I in Hi-C images, but a careful inspection revealed plasma outflows from the bright dot-like location, therefore we moved it to type III.}
	\tablenotetext{g}{In surges by bidirectional flows we refer to when we see cool plasma outflow from the surge-base/source and then inflow as well after a while. We call the flow as unidirectional when we see plasma flow (or it could be intensity propagation, see e.g., \cite{depo17}) from one foot of the loop towards the other, e.g., when in surges we only see outflow near the base of a surge and do not see inflow (following outflow). Note the flows in AIA 171, 304, 211 \AA, and in IRIS 2796, 1400, and 1330 \AA\ movies.}
	\tablenotetext{h}{This loop-like event has no evidence of opposite-polarity field. This event shows clear outflow (or outward intensity propagation). It is ambiguous whether it should be considered a type III event instead of a type II event. }
    \tablenotetext{i}{In this case only outflow (blueshift) from the surge-base is captured by the IRIS slit. In other five surges both inflow (redshift) and outflow (blueshift) are captured by the IRIS slit.}
   \tablenotetext{j}{Measurable flux cancellation. In all the cases when flux convergence is ``yes" but measurable flux cancellation is ``No" there is prospective flux cancellation that cannot be reliably measured because the cancelling flux of either polarity cannot be isolated well enough.}
	\tablecomments{The uncertainties in the lifetimes are based on the temporal cadence, thus depending on the event caught by Hi-C, IRIS or AIA the uncertainties are smaller or larger. The maximum uncertainty estimated is from two image frames of each instrument.}
\end{deluxetable*}

We created three movies from Hi-C, IRIS and AIA images. The first movie ``hic\_iris\_sdo.mp4" contains eight panels: Hi-C 172, AIA 171, 304, hot 94, IRIS 2796, 1400, 1330 slit-jaw images and SDO/HMI line of sight (LOS) magnetograms, with LOS magnetic contours (of level $\pm$25 G) plotted on each image. Hot 94 was calculated by removing warm components from AIA 94 by using the method of \cite{warr12}. The second movie ``iris\_long.mp4" contains six panels: IRIS 2796, 2832, 1400, 1330 SJ images, AIA 171 images, and HMI LOS magnetograms with the magnetic contours over plotted on each frame as in the first movie. The third movie ``sdo\_long.mp4" contains six panels: AIA 171, 304, 193, 211, hot 94 images and HMI LOS magnetograms, with the magnetic contours over plotted on each image frame. While the first movie spans only the Hi-C observation time, the second and third movies are for about two hours, and two and an half hours, respectively, covering the five minutes of the Hi-C observations in their middle.  
We have also created a Dopplergram movie (``doppler.mp4") from the spectral rasters of IRIS for the \MgII\ k, \CII\ and \SiIV\ lines to check the Doppler flows in the covered parts of the events. 

\subsection{Type I -- Dot-like brightening events}\label{sec_t1}
\begin{figure*}[h]
	\centering
	\includegraphics[trim=0cm 2cm 0cm 8.8cm,clip,width=\textwidth]{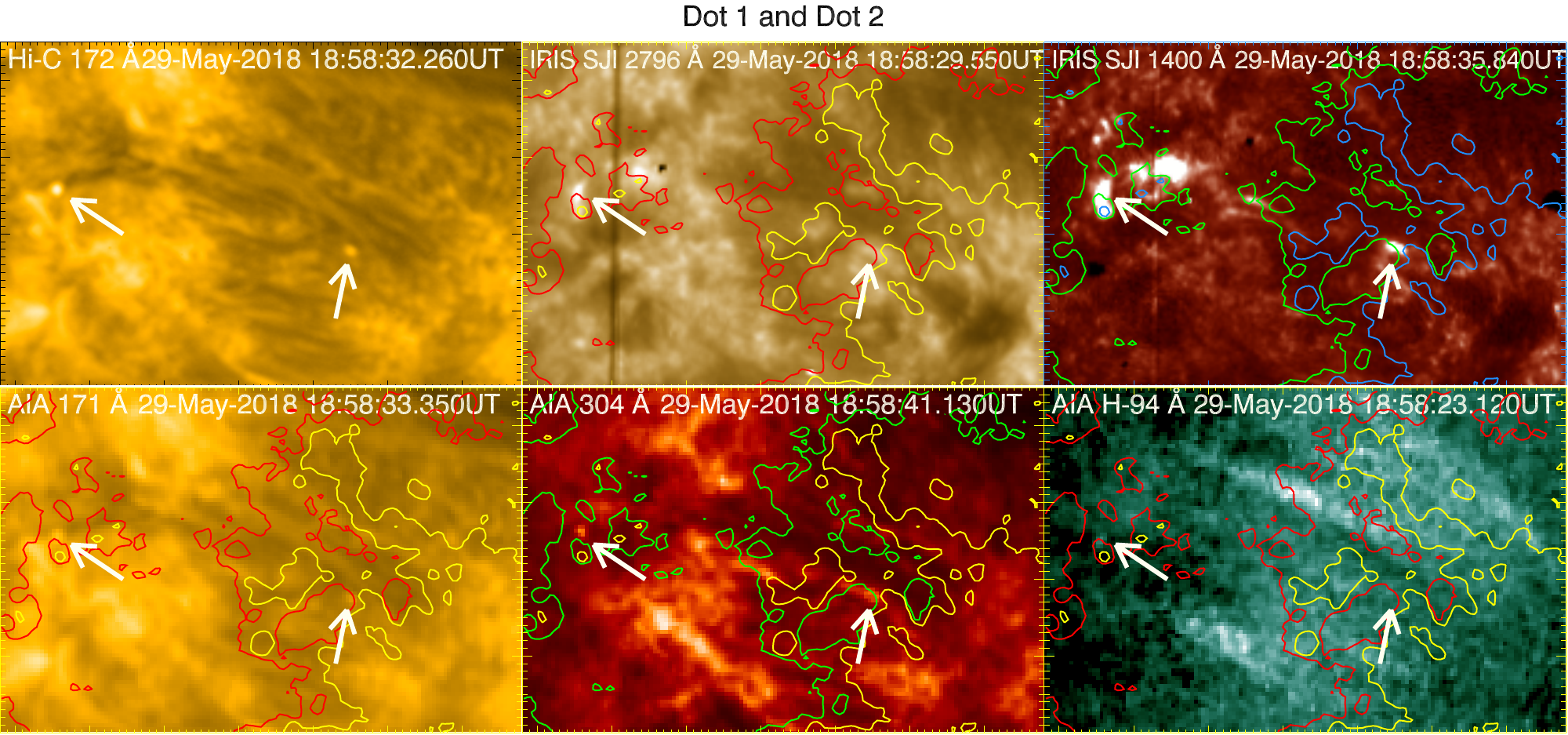}
	\caption{The two type I (dot-like) brightening events (Dot 1 in the right/Solar-West, and Dot 2 in the left/Solar-East), pointed to by arrows in the Hi-C image. The same locations are pointed to by similar arrows in the IRIS SJI 2796, IRIS SJI 1400, AIA 171, AIA 304, and AIA hot 94 \AA\ images. The red and yellow contours are for positive and negative LOS magnetic field at a level of $\pm$25 G. For better visibility red color is replaced by green for contours on IRIS SJI 1400 and AIA 304 \AA\ images, and yellow color is replaced by blue for contours on IRIS SJI 1400 \AA\ images.}
	\label{dot}
\end{figure*}

In Figure \ref{dot} we display the two dot-like round-ish events, listed in Table \ref{t1}, appearing in the same Hi-C frame. Although we display the image in Figure \ref{dot} for the time when both dots appear in the same frame, their peak brightness times, as listed in Table \ref{t1}, are slightly different. The dot on the right (in the solar West) is named Dot 1 as its intensity peaks slightly before the dot on the left (in the solar East), which is named Dot 2. 

\begin{figure}[h]
	\centering
	\includegraphics[trim=2cm 3.8cm 5.6cm 10.41cm,clip,width=\columnwidth]{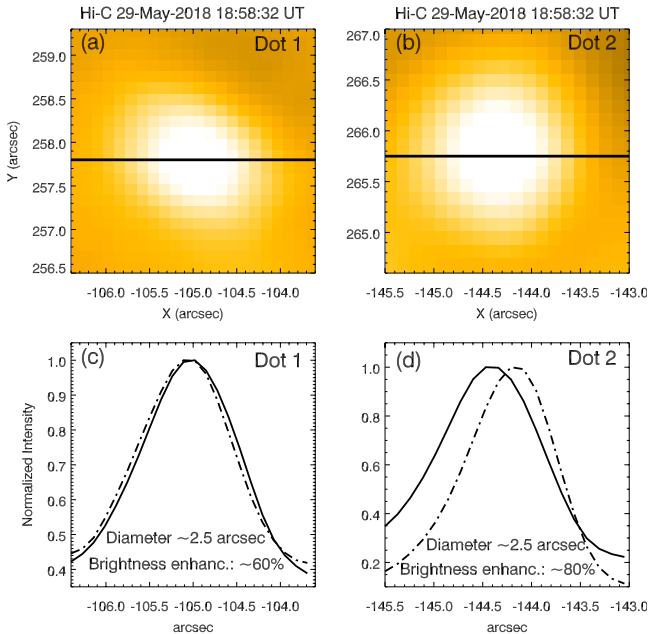}
	\caption{A close up look at the two Hi-C dot-like brightening events. The solid and dash-dotted lines in panels (c) and (d) are two plots at nearby different times (solid line at 18:58:32, dash-dotted line at 18:58:58) when both dots seem to be apparently brightest. Slight shift in the Dot 2 intensity most probably is a real shift in the location of the peak intensity of the dot in the given time difference, but might be due to smearing of the Hi-C data \citep{rach19}. Approximate diameter and brightness enhancement with respect to the background of the dots are also given. The diameter of dots ($\sim$2.5 arcsec) is several times larger than the diameter of Hi-C point spread function \citep[$\le$0.4 arcsec:][]{koba14}.}
	\label{dot_size}
\end{figure}

The size and brightening enhancement of each of the two dots are estimated and given in Figure \ref{dot_size}. The average diameter is 2.5 arcsec and intensity enhancement is 70\% with respect to the background. These numbers are similar to those for penumbral bright dots \citep{alp16}, EUV bright dots \citep{regn14}, and IBs \citep{pete14}.  

After we followed closely the Dot 2 event by combining IRIS movies with Hi-C we found that the base of Dot 2 event is located farther south, on the PIL of cancelling opposite-polarity magnetic field. A loop like structure extends towards the north from the PIL and Dot 2 in Hi-C 172 \AA\ images appears at the peak time of the loop in IRIS (see, e.g., 1400 \AA\ panel in Figure \ref{dot}). Therefore dot-like events apparently are closely connected to loop-like events, described later. However, the fact that the Dot 2 event is apparently driven from the cancellation PIL (an obvious plasma flow to the north from the PIL is seen in the IRIS 1400 \AA\ SJ movie), Dot 2 shares some properties of type III events, see Section \ref{sec_t3}.

\begin{figure*}[h]
	\centering
	\includegraphics[trim=0cm 13.6cm 0cm 2.2cm,clip,width=\textwidth]{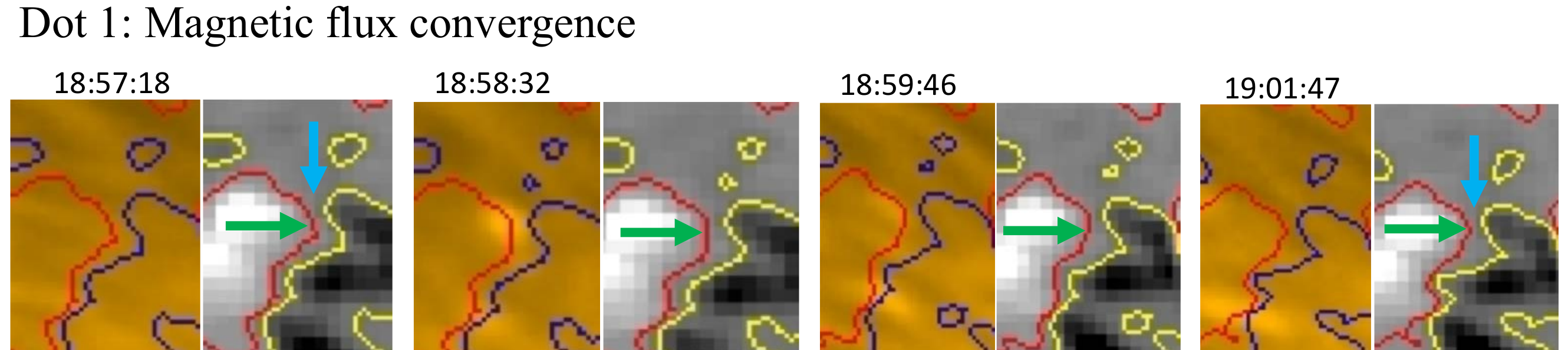}
	\includegraphics[trim=1.5cm 1.7cm 1.5cm 4cm,clip,width=0.4\textwidth]{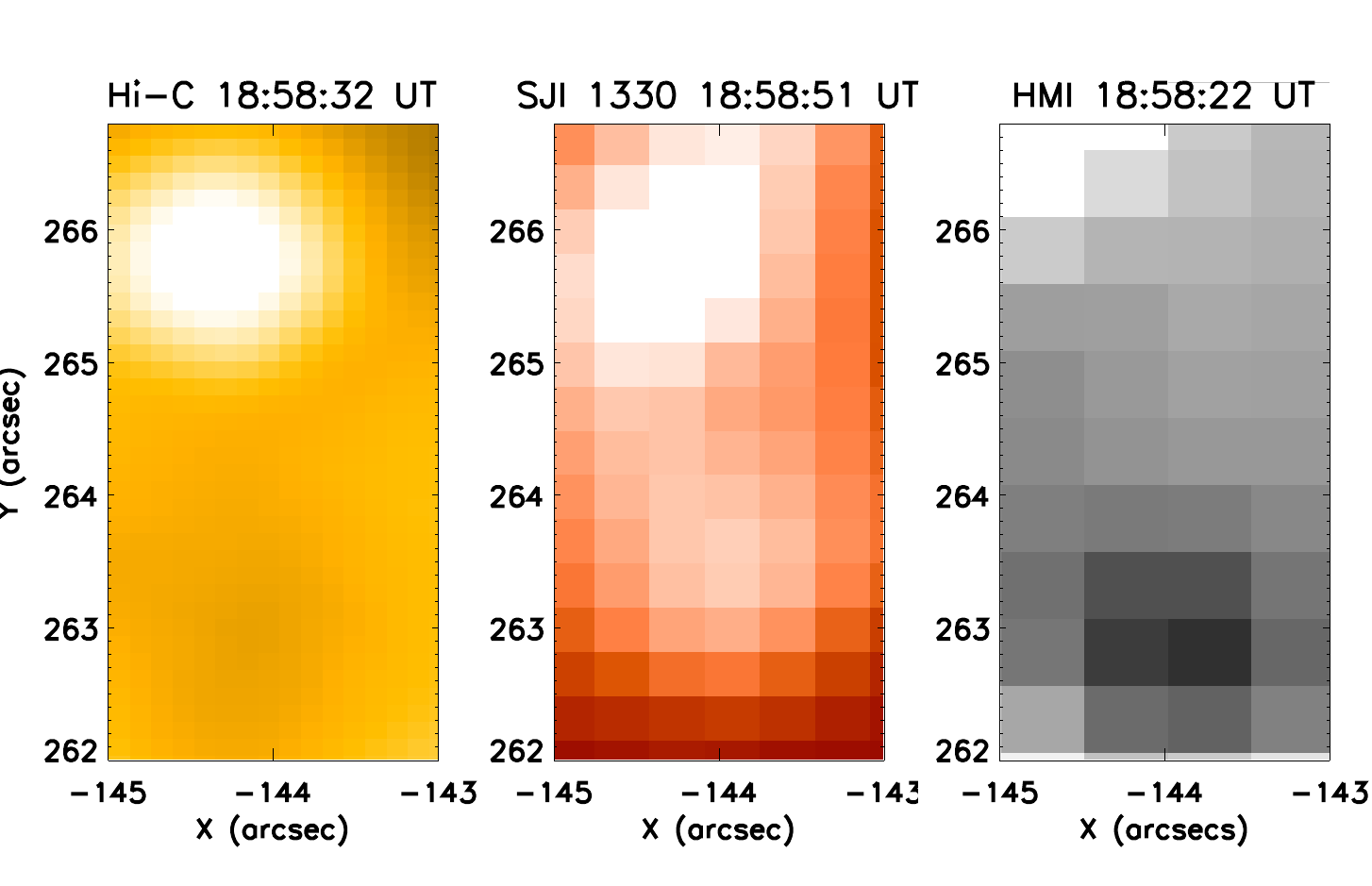}
	\put(-200,170){\Large Dot 2: Magnetic flux} \put(-200,150){\Large convergence and cancellation}
	\includegraphics[trim=0.7cm 0.4cm 0.6cm -1cm,clip,width=0.49\textwidth]{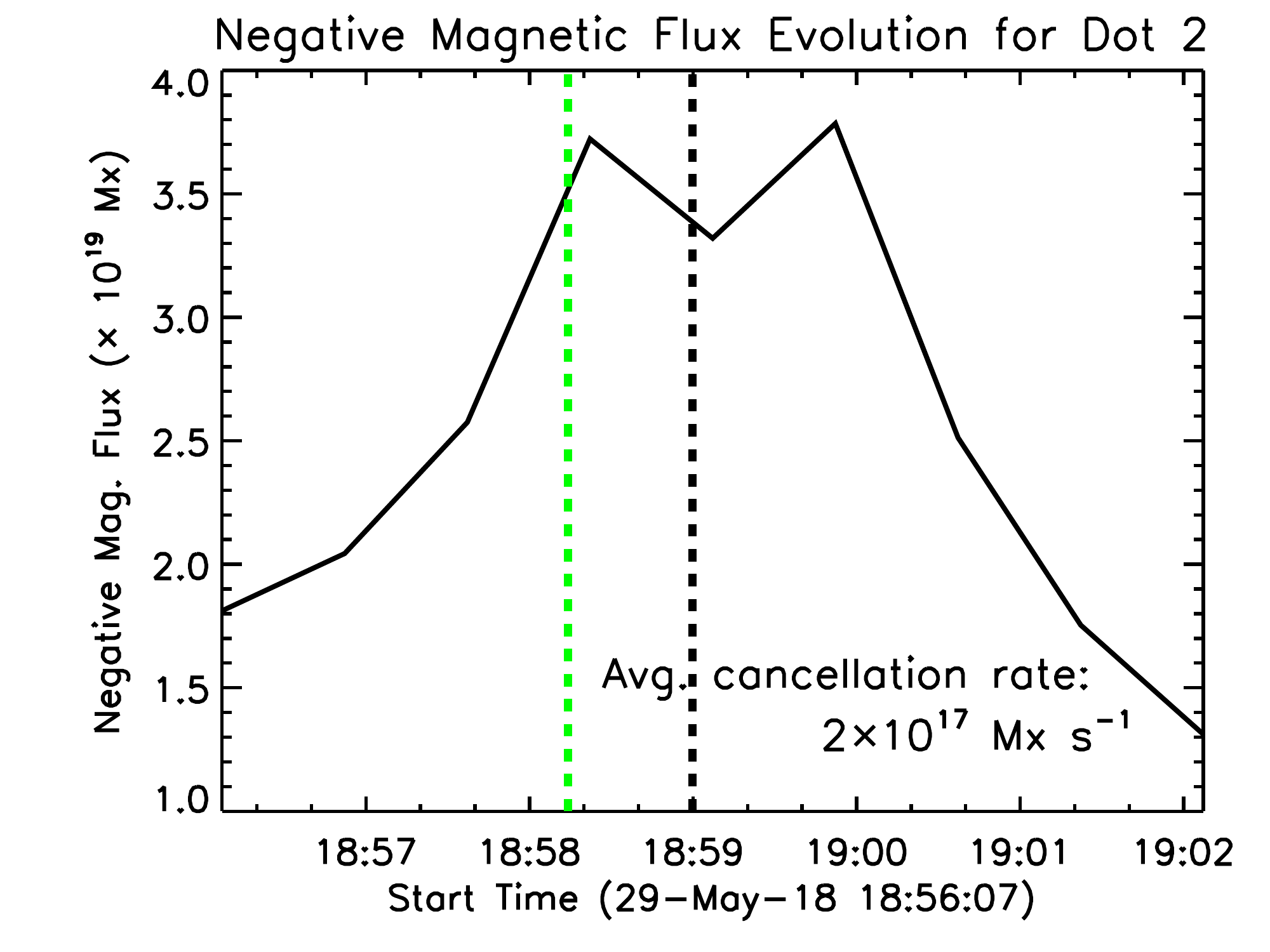}
	\includegraphics[trim=0cm 15.2cm 0cm 2.6cm,clip,width=\textwidth]{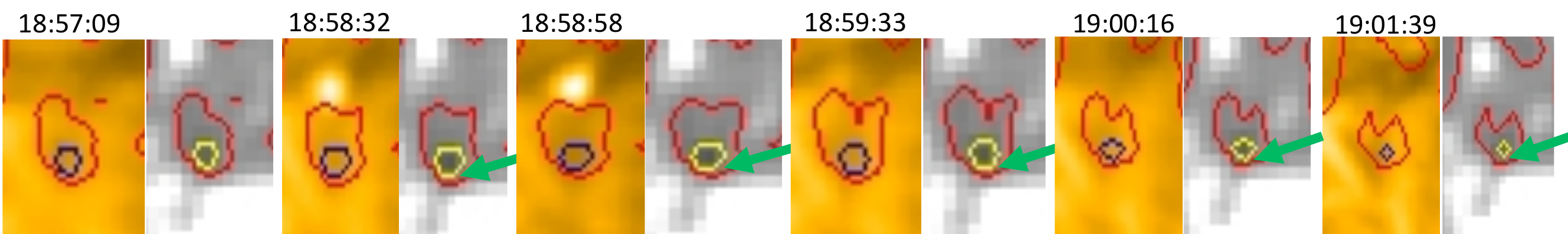}
	\caption{Images showing magnetic flux evolution for Dot 1 and Dot 2. A convergence can be noticed at the places pointed to by green and blue arrows in the top row for Dot 1. These are a small FOV taken from a few frames of the movie ``hic\_iris\_sdo.mp4" -- the convergence can be more closely followed in the movie. Calculation of flux cancellation rate is not reliably possible here due to difficulty in isolating the flux patch of interest. Probably convergence-driven cancellation that triggers Dot 1 is happening at the PIL, and triggers a few other fainter brightenings at the location of Dot 1 (see for example the faint brightening in the third frame of the Hi-C image at 18:59:46). Similar flux evolution for Dot 2 is displayed in the lowest row. Both flux convergence and flux emergence are visible in the images. The green arrows in the lower panel of stacked Hi-C and HMI images point to one of the locations where flux convergence is happening. The contours converge on the PIL south of the Dot 2 as the minority polarity decreases, that is, as the area of minority polarity flux encircled by its contour decreases. We also make a plot showing negative magnetic flux evolution for Dot 2, shown in the right of the middle row. The FOV used to calculate negative magnetic flux evolution is shown in the left three panels of the middle row. In the flux evolution plot the Hi-C peak time of Dot 2 is marked by a dashed black vertical line. The vertical green dashed line marks the time when the event starts appearing in IRIS SJI 2796 \AA. The flux is integrated over the area south of Dot 2 because from IRIS SJI 1400 and IRIS SJI 1330 images the southern part is evidently linked with the Dot 2 brightening -- IRIS SJI show that the Dot 2 is near the middle or foot of an extended loop that starts before and ends after the 172 \AA\ Dot 2 disappears. The flux cancellation rate is mentioned on the plot. Note that magnetic flux is emerging when the event Dot 2 is triggered -- probably emergence-driven cancellation is happening at the PIL.}
	\label{fcr_dots}
\end{figure*}

We show in Figure \ref{fcr_dots} the magnetic flux evolution of each of our dots. Careful inspection of contours of opposite-polarity magnetic field near each dot shows flux convergence at a ``sharp" neutral line (a PIL interval along which the positive-flux 25 G contour is within a few pixels of the negative-flux 25 G contour), marked by arrows in each case. In Dot 1 a small positive polarity flux patch, crossed by the green arrow, is cancelling at 18:58:32. The overall convergence continues along the neutral line afterwards; see the location pointed to by the blue arrow at 18:57:18 and 19:01:47. For Dot 2 a clear emergence of negative flux can be seen, which cancels with the encountered ambient majority positive magnetic flux on its south side. The location of its convergence with the positive polarity magnetic flux is marked by green arrows in lowest row of Figure \ref{fcr_dots}. Follow the evolution of these dots in the movie hic\_iris\_sdo.mp4. 
     
Because we could isolate the minority-polarity negative magnetic flux in the extended base of Dot 2 event, we made  a plot of the time evolution of flux in that negative patch (Figure \ref{fcr_dots}, middle right panel). Flux increase (emergence) is followed by flux decrease (cancellation). We estimate the flux cancellation rate to be 2$\times 10^{17} Mx~s^{-1}$. This and any other flux evolution rates that we have provided in this paper are crude (order of magnitude) estimates and should be taken with caution. With both emergence and cancellation happening at the same time, which is often the case in our present study, it is not possible to reliably estimate either the cancellation rate or the emergence rate from a flux-time plot.

\begin{figure*}[h]
	\centering
	\includegraphics[trim=1.5cm 1.4cm 1.2cm 2.4cm,clip,width=\textwidth]{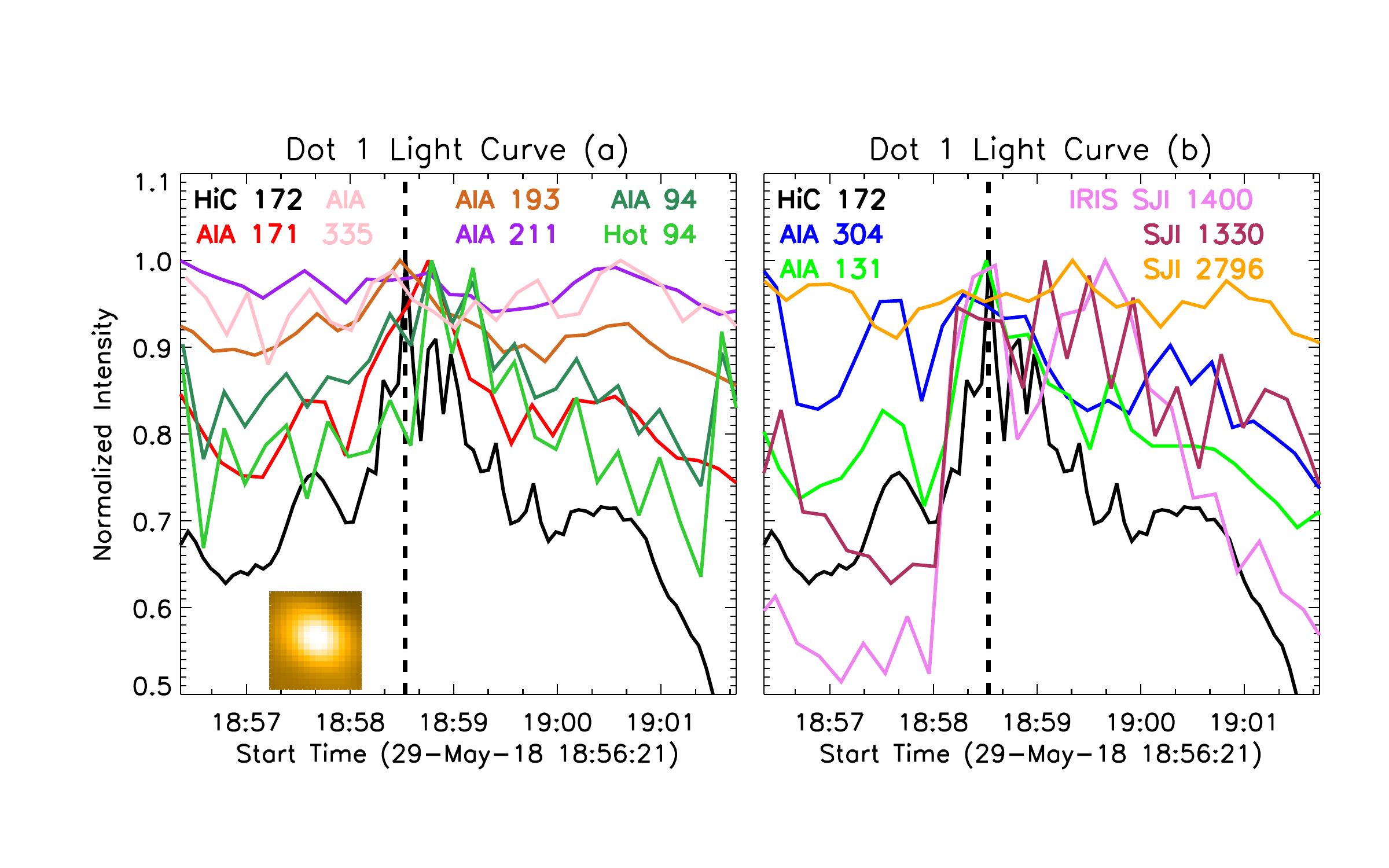}
	\includegraphics[trim=1.5cm 1.4cm 1.2cm 2cm,clip,width=\textwidth]{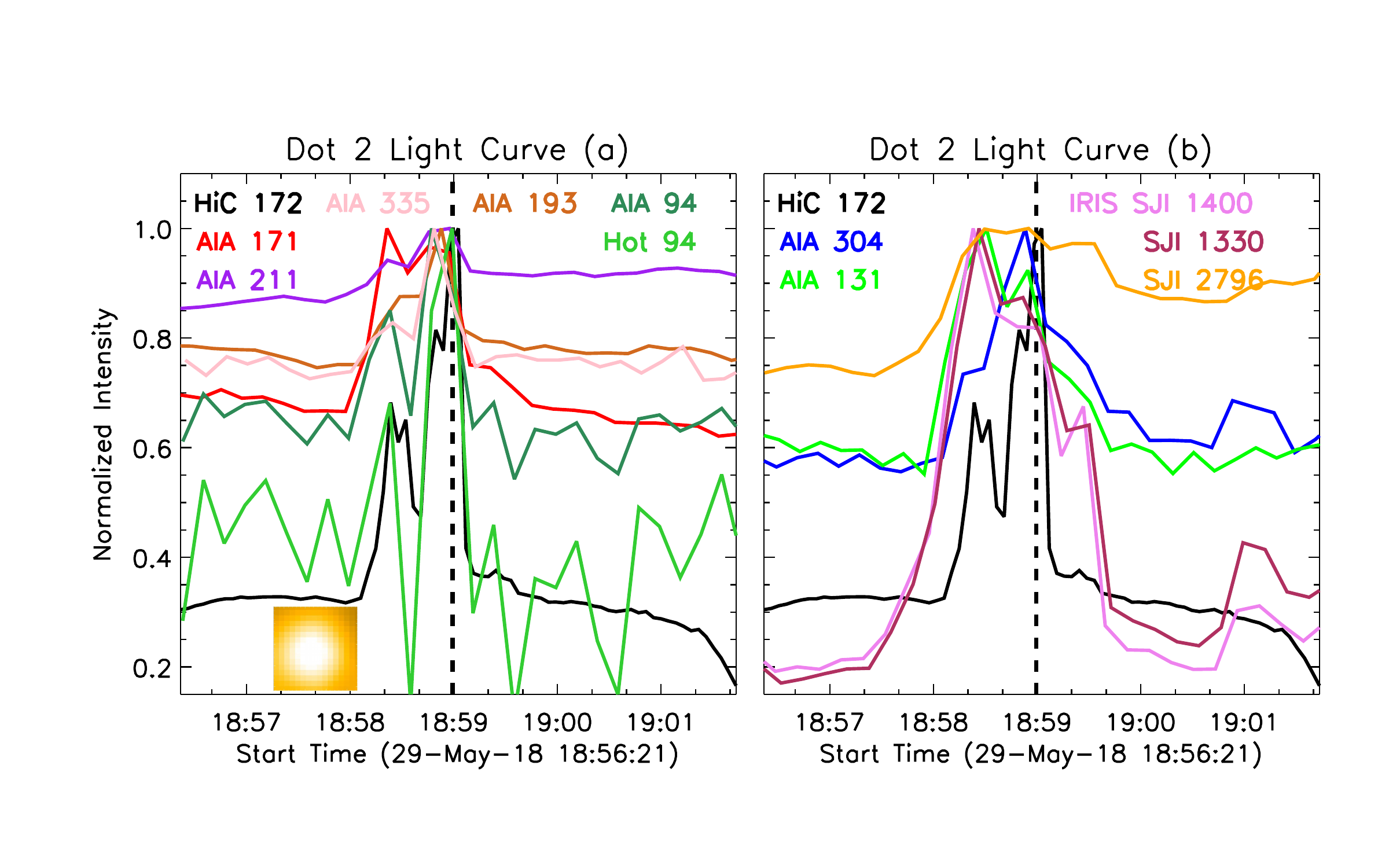}
	\caption{Light curves from Hi-C, AIA and IRIS intensity images over Hi-C time for the two type I (dot-like) events pointed to by arrows in Figure \ref{dot}. To avoid confusion due to overlaps, light curves for each dot (and for all other events discussed later) are plotted in two panels. The Hi-C area selected for making light curves is displayed as a small inset in the left panel for each dot during its peak intensity time in Hi-C. The vertical dashed line in each panel marks the peak time of the event in Hi-C 172 \AA, as listed in Table \ref{t1}.  }
	\label{lc_dots}
\end{figure*}

To investigate formation temperature of each Dot we made light curves of all AIA EUV and IRIS SJ wavelengths. In Figure \ref{lc_dots} we display light curves (intensity integrated over $\sim2\times2$ arcsec$^2$ $\sim16\times16$ Hi-C pixel$^2$ $\sim4\times4$ AIA pixel$^2$) from different AIA channels, IRIS wavelengths, and Hi-C images. The area for making light curves is selected during the peak intensity time, and is shown as insets on the light curve images (also true for type II and III events).

All of the AIA and IRIS light curves that peak for Dot 1 and Dot 2 peak nearly simultaneously in Figure \ref{lc_dots}. Hi-C 172 and AIA 171 have a double peak for Dot 2, which is compatible with similar two peak behaviours seen in several other wavelengths. This behaviour is similar to some of the EUV bright dots found in moss regions (at the edge of an AR) \citep{regn14}. We also calculated error bars (not shown here) for AIA 94 channel to verify the reliability of their light curves. In  particular, we verified that AIA 94 \AA\ intensity peaks are above noise, and are real. Although the light curves in AIA 94 peak slightly after Hi-C 172/AIA 171 in Dot 1, these are still near simultaneous.    

Although most light curves peak closely together for both dots, AIA 335, 211, 304, and IRIS SJ 2796 do not show significant peaks for Dot 1. Because there is no indication that the dots systematically appear in the hotter passbands (such as hot 94 or AIA 94 \AA\ shown in green) before the cooler ones (such as AIA 171 \AA\ shown in red) these events are different from a typical coronal flare. This behaviour of dot-like events (that the peak happens in all passbands at the same time without significant delays) is closely similar to the low-lying Hi-C 193 \AA\ loop nanoflare events studied by \cite{wine13}. These events are thus evidently at transition-region temperature.
For comparison, an example of a sub-flare is shown in Appendix \ref{app_lc_flare}, showing the typical cooling behaviour observed in coronal flares.  

Dot 1 does not show a response or peak in AIA 211 \AA, and Dot 2 shows only a weak response, probably because AIA 211 detects plasma emission from the overlying hotter atmosphere (at 2 MK) and has a response an order of magnitude lower to the 700,000 K plasma seen by the Hi-C 172 \AA\ filter and AIA 171 \AA\ filter. 

Note that, although hot 94 calculation works relatively well for hotter AR loops, it may not work so accurately for tiny, cooler events such as our Hi-C dots due to a rather complicated thermal response of AIA 94 \AA\ \citep[see e.g.,][]{schm11,asch11,delz11,fost11,test12_atomic,delz13}. Thus, the appearance of a dot in the hot 94 image may not suggest dot's true temperature. This caveat is also valid for type II and type III events, explored in next two subsections. 

We carefully inspected for any dark/bright plasma flows linked to type I events. We found no apparent outflows (plasma flowing away from dots) or inflows (plasma flowing towards dots) within either of these two dot events in the Hi-C 172 \AA\ and AIA 171 \AA\ images. The IRIS SJ images, however, show plasma upflow in the Dot 2 event, from the brightening (prospective magnetic reconnection) site south of the dot. 

Each event is more elongated in IRIS 1400 SJ images than in the Hi-C 172 and AIA 171 \AA\ images, with the bright dot seen in Hi-C 172 and AIA 171 being nearly in the middle of the elongated brightening seen by IRIS. A cartoon diagram depicting a possible formation mechanism of type I events is shown in Figure \ref{cartoon1}. We repeat that the ``dot-like" nature only applies to Hi-C 172 \AA\ or AIA 171 \AA\ images since IRIS (SJI 1400 and 1330 \AA) images show a loop-like feature (covering the Hi-C dot in the middle or slightly farther north of the feature). Therefore, as discussed later, the true magnetic structure of dots might resemble that of either type II events or type III events.

\subsection{Type II -- Loop-like events: Elongated brightenings in small magnetic loops}
\begin{figure*}[h]
	\centering
	\includegraphics[trim=0cm 2cm 0cm 8.8cm,clip,width=\textwidth]{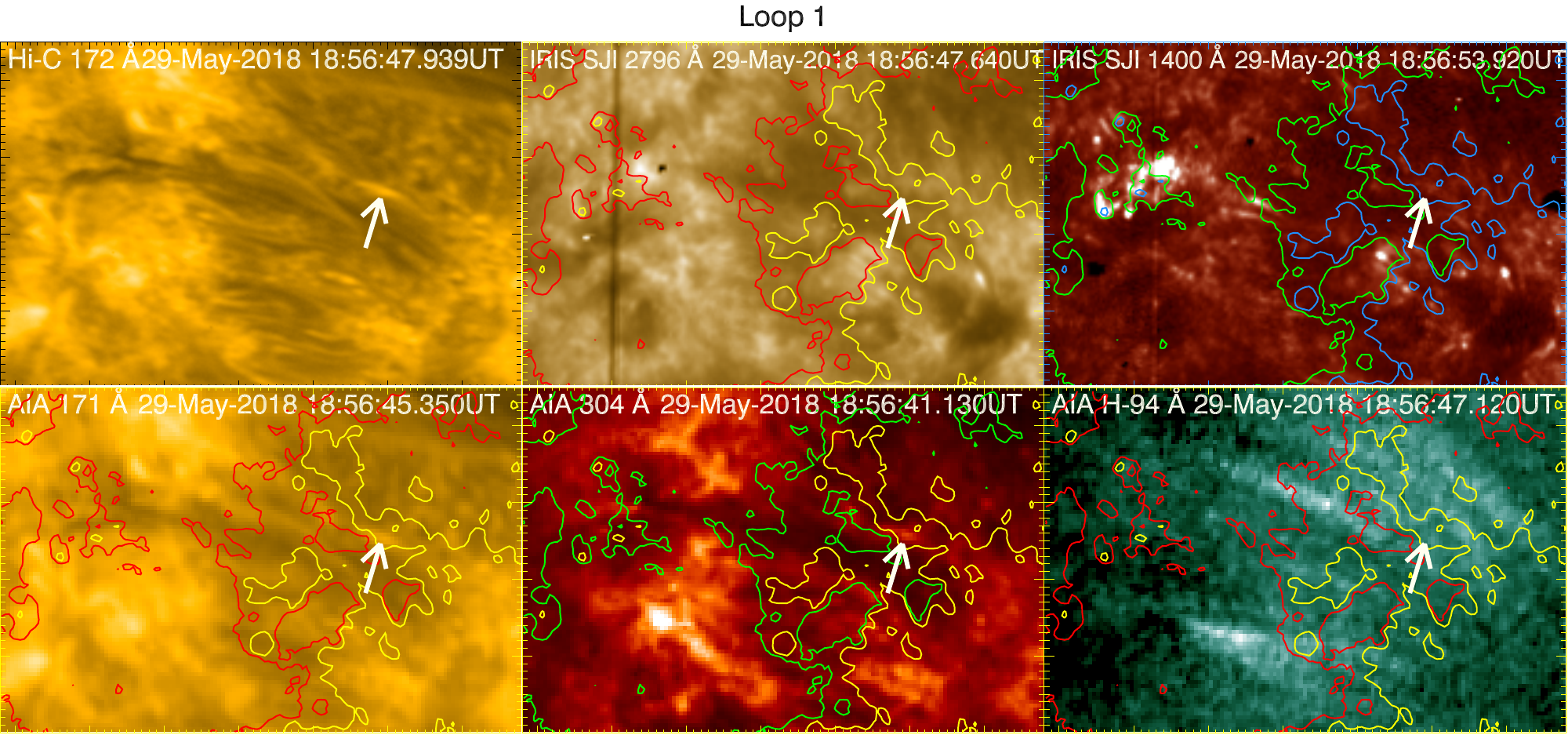}
	\includegraphics[trim=0cm 2cm 0cm 8.5cm,clip,width=\textwidth]{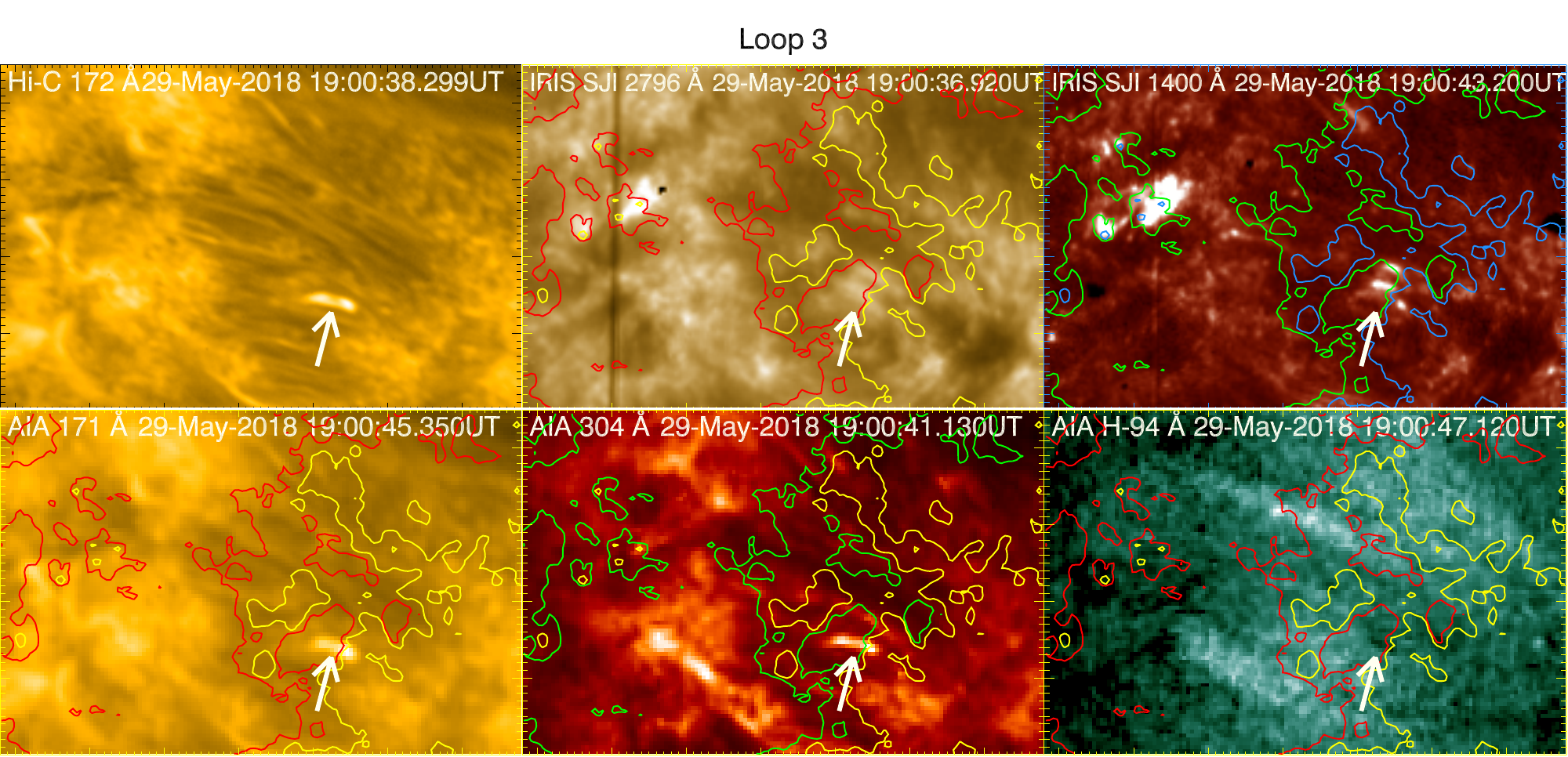}
	\caption{Two examples of type II (loop-like) events pointed to by arrows in the Hi-C images. These events are Loop 1 and Loop 3 in Table \ref{t1}. The location of the loop in each case is also pointed out by the white arrow in IRIS SJI 2796, IRIS SJI 1400, AIA 171, AIA 304, and AIA hot 94 \AA\ images. The red and yellow contours are for positive and negative LOS magnetic field at a level of $\pm$25 G. For better visibility red color is replaced by green for contours on IRIS SJI 1400 and AIA 304 \AA\ images, and yellow color is replaced by blue for contours on IRIS SJI 1400 \AA\ images.}
	\label{loop}
\end{figure*}

We noticed several brightening events that are elongated in Hi-C 172 and/or AIA 171 \AA\ images and look like small magnetic loops (see Table \ref{t1}). In most of these cases the brightening starts from one end and extends to the other end. Two of the loop-like brightening events are shown in Figure \ref{loop}. Other type II events can be noticed in the movies marked by arrows. Each event listed in Table \ref{t1} is marked by an arrow in Appendix \ref{app_all_events} (Figure \ref{all_events}).

\begin{figure*}
	\centering
	\includegraphics[trim=0.1cm 9.7cm 1.3cm 1.4cm,clip,width=\textwidth]{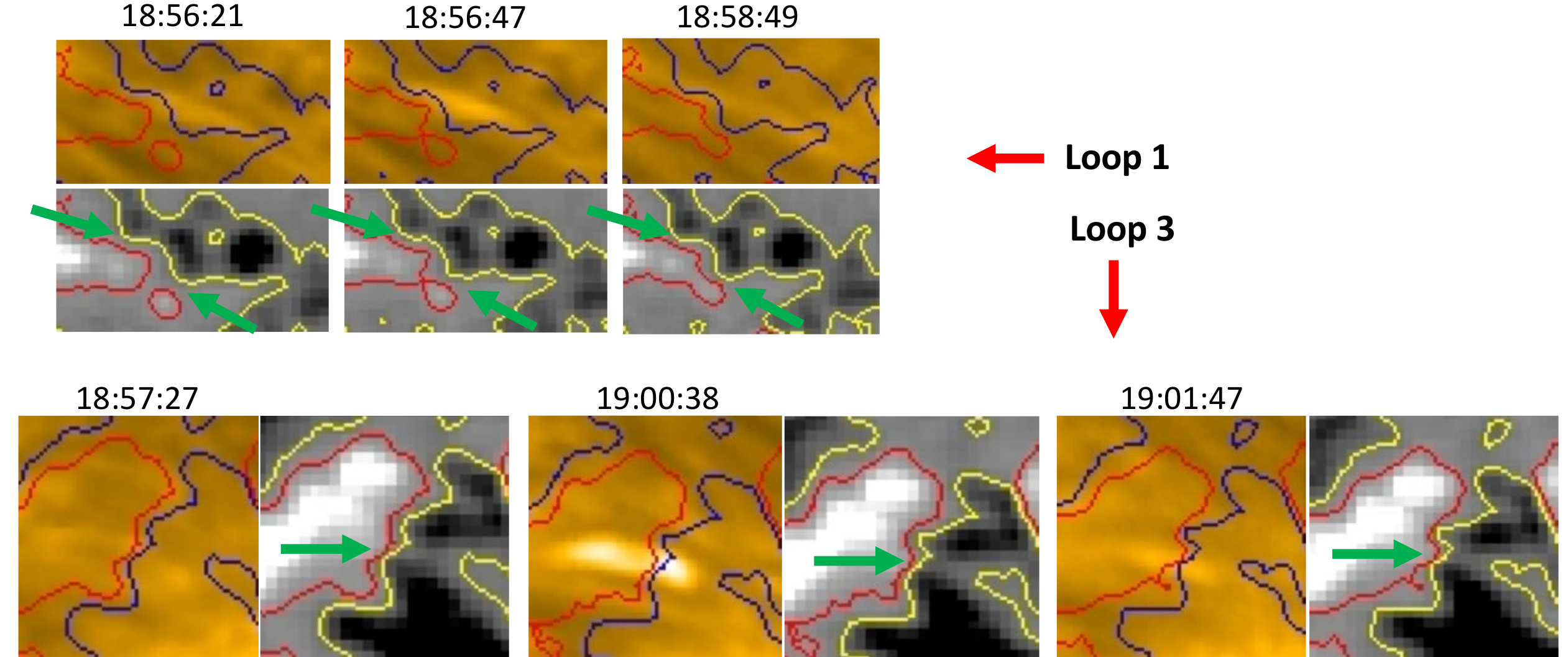}
	\caption{Three image frames of Hi-C images and HMI LOS magnetograms showing closely the magnetic evolution for Loop 1 (upper panel) and Loop 3 (lower panel). The green arrows point to the locations where flux convergence is more obvious, although convergence is happening all along the neutral line in longer time span (see the movies hic\_iris\_sdo.mp4, and sdo\_long.mp4). Contours are the same as in other images and movies.}
	\label{fcr_loop1+3}
\end{figure*}

The events (Loop 1 and Loop 3 in Figure \ref{loop}) are located on obvious sharp neutral lines, and the long AIA movie (sdo\_long.mp4) shows a trend of flux convergence in general over the time. We show in Figure \ref{fcr_loop1+3} flux convergence at the neutral lines of Loop 1 and Loop 3,  suggesting flux cancellation being involved in triggering these events. However it is difficult to isolate either of the magnetic polarities here and therefore a reliable estimate of magnetic flux cancellation rate is not possible in these cases. The same is true for the other loop-like events. Therefore, we can only infer the possibility of flux cancellation in these cases in which the cancelling flux cannot be isolated well enough to reliably measure the amount of flux cancellation.

Interestingly, the Loop 2 (at 18:59:02) does not show a neutral line in the $\pm$25 G level contours. Nonetheless because the Loop 2 event is a `flare-like' explosive energy release (similar to all of our events) the magnetic flux presumably has a neutral line.  In any case, other mechanisms (than flux cancellation) are possible in each of our type I and type II events. One such mechanism could be the convective driving of braiding from the feet of the loop leading to the event \citep{parker83a,parker88,tiw14}. Another possibility is that braiding from the feet built up the free-energy in the loop and then the event was triggered by waves produced from photospheric convection and p-mode oscillations \citep{ning04,mori04,chen06,hegg09}. Alternatively, wave dissipation without the presence of any braiding in the loops can also lead to transient heating events \citep{oste61,heyv83}.  

Although proposed for coronal heating in quiet solar regions, flux tube  tectonics heating model by \cite{prie02} may be equally valid in the closed loop system of the AR core. Any lateral motions of the surface magnetic flux in such a closed loop system can drive transient heating in the chromospheric/coronal separatrix surfaces of current sheets by fast reconnection (or in a turbulent manner, see e.g., \cite{zank18}).

Because there is a hint of weak negative magnetic flux (below $\pm$25 G) at the right/West end of this event (Loop 2), we can not rule out the possibility of this event being a very tiny surge-like event (type III), discussed in Section \ref{sec_t3}. 

\begin{figure*}
	\centering
	\includegraphics[trim=1.5cm 1.4cm 1.2cm 2.4cm,clip,width=\textwidth]{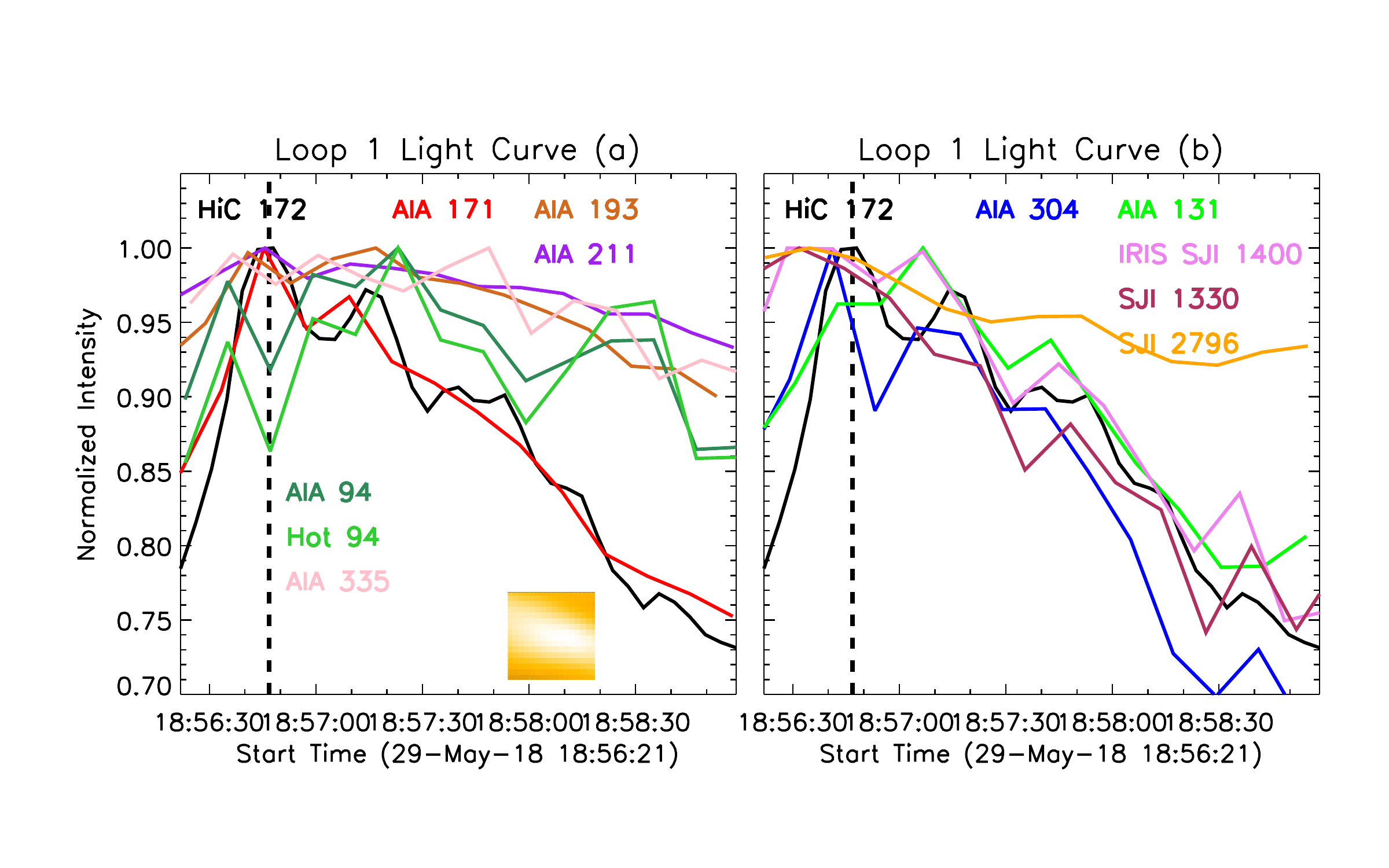}
	\includegraphics[trim=1.5cm 1.4cm 1.2cm 2cm,clip,width=\textwidth]{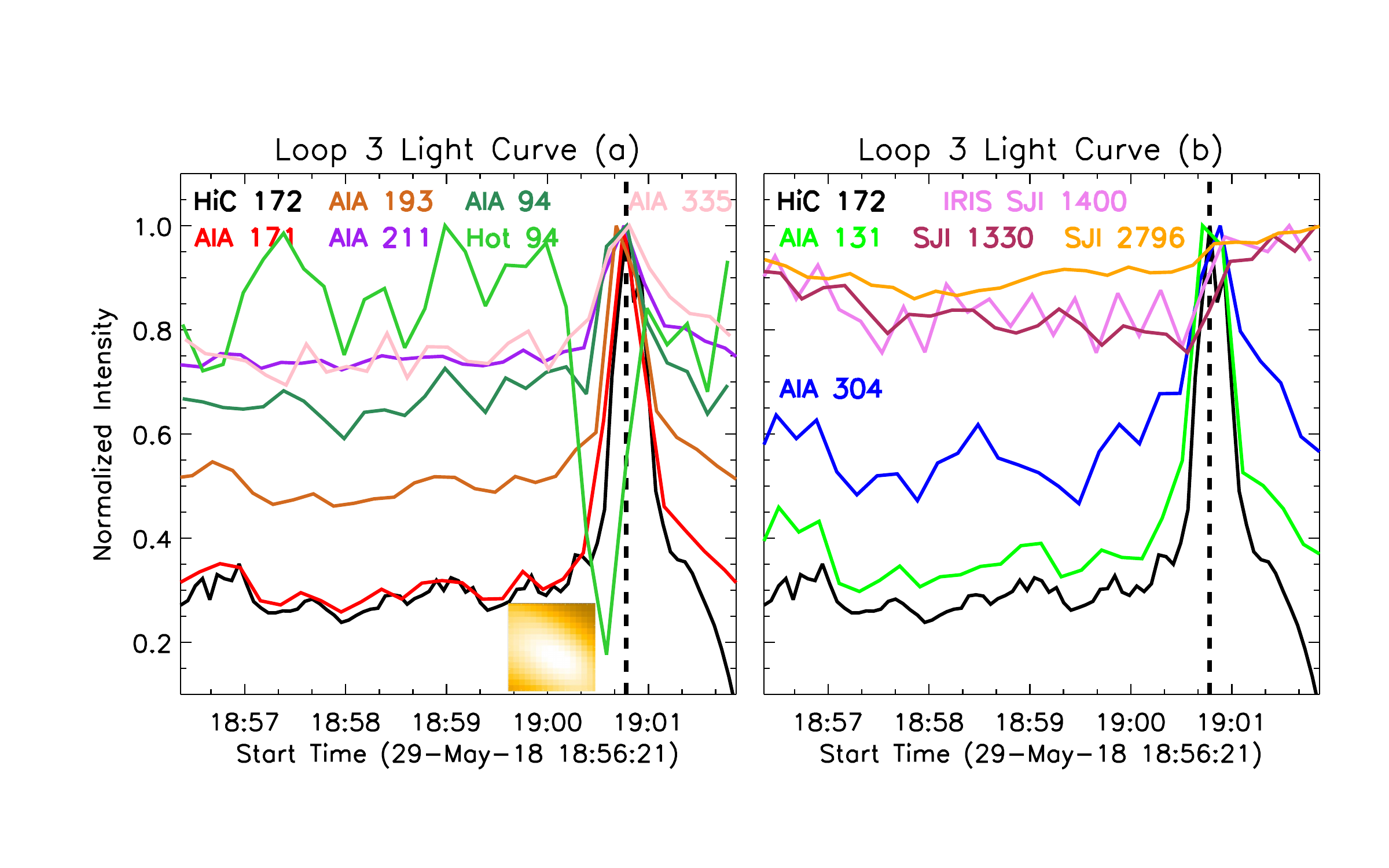}
	\caption{Light curves from Hi-C, IRIS slit-jaw (SJ), and AIA images during Hi-C time for two type II (transient bright loop) events, namely Loop 1 and Loop 3, pointed to by arrows in Figure \ref{loop}.  The Hi-C area selected for making light curves is displayed as a small inset in the left panel for each loop during its peak intensity time in Hi-C. The peak times of the events in Hi-C are marked by vertical dashed lines. The light curves for Loop 1 peak in the early phase of the Hi-C observing period. Due to the integrated area of the SJI covering a few (dark) pixels from a dust patch some of the IRIS light curves show repeated fluctuations.}
	\label{lc_loops}
\end{figure*}

In Figure \ref{lc_loops} we display light curves (intensity integrated over $\sim2\times2$ arcsec$^2$ $\sim16\times16$ Hi-C pixel$^2$ $\sim4\times4$ AIA pixel$^2$) of Loops 1 and 3 in different AIA channels, IRIS wavelengths and Hi-C images. Light curves for other loops listed in Table \ref{t1} can be found in Appendix \ref{app_lightcurves}. Most of the plots peak nearly simultaneously. Note that neither hot 94 nor any of the IRIS light curves show a peak in the Loop 3 event, which is a particularly prominent loop. The absence of Loop 3 in IRIS SJI and in hot 94 together suggests this loop forms in the transition-region. Consistently, \cite{pete19} show that in this event there are simultaneous cool loops in IRIS but they are not co-spatial -- there is a small  but significant offset between the warm/hot Hi-C loop and cool IRIS loops, as can be seen in Figure \ref{loop}.     

We do not find the systematic cooling sequence in any of the loop events (an example of such sequential cooling in a small flare is shown in Appendix A). Thus similar to type I events, type II events also do not behave like coronal flares, and are cooler (of transition-region origin).     

In most of type II events brightening starts from one end and moves to the other end. In a few cases both outflow (plasma flowing away from the bright end) and inflow (plasma flowing towards the bright end) signatures of hot (bright in most channels) plasma can be noticed. However it is difficult to conclude from the images if these are plasma flows or only apparent motions (e.g., heating fronts, propagation of shocks, etc). 

We explain via a cartoon in Figure \ref{cartoon1} how type I and type II could be similar in magnetic configuration and reconnection, and how type Is could be either symmetrically heated (true dots) or actually asymmetrically heated as in type II, depending on the visibility of heated loops.

\subsection{Type III -- Surge-like eruptions}\label{sec_t3}
\begin{figure*}
	\centering
		\includegraphics[trim=0cm 2cm 0cm 8.8cm,clip,width=\textwidth]{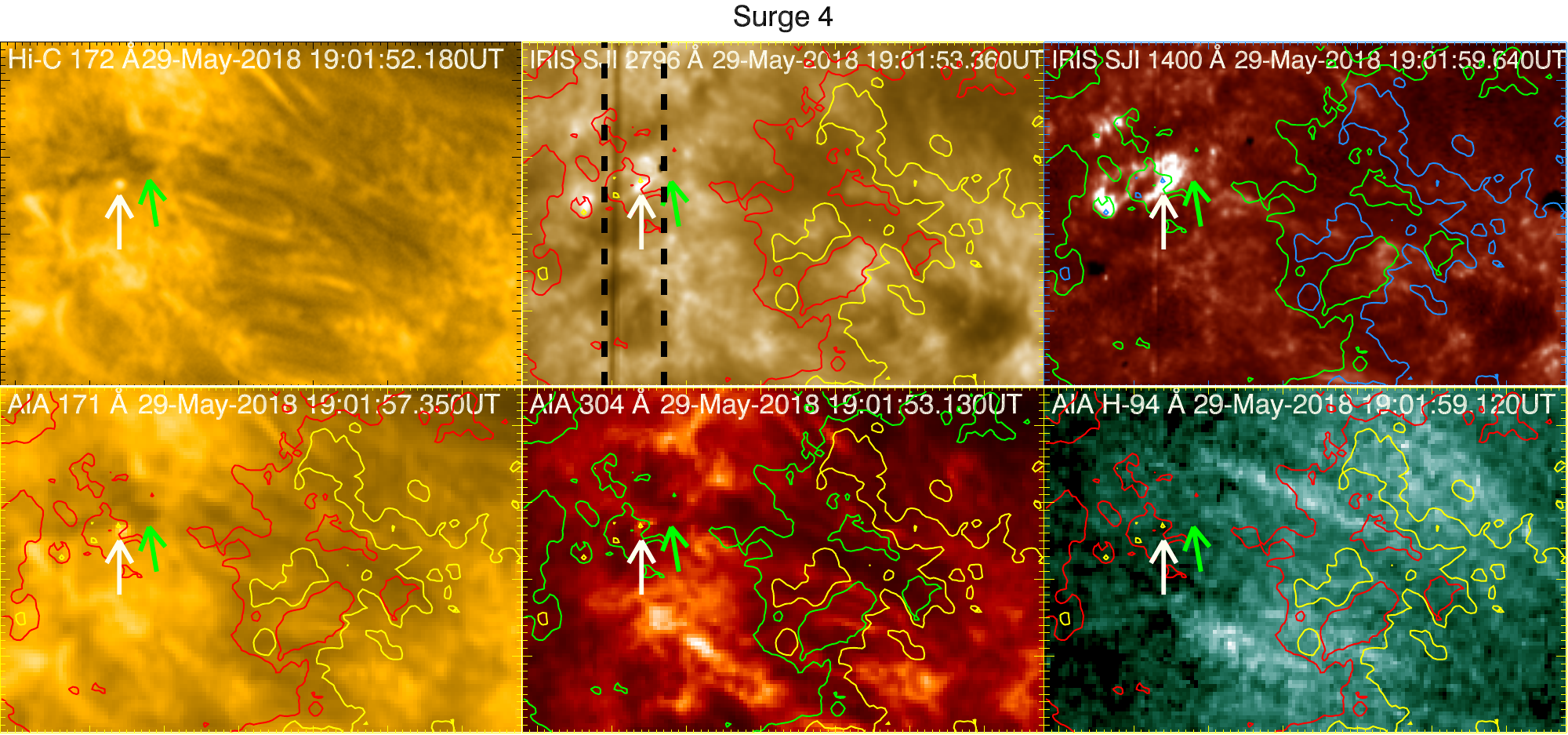}
	\includegraphics[trim=0cm 2cm 0cm 8.5cm,clip,width=\textwidth]{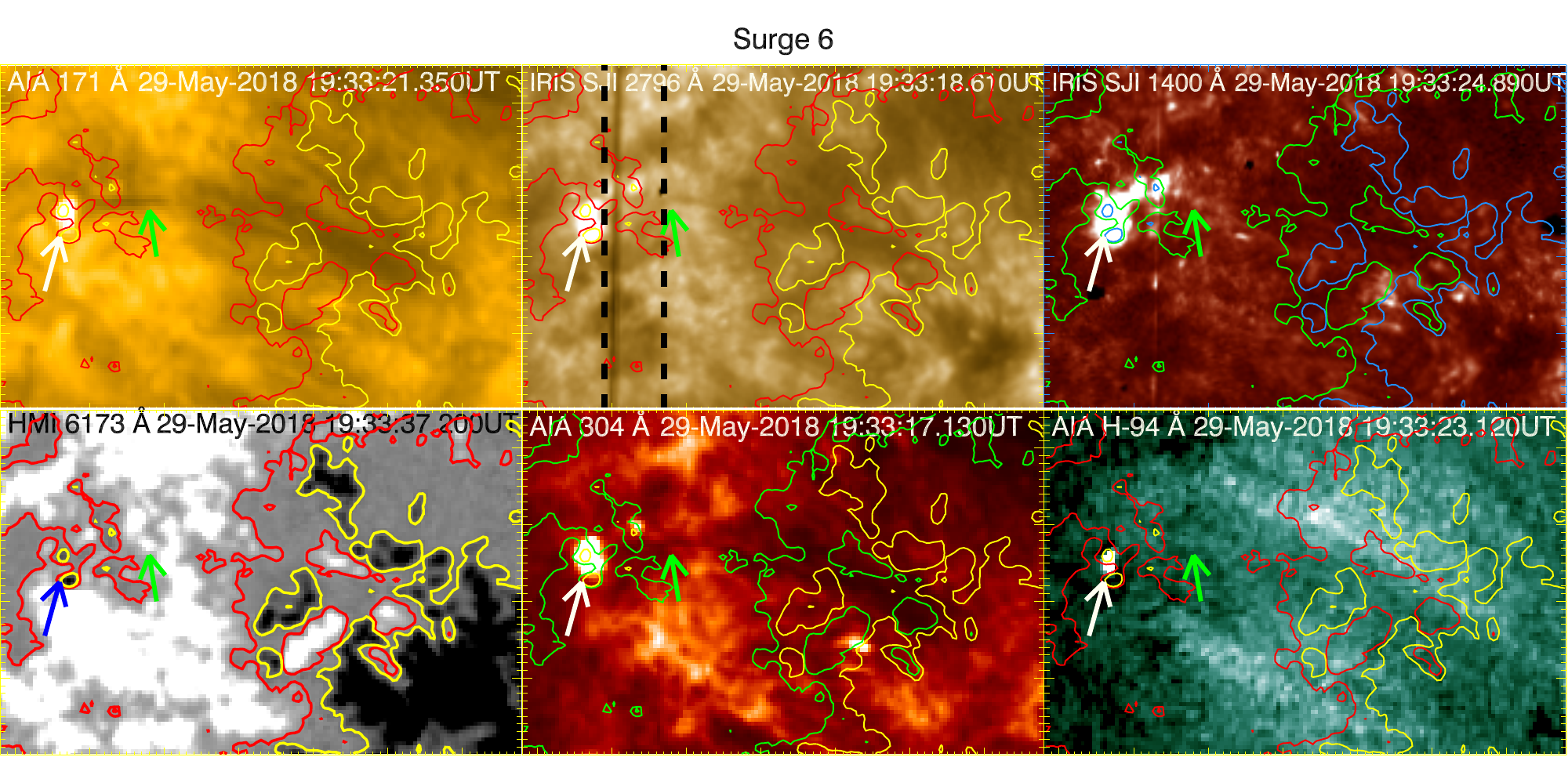}
	\caption{Two examples of type III events (Surge 4 and Surge 6) each pointed to by white arrows in each panel: Hi-C, IRIS SJI 2796, IRIS SJI 1400, AIA 171, AIA 304, and AIA hot 94 \AA\ images. Note that Hi-C data is not available for Surge 6 and HMI LOS magnetogram is used for a panel, instead. Green arrows point to the cool plasma (most clearly visible in AIA/Hi-C 171/172 \AA) shooting outward. The red and yellow contours are for positive and negative LOS magnetic field at a level of $\pm$25 G. For better visibility red color is replaced by green for contours on IRIS SJI 1400 and AIA 304 \AA\ images, and yellow color is replaced by blue for contours on IRIS SJI 1400 \AA\ images. Two vertical dashed lines in SJI 2796 mark the East-West boundary of IRIS slit scans, Dopplergrams for which are available as a video -- doppler.mp4.}
	\label{surge}
\end{figure*}

We found another (third) dot-like event in the Hi-C images (at 19:01:56 UT; named as Surge 4 in Table \ref{t1}), but careful inspection showed cool plasma outflow from the event followed by a weak inflow. This behaviour is similar to a weak surge/jet activity. When we followed AIA and IRIS movies in time beyond the Hi-C time range, in the core of the Hi-C AR, we found several other surge-like activities in that plasma shoots up, travels along a long loop, and then sometimes drains back. The IRIS slit has covered the shooting end of these events in some cases allowing us to create Dopplergrams to detect the flows. Two example type III events are shown in Figure \ref{surge}. The base of the surge-like activity is pointed to by a white arrow in each panel. A cool plasma outflow from the base is also marked by a green arrow in each panel -- plasma outflow is most clearly visible in Hi-C 172, AIA 171, and AIA 304. The surge extends to the right and then (more clearly visible in AIA 304, 171, and 211) drains back towards the base (see movies sdo\_long.mp4, and iris\_long.mp4).

\begin{figure*}
	\centering
	\includegraphics[trim=1.5cm 1.9cm 1.5cm 4.6cm,clip,width=0.5\textwidth]{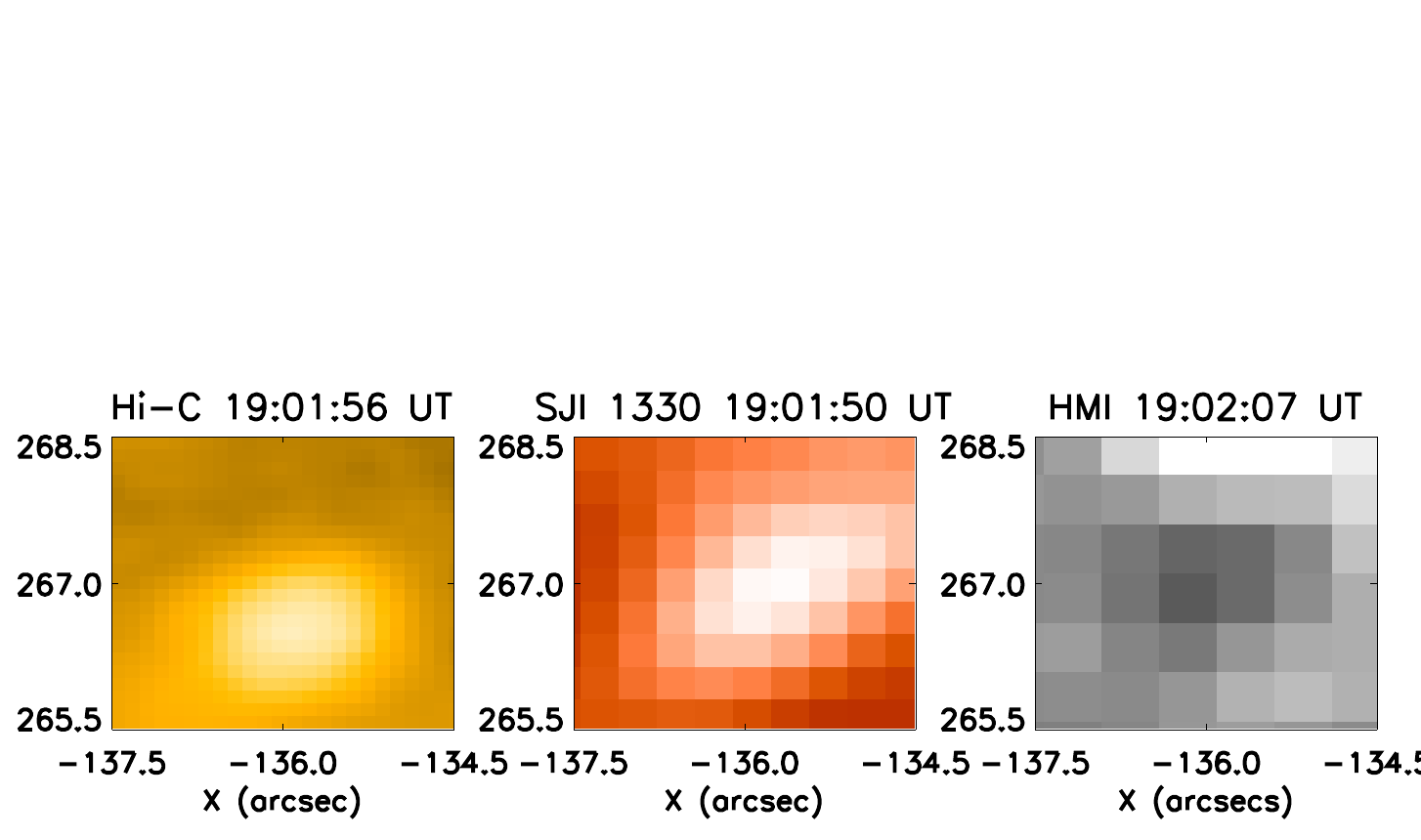}
	\put(-200,175){\Large Surge 4: Magnetic flux} \put(-200,150){\Large convergence and cancellation}
	\includegraphics[trim=0.7cm 0.4cm 0.6cm 0cm,clip,width=0.49\textwidth]{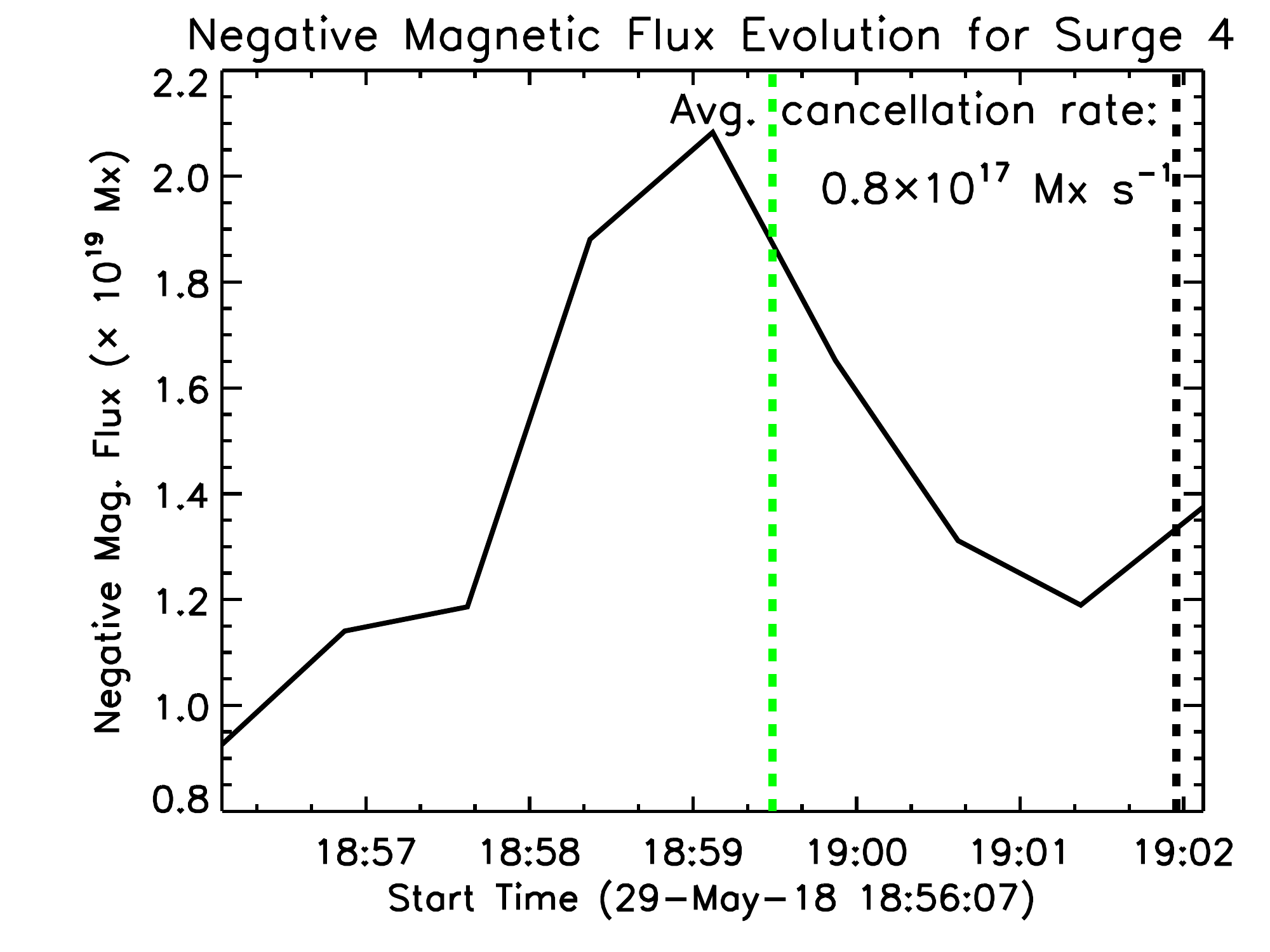}
	\includegraphics[trim=0cm 14.8cm 0cm 2.6cm,clip,width=\textwidth]{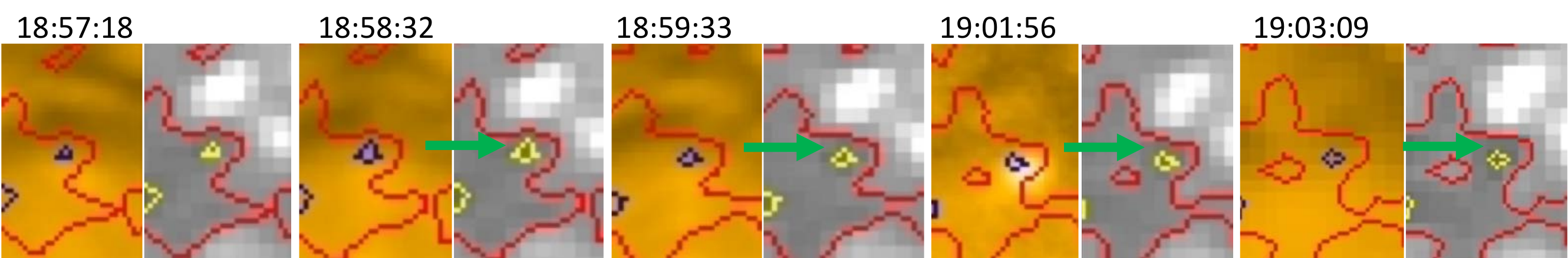}
	\includegraphics[trim=0.3cm 2cm 0.38cm 4cm,clip,width=0.5\textwidth]{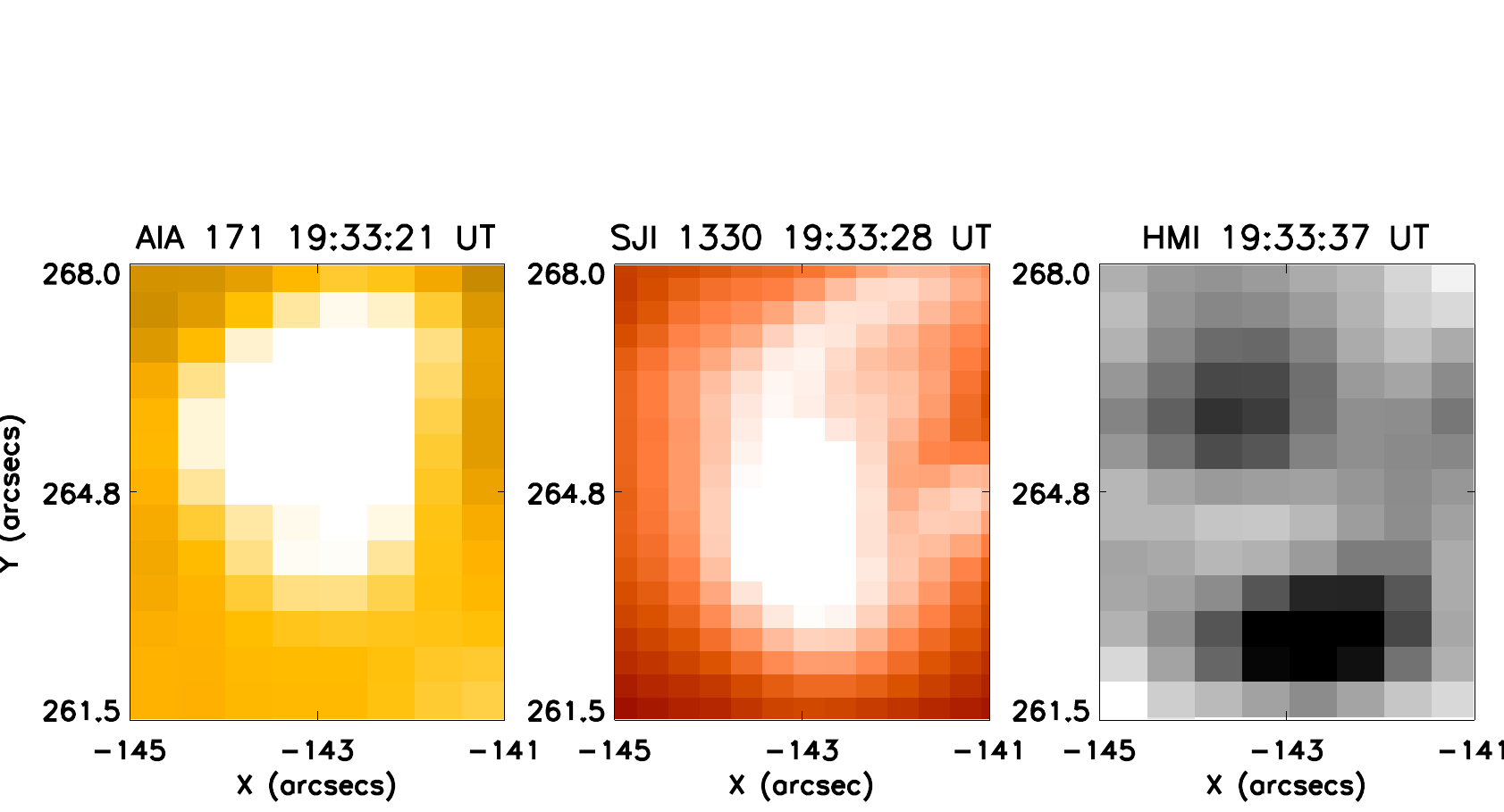}
	\put(-200,175){\Large Surge 6: Magnetic flux} \put(-200,150){\Large convergence and cancellation}
	\includegraphics[trim=1cm 0.4cm 0.6cm -1cm,clip,width=0.49\textwidth]{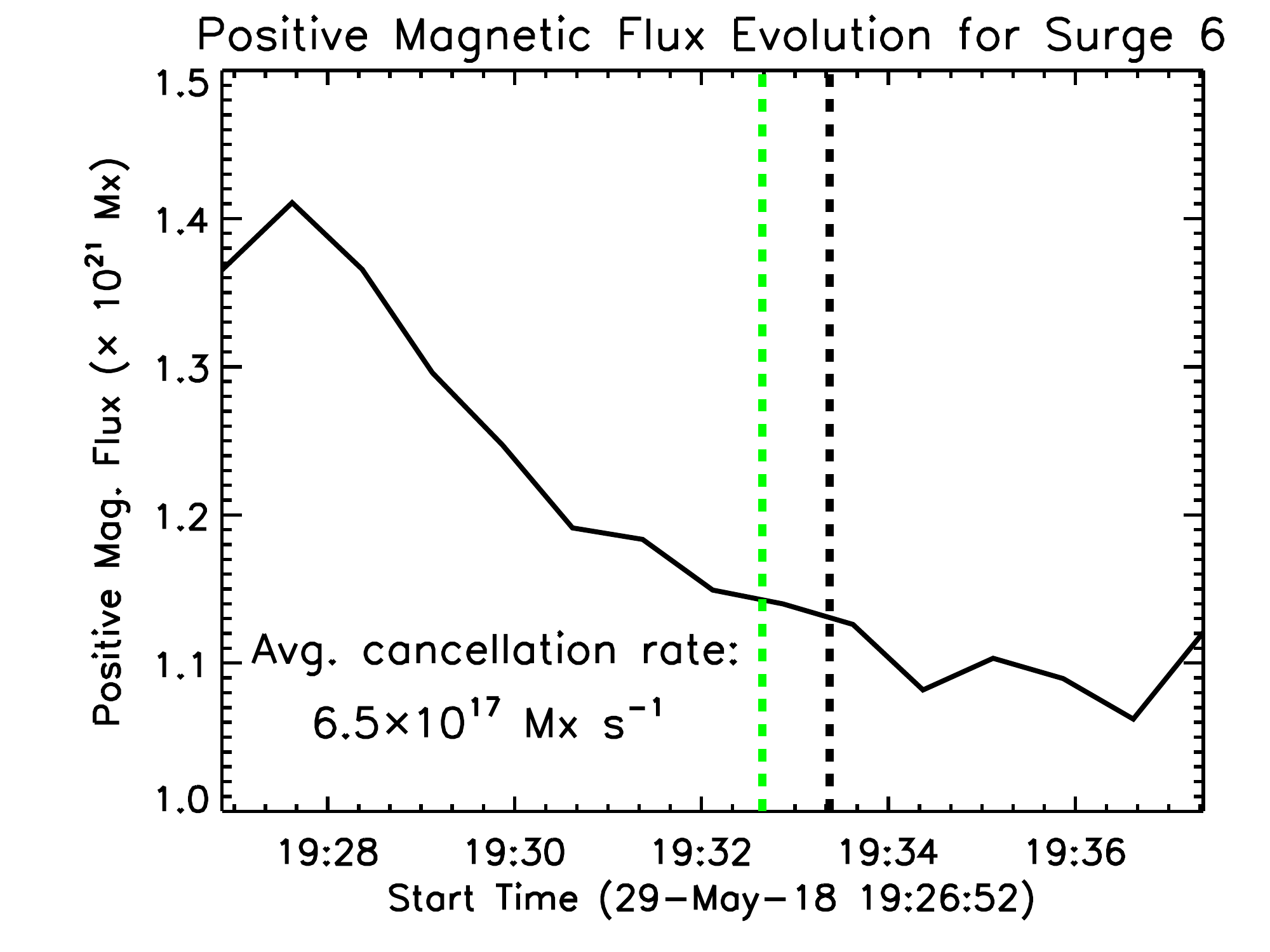}
	\includegraphics[trim=0cm 14.1cm 0cm 3.1cm,clip,width=\textwidth]{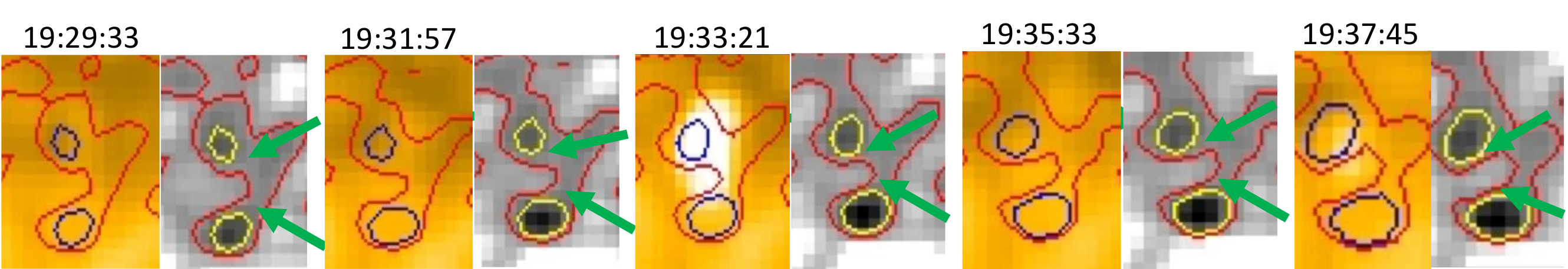}
	\caption{Magnetic flux convergence and cancellation in Surge 4 and Surge 6. Small FOV covering the base of Surge 4 and Surge 6 (Hi-C, SJI 1330, and LOS magnetogram) are shown in the upper left panels for both Surge 4 and Surge 6 -- same FOV is used to calculate flux evolution plots (negative flux for Surge 4; positive flux for Surge 6) shown in the upper rightmost panels for each of these surges. The peak time of the event is marked by a dashed black vertical line. The vertical green dashed line marks the time when the event starts appearing in SJI of \MgII\ 2796 \AA. The emergence, convergence and cancellation are also visible for each event in the movie hic\_iris\_sdo.mp4. The flux cancellation rate is mentioned on the plots. Evidently magnetic flux emergence (and convergence, see contours at neutral lines marked by green arrows) - driven cancellation at the PIL triggers these events.}
	\label{fcr_surge4+6}
\end{figure*}

In all type III events the presence of mixed-polarity flux, flux emergence, convergence and cancellation are clearly visible. In Figure \ref{fcr_surge4+6} we show flux emergence, and convergence-driven cancellation taking place in Surge 4 and Surge 6.  Although convergence-driven flux cancellation seems to be clearly responsible for triggering some type III events, there are some clear examples of type III events happening during the emergence of the minority polarity flux. We show in Appendix \ref{app_flux_evoltion} flux evolution plots for the four other type III events. Because in each of the type III (surge)  events plasma first shoots up (or to the right along the magnetic field lines) and there is flux cancellation going on at the base, these together suggest that the flux cancellation prepares and triggers the eruption that drives the plasma outflows.

\begin{figure*}
	\centering
	\includegraphics[trim=1.5cm 1.4cm 1.2cm 2.4cm,clip,width=\textwidth]{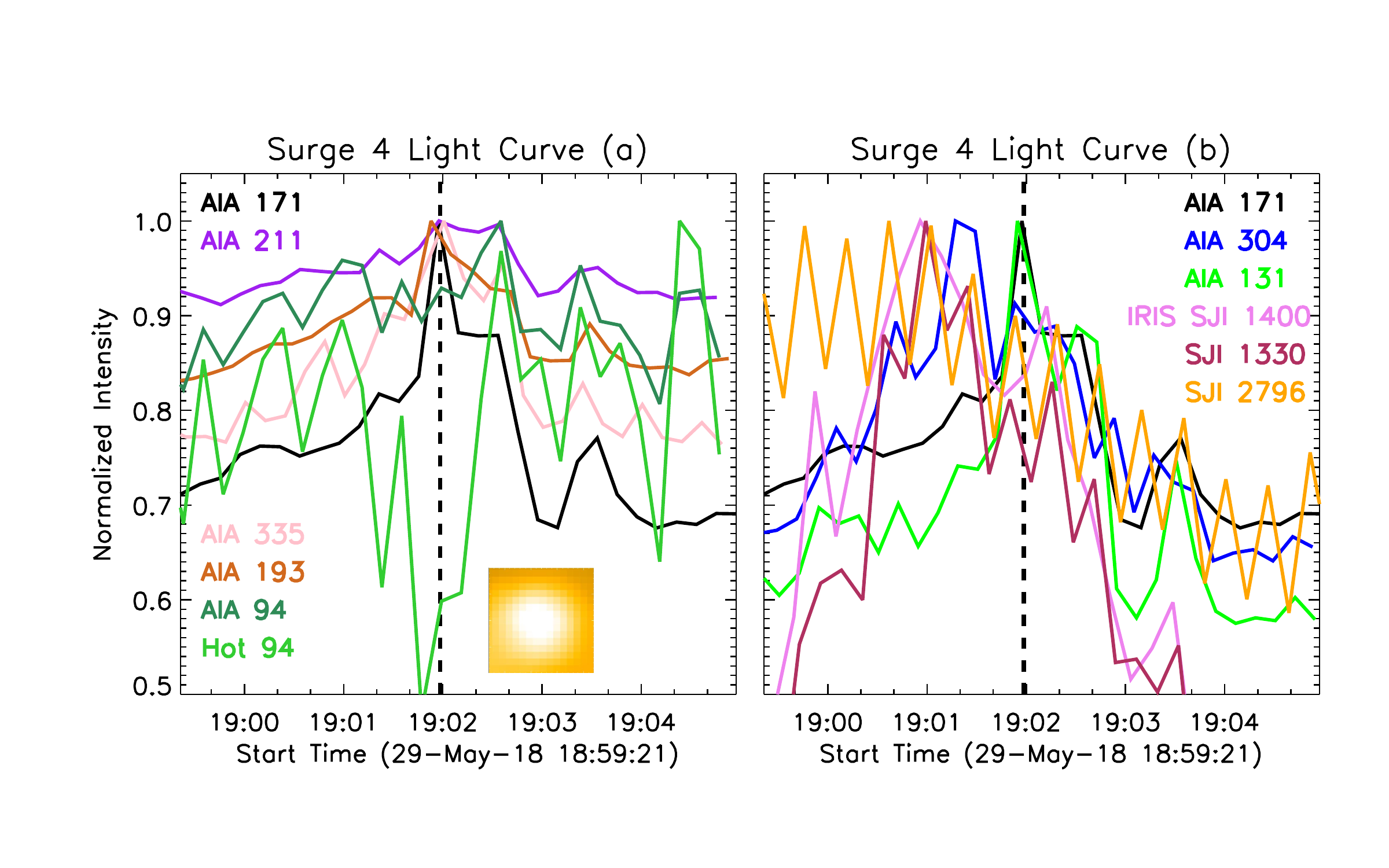}
	\includegraphics[trim=1.5cm 1.4cm 0.85cm 2cm,clip,width=\textwidth]{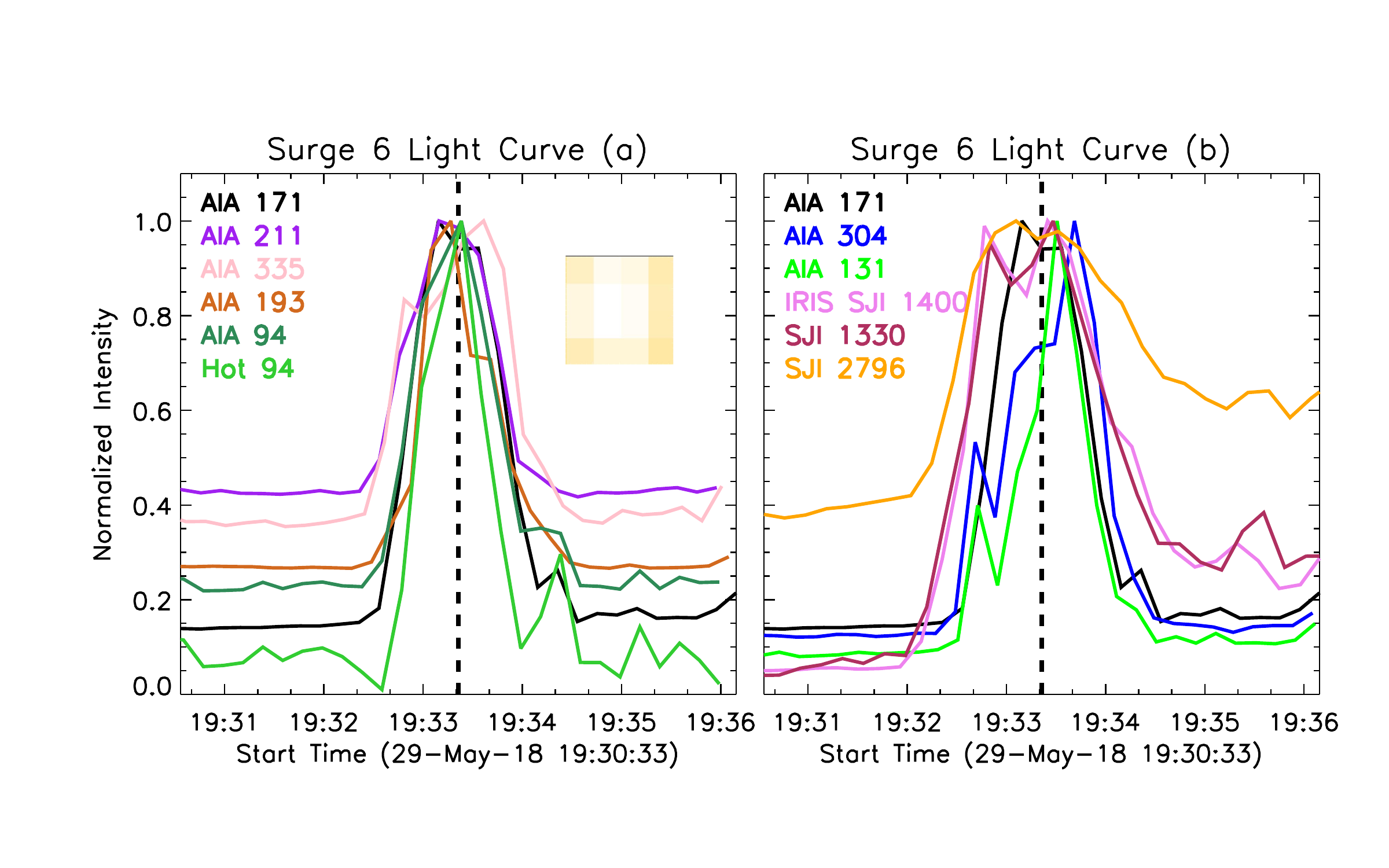}
	\caption{Light curves from AIA and IRIS intensity images for two Type III (surge/jet like) events: Surge 4 and Surge 6 pointed to by arrows in Figure \ref{surge}. The Hi-C area for Surge 4, or the AIA area for Surge 6, selected for making light curves is displayed as a small inset in the left panel for each surge  during its peak intensity time in AIA 171. Because the relative intensity of the Hi-C images drops quickly after 19:01:20 UT (e.g., see Figure 7 of \cite{rach19}), the images after this time are not usable for making light curves and thus the Hi-C light curve is not plotted for Surge 4. The vertical dashed lines mark times for the peak brightness of the events in AIA 171 \AA. Due to the integrated area of the SJI covering a few (dark) pixels from a dust patch some of the IRIS light curves show repeated intensity fluctuations. }
	\label{lc_surges}
\end{figure*}

The light curves (intensity integrated over $\sim2\times2$ arcsec$^2$ $\sim16\times16$ Hi-C pixel$^2$ $\sim4\times4$ AIA pixel$^2$) of different AIA channels, IRIS wavelengths for Surges 4 and 6 are plotted in Figure \ref{lc_surges}. Light curves for other surges listed in Table \ref{t1} can be found in Appendix \ref{app_lightcurves}. Similar to as for type I and type II events all the light curves for type III peak nearly simultaneously suggesting that we see the cooler plasma detected by the hotter channels, so that these events are cooler/chromospheric/transition-region events \cite[as in][]{wine13}. 

Note that hot 94 does show intensity enhancement in Surge 6, but not in Surge 4. Nonetheless neither of the two examples nor any of the other type III events show a systematic cooling pattern. Thus they are likely not coronal flare-like events. However, some of their appearance in hot 94 (\FeXVIII\ emission) suggests that these events might be heated up to 6 MK or more. In those cases (because intensities peak together with cooler wavelengths) the cooling must be very fast so that the heating can be balanced merely by radiative cooling (probably conduction does not play a role, thus no significant time lag is seen). However, such a scenario would require very high plasma density, which is not estimated in the present work. Further, the calculation of hot 94 emission might have uncertainties, particularly during a flare \citep{warr12}.

We can not rule out the possibility of some of these events being multi-thermal. The emission at 6 MK may become weak very fast due to the expansion, or because it is obscured by EUV absorption.

\begin{figure*}[htp]
	\centering
	\includegraphics[trim=0cm 4.7cm 0cm 2.5cm,clip,width=\textwidth]{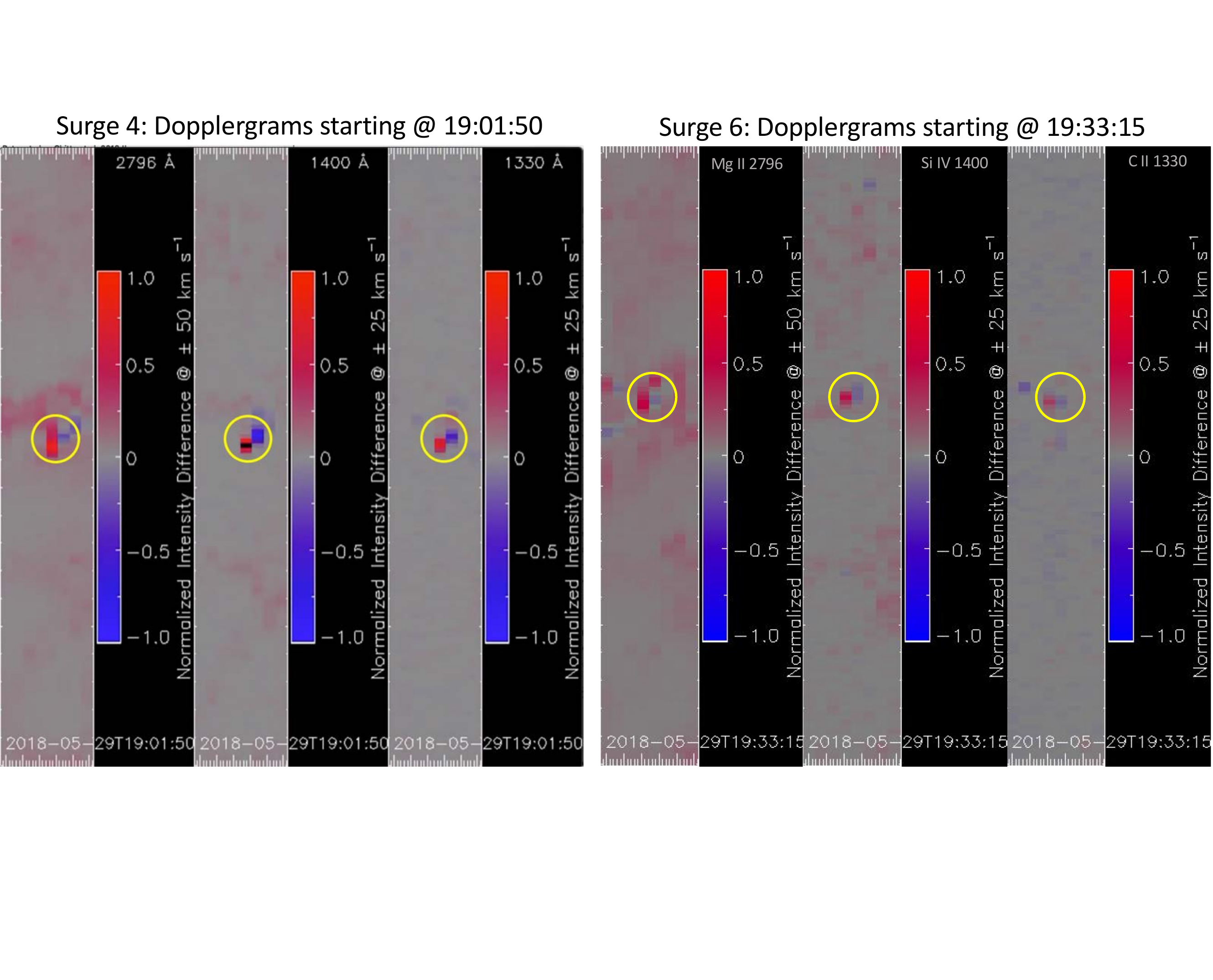}
	\caption{Dopplergrams of \MgII, \SiIV, and \CII\ lines during the peak of the events Surge 4 and Surge 6 displaying blueshift and redshift (plasma flow patterns) at the event locations. See the movie ``doppler.mp4" to follow these events in time. The black saturated is redshift, and white saturated is blueshift in the image and in the movie. Similar to that in the movie, the time of first slit position in each raster is given on each panel of the image. The yellow circle is centered on the base in Surge 4 (white arrow in Figure \ref{surge}), but is centered beside (west of) the base in Surge 6 (white arrow in Figure \ref{surge}). The spectra along two slit positions is shown in Figure \ref{spectra_surge4+6}. }
	\label{doppler_surge4+6}
\end{figure*}

\begin{figure*}[htp]
	\centering
	\includegraphics[trim=0.8cm 5.6cm 1.9cm 0.3cm,clip,width=0.498\textwidth]{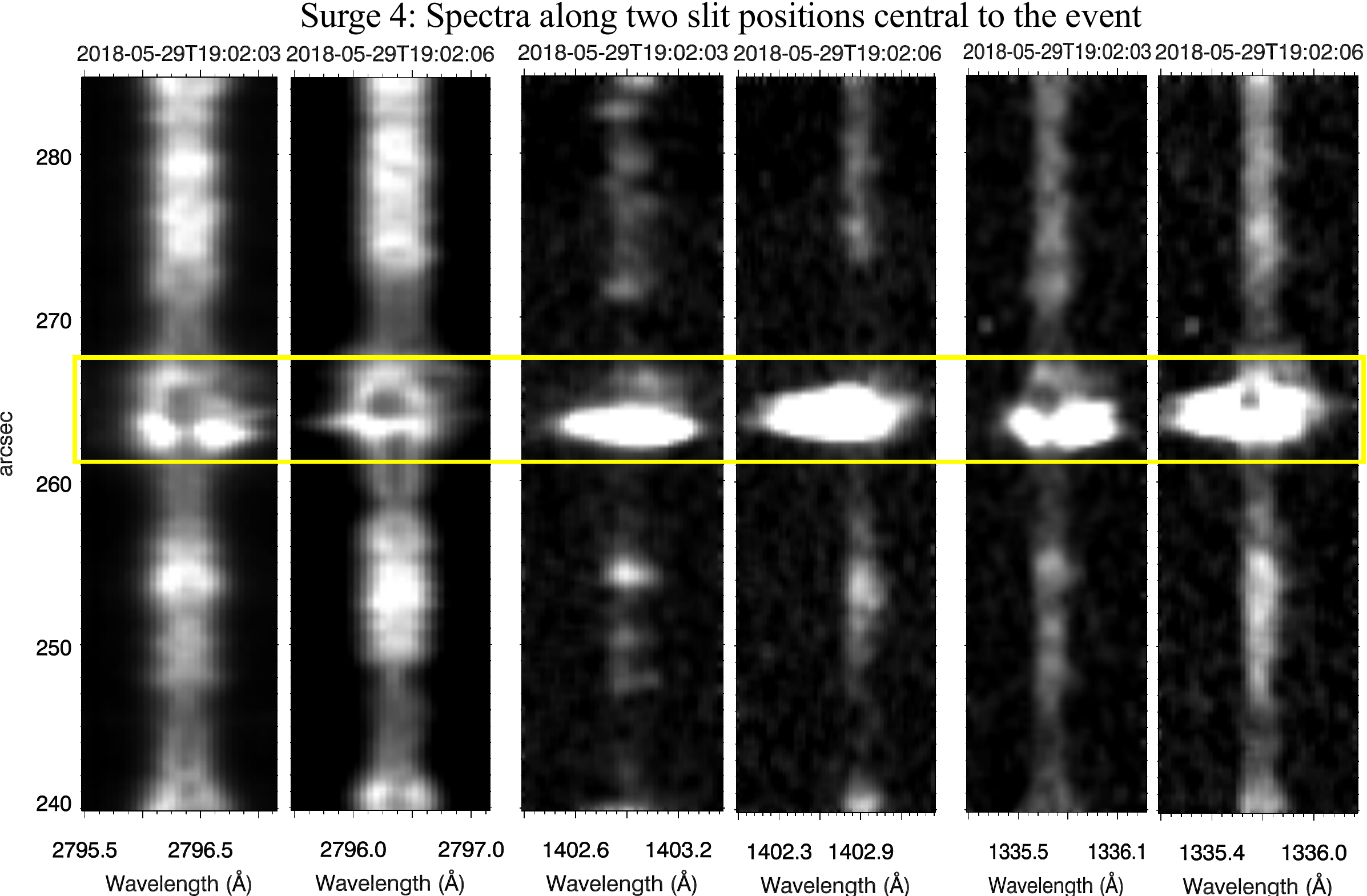}
	\includegraphics[trim=1.4cm 5.6cm 1.8cm 0.3cm,clip,width=0.490\textwidth]{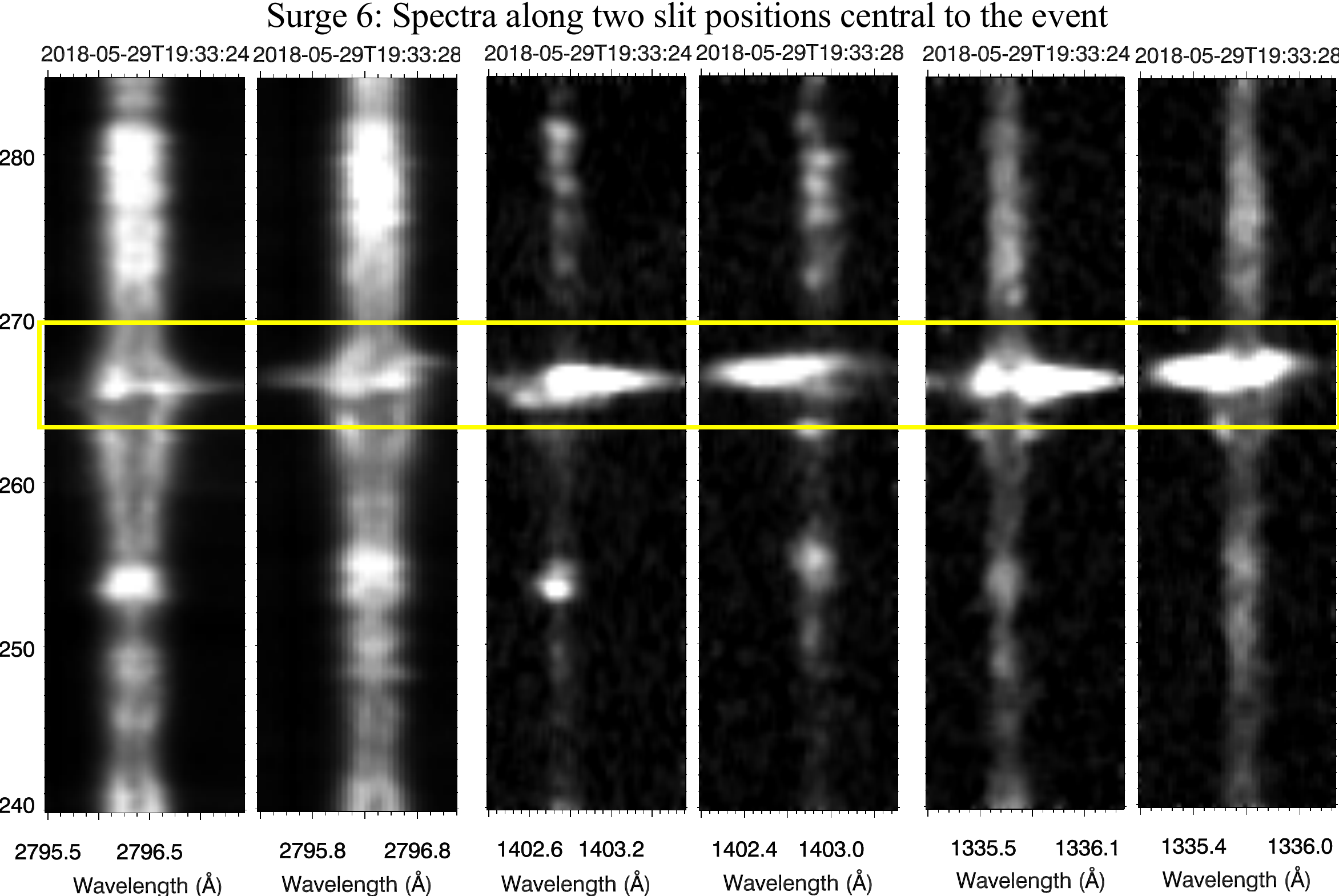}
	\caption{\MgII, \SiIV, and \CII\ spectra along two consecutive slit-positions for Surge 4 event (at 19:01:56 UT), and Surge 6 event (at 19:33:21 UT). Corresponding Dopplergrams displaying blue and red flow patterns during the event are shown in the Figure \ref{doppler_surge4+6}. The yellow lines outline the North-South boundary of the events, outlined by circles in Figure \ref{doppler_surge4+6}.}
	\label{spectra_surge4+6}
\end{figure*}

The IRIS spectra did not cover any of the type I or II events but did cover bases of most of the Type III events. Thus, it is possible that type I and II events also had outflows (though unobserved). Therefore, the idea that they could also be due to some kind of unresolved surge-like eruptive process cannot be ruled out. We made Dopplergrams to verify upflows/outflows (plasma flowing away from the base of the surge) and/or downflows/inflows (plasma flowing towards the base of the surge) at or near the base of surges and to see if there is a twisting of the magnetic field, similar to jets \citep{schm13,cheu15,moor15,pane16qr,pane17,ster17,tiw18}, which is generally expected in flux rope eruptions.

The Dopplergrams (Figure \ref{doppler_surge4+6}) near the base of surges for each of the three wavelengths (IRIS 2796, 1400, 1330 \AA) often show simultaneous redshift and blueshift next to each other, which reveals simultaneous upflow and downflow patterns during the onset of surge events. In most cases a clear outflow is evident near the bright source/base of the surges, consistent with similar blueshift found in H$\alpha$ surges by \cite{canf96}. Because the redshift and blueshift are not on the top of each other (across a surge/jet, which could then suggest twisting motion, see e.g., \cite{tiw18}), rather they are side by side (along a surge), we interpret this as plasma downflow and upflow along surge. Thus, no clear Doppler signatures of twisting motions are found (in which across the elongated axis of the surge/jet the LOS velocity on one side is toward the observer and the LOS velocity on the other side is away from the observer \citep[\eg][]{tiw18}).

In the Dopplergrams for the Surge 4 event (Figure \ref{doppler_surge4+6}), the strong red-shifted point sits on the AIA 171 dot, and the blue-shifted feature is on the outflow site seen in the AIA 171 movie. These Dopplergrams show no evidence of spin in the outflow, but show only the component of the outflow velocity along the line of sight. Thus, the Hi-C 172 dot-like brightening in this case is a jet-base bright point, and it is compatible with the idea of surge formation in Figure \ref{cartoon2}, that it should have downflow in it giving the redshift in the IRIS spectra. In Figure \ref{spectra_surge4+6} spectra along two slit positions during the peak of events Surge 4 and Surge 6 are displayed, which, consistent with the Dopplergrams, show redshifts and blueshifts.

\section{Discussion} \label{diss}
We report on three types of small-scale explosive energy release, sudden brightening events, in the core of an active region observed by a unique combination of instruments -- Hi-C, IRIS, and SDO/AIA. We first characterize the transient brightening activity that we noticed in the Hi-C 172 \AA\ images: type I -- a confined dot-like brightening event, never reported before in the core of an AR, type II -- an elongated brightening in and along a short magnetic loop. We then investigate a third type of event occurring in the same region -- type III -- a surge/jet-like eruption. 

Each of our events (with one exception) shows the presence of mixed-polarity magnetic field (with sharp neutral line/s shown by $\pm$25 G contours) at the base, often with ongoing flux convergence. We show quantitative evidence of flux cancellation in seven cases, and infer the presence of flux cancellation in other cases based on the observed flux convergence. Although we do not rule out other possibilities, the observations of flux convergence at the base of events suggests that flux cancellation could play an important role in triggering several of these events, in accord with many recent similar findings of flux cancellation leading to jet eruptions \citep{huan15,pane16qr,pane17,tiw16,tiw18,ster17,ster18,pane18,lope18}. As was first proposed by \cite{van89} and \cite{moor92}, and has been observationally confirmed \citep[e.g.,][]{pane16qr,pane17,pane18,pane18a,tiw18,ster18,chin19}, the process of flux cancellation (driven by converging photospheric flows) can prepare and trigger the magnetic field that explodes in a flare eruption. The magnetic explosion is either confined (does not produce a surge, jet, or CME) or ejective (produces a surge, jet, or CME) \citep[e.g.,][]{mach88,moor01}. Some of our small-scale events occur during flux emergence suggesting that emergence-driven cancellation prepares and triggers some of these events, the preparing again being by flux cancellation in the manner of \cite{van89}, and the triggering again being by flux cancellation in the manner of \cite{moor92}.       

Although observed in a different wavelength, our dot-like brightening events have much visual similarity with IRIS bombs \citep{pete14} and Ellerman bombs \citep{rutt13}. However, EBs are much cooler ($<$10,000 K), have much longer lifetimes, and are more stable/continuous/repetitive brightenings than our dot-like small-scale events reported here. Thus, the observed dot-like events are not simply EBs. IRIS bombs \citep{pete14} have more visual similarities with our dot-like events -- they are hotter than EBs, and have a similar size and intensity enhancement with respect to background as our dot-like events, and they are all found near mixed-polarity flux and sharp neutral lines. However both EBs ($\sim$560 s: \cite{wata11} ) and IBs ($\sim$5 minutes) have much longer lifetimes than of our Hi-C dot-like events ($\sim$70 seconds). A caution with this interpretation is made at the end of next paragraph.

Further, IBs and our dot-like events are apparently seen at different temperatures.
\cite{pete14} found that IBs show no brightenings in AIA 171 \AA,  whereas our dot-like events are seen in Hi-C 172 \AA. Thus, our dot-like events (Type Is) are apparently much hotter and briefer explosions than those of IRIS bombs. A caveat is that AIA 171 channel emission might suffer with bound-free absorption, which could lead to a shorter lifetimes of our dot-like events. Moreover, the Hi-C passband covers O V/VI lines, which form at a much lower temperature. Thus, it is possible that we see cool transition-region contamination in the AIA and Hi-C passbands. This subject thus remains open for further investigation.  

Note that type I events are also visible in AIA 171 \AA\ but not as outstandingly as in Hi-C, and so remained unnoticed earlier and were not reported before in the core of ARs. There are similar bright dots reported in the past elsewhere in the solar atmosphere e.g., in sunspot penumbra using IRIS data \citep{tian14} and Hi-C 1 (in 193 \AA) data \citep{alp16}, in the surroundings of the Hi-C 1 AR \citep{regn14}. The `sparkling' bright dots in the moss region (at the edge of the AR) of Hi-C 1, studied by \cite{regn14} have shorter lifetimes (25 s) and are smaller (700 km) than our dots. These sparkling bright dots form in EUV corona, having a temperature of 1 -- 1.5 MK, similar to that of our dots.
   
The moving bright dots in sunspot penumbra were proposed to form due to impact of strong downflows from the corona into the diverse-density chromosphere/ transition-region, or by magnetic reconnection in two field lines inclined at different angles \citep{alp16}. The dots in plage area surrounding ARs were proposed to be a result of nanoflares high in the moss loops \citep{regn14}. The formation mechanism of the present dot-like brightening events seem to be different -- these are located at or near sharp PILs and so are plausibly triggered by flux cancellation or by flux emergence (that drives flux cancellation on its outside: \cite{moor92}), which was not the case in the EUV bright dots at the edge of Hi-C 1 AR, or in penumbral moving bright dots.

Most of the type II loop-like events also have mixed-polarity magnetic field (with sharp neutral lines) on the photosphere but flux cancellation is not as clearly visible as in type I or in type III events.  However a careful inspection reveals the presence of flux convergence along the sharp neutral line, plausibly driving cancellation, accompanied by the loop brightening. The presence of mixed-polarity field and/or flux cancellation has been recently reported to play an important role in coronal loop heating and is proposed to be present at least at one footpoint of a bright coronal loop \citep{tiw14,tiw17,chit17_sunrise,prie18}.
Here we show smaller loop events than earlier reported ones but some of these might share the heating mechanism with those coronal loops with mixed-polarity field at least at one foot.

We note the following caveat: The presence of a neutral line in short loops does not necessarily mean there must be flux cancellation. Short, low-lying loops, obviously have opposite-polarity magnetic field in proximity and thus occur close to the neutral line, and thus may have flux cancellation. Therefore, whether most of the bright loops form because of flux cancellation (that results from submergence of short loops made by magnetic reconnection of the legs of adjacent sheared loops driven together at the PIL by convection) remains elusive.

Several alternative mechanisms are plausible to generate types I and II events. E.g., random footpoint shuffling of magnetic loops can braid the loops, which can lead to the events by reconnections in the form of nanoflares (sudden current dissipation) \citep{parker83a,parker88}. The heating could also be caused by wave dissipation \citep[\eg][]{oste61,heyv83}. Reconnection events could be also triggered by waves \citep[produced from the photospheric convection at loop-foot:][]{hegg09}, or by external triggering of loops \citep{tiw14}. As discussed before, flux tube tectonics heating model also predicts low-lying smaller loops to possess enhanced heating \citep{prie02}.

Because in most of type II events brightenings start at one end and move towards the other end along the loop, it is possible that type II events are formed in the same way as type III events.

If they peak, the light curves from all AIA and IRIS channels (for all of these events) peak nearly at the same time. Note that in a few cases, e.g., in Loop 3 in Figure \ref{lc_loops}, IRIS SJ intensity do not show a consistent peak in the light curves. Similarly, although many of our events do, some do not show a peak in hot 94. However, none of our events display a systematic cooling pattern as seen for typical coronal solar flares. Thus, our events are either cool (at chromospheric/transition-region temperature) i.e., a cool contamination to coronal passbands, or they are broadly iso-thermal (in sub-structures/strands), similar to the low-lying loop nanoflare events reported by \cite{wine13}. If the latter is true, then the cooling time in each strand might be so short that the heating is balanced by radiative cooling.

Type III events show clear flux cancellation and plasma outflow from the source region, often followed by plasma inflow. These events have a dot-like structure at the source region in Hi-C 172 \AA\ (when available) and AIA 171 \AA\ images. Therefore, if surges are very small and do not show a clear outflow they can be mistaken to be type I events. We suspect Dot 2 is a type III event. Thus, Dots 1 and 2, and any other dot-like events, might be made in the same way as a type II or a type III event. Similarly, some type II events can be interpreted as small type III events, as we suspect in the case of Loop 2.    

Dopplergrams of type III events provide confirming evidence of plasma outflows along the field during its initial phase and inflows during the later phase. The eruption triggered by flux cancellation (due to submergence of lower reconnected loop) evidently drives outflows -- if similar flux cancellation were to occur at both feet of a loop system these could drive simultaneous bi-directional flows, which will be similar to the well known counter-streaming flows in large classical filaments \citep[\eg][]{alex13}. However this remains only speculation in absence of a clear evidence of such counter-streaming flows in the core of the Hi-C AR studied here.

In Table \ref{t1} only one event (type II, Loop 2) does not contain a clear neutral line shown by $\pm$25 G contours. This could be either due to the  absence of mixed-polarity field, or the minority polarity flux may be below the detection limit of the SDO/HMI instrument. 
A more detailed future study with higher resolution vector magnetograms e.g., obtained by DKIST \citep{rimm18} or other new generation solar telescopes would confirm or deny the proposed scenario of the formation of loop-like events.

\subsection{Proposed Configuration and Reconnection of the Magnetic Field in Each Event Type} \label{formation}

\begin{figure}
	\centering
	\includegraphics[trim=0cm 0cm 0cm 0cm,clip,width=\columnwidth]{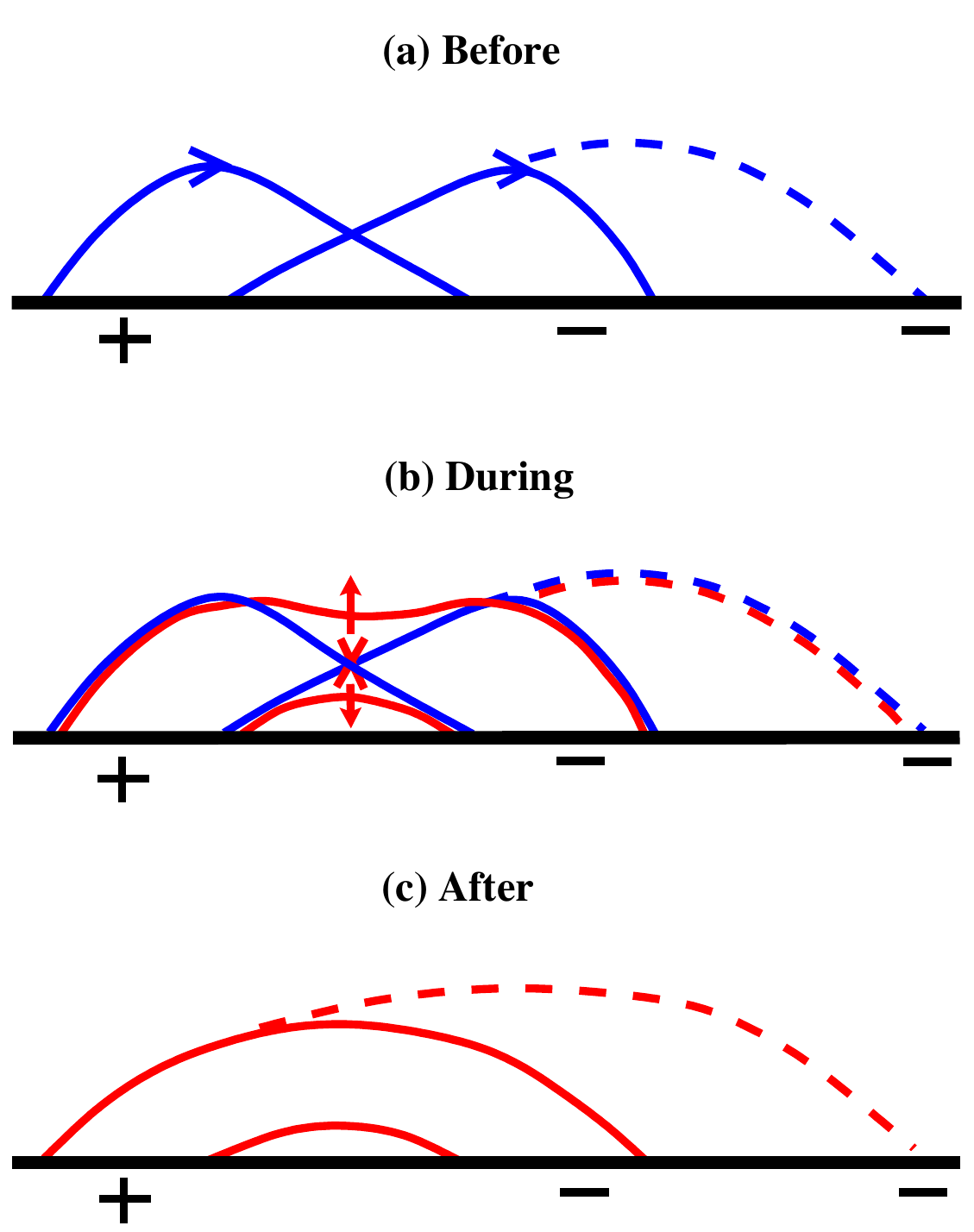}
	\caption{Schematic depiction of the proposed configuration and reconnection of the sheared and twisted bipolar magnetic field in fine-scale explosive energy release events of type I (dot-like) and type II (loop-like). The thick black line is the photospheric surface.  The plus and minus signs give the polarity of the photospheric magnetic flux.  The curves represent crossed field loops projected on a vertical plane perpendicular to the polarity inversion line (PIL).  The PIL lies along the view direction and is midway between the plus sign and the first minus sign to the right of the plus sign.  Blue curves are for field that has not yet undergone reconnection.  Red curves are for reconnected field.  In drawings (a) and (b), the right leg of the left blue loop is in front of the left leg of the right blue loop.  In drawing (b), the red X marks the site of ongoing reconnection between those two loop legs, the two solid-line loops are for a type I event, the dashed curve depicts that the right loop has a longer rightward reach in a type II event, and the red arrows denote the outflow of the upper and lower reconnected field loops and their plasma.  In drawing (c), the lower solid curve is the lower reconnected field loop in a type I event as well as in a type II event, the upper solid curve is the upper reconnected field loop in a type I event, and the dashed curve depicts that the upper reconnected field loop as a greater rightward reach in a type II event. Note that this depiction is a possibility for type I and type II events whether or not the pre-event magnetic field is prepared and triggered by flux cancellation at the PIL. The pre-event field might instead be twisted by convection in the loop feet and perhaps triggered by the same convection or p-mode oscillations. }
	\label{cartoon1}
\end{figure}


\begin{figure}
	\centering
	\includegraphics[trim=0cm 0cm 0cm 0cm,clip,width=0.58\columnwidth]{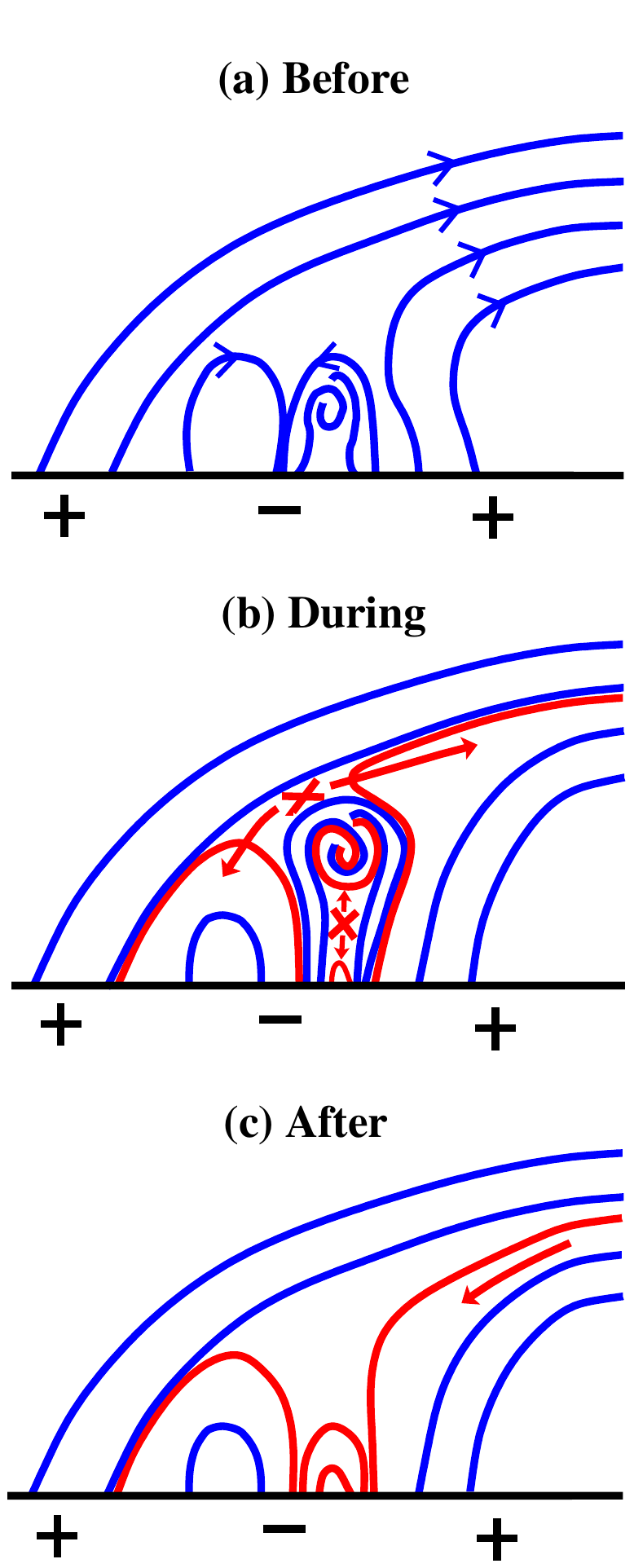}
	\caption{Schematic depiction of the proposed configuration, eruption, and reconnection of the magnetic field in fine-scale explosive energy release events of type III, each of which is a surge/jet-like eruption from a fine-scale island of minority-polarity (negative) flux that is undergoing cancellation with the majority-polarity (positive) flux in the east end of the arch filament system.  The style and meaning of the symbols, lines, curves and color are the same as in Figure \ref{cartoon1}.  Here, each curve is either a field line in – or the projection of a field line onto – a vertical plane through the center of a negative-flux island in surrounding positive flux.  Rightward is heliographic west; leftward is east.  In drawing (a), the curled field straddles the PIL on the west edge of the island and is in a twisted flux rope (viewed end-on from the south) that has been built by flux cancellation driven at the PIL by convergence of convection flow in and below the photosphere.  This flux rope is the core of a sheared magnetic arcade that straddles the PIL and that, as a result of further flux cancellation at the PIL, is triggered to start erupting in the time between drawing (a) and drawing (b).  In drawing (b), the erupting arcade is driving: (1) external reconnection with encountered far-reaching field that reaches to the west end of the arch filament system, and (2) internal reconnection of the legs of the erupting arcade.  The external reconnection drives westward plasma flow out along the reconnected far-reaching field.  The internal reconnection builds a miniature flare arcade that is seen in Hi-C and AIA coronal EUV images (registered with HMI magnetograms) as a bright point located on the cancellation PIL of the magnetic island.  In drawing (c), the eruption and reconnection have ended and some of the previously ejected plasma is draining back to the foot of the reconnected far-reaching field. }
	\label{cartoon2}
\end{figure}

Here we present simplistic 2D schematic drawings for the magnetic field and its reconnection that could produce the events studied here. As discussed earlier, type I could be the same as type II -- both seem to be often elongated (particularly visible when seen in IRIS SJI) -- at some phase of these events a loop like elongated structure can be noticed. This suggests that dot-like events are basically similar to loop-like events but with more confined extension of a reconnection-resultant loop. Plausibly IRIS sees both the upper and lower loops that are made and heated by magnetic reconnection and the Hi-C 172 and AIA 171 images show only the upper loop because the upper loop is hot enough to show in Hi-C 172 and AIA 171 \AA\ images, but the lower loop is not hot enough. Both of these are found mostly located at or near mixed-polarity field/neutral lines. 

The cartoon diagram shown in Figure \ref{cartoon1} proposes a possible formation mechanism for type I and II brightening events. The magnetic reconnection (indicated by the red X in panel b) occurs between the legs of two sheared field loops (one in front of the X, the other behind) that are perhaps sheared and pushed together by photospheric shearing convection merging at the PIL. We note that this picture (presented in Figure \ref{cartoon1}) is a speculation based on the magnetic settings and evolution of magnetic fiux in the photosphere -- to the best of our (and an anonymous referee's) knowledge no theoretical model or computational simulation that tests this specific scenario is available.  However, we note that the magnetic topology and reconnection depicted in Figure \ref{cartoon1} and their rationale are essentially those for any one of the nanoflare reconnection events proposed by \cite{parker83b,parker83a,parker88} for coronal heating in a closed magnetic loop, each nanoflare burst of reconnection occurring at a current sheet built in the body of the loop by photospheric convection in the loop's feet.

Similar to type II events, type Is are confined (no obvious ejective outflow), but unlike type IIs type I events are brightest in the middle, not brightest on one end.  Thus, type Is could be the same as type II (loop) events except they are shorter and more symmetrically heated.

An alternative possibility of the formation of type II loop-like events (and possibly of dot-like events) is given in the following. A few of these cases, particularly in the type II (loop eruptions with mostly unidirectional flow), have flux emergence before flux cancellation, thus suggesting a loop-loop-interaction scenario (with three legged magnetic field configuration), as suggested by \cite{hana97} for flares, jets, and surges. The loop-loop interaction may cause component reconnection of crossed flux tubes rooted in the same-polarity magnetic flux. 

Type III events most clearly show flux emergence and cancellation at the driver end. Plausibly, the flux-emergence-driven cancellation at the neutral line prepares and triggers a fine scale core-magnetic-field structure (a small sheared/twisted core field or flux rope along and above the cancellation line) to explode. A cartoon diagram depicting this formation scenario for type III surge-like events is shown in Figure \ref{cartoon2}. The flow patterns (redshift in the left/East and blueshift in the right/West) in the panel (b) is consistent with that observed in Dopplergrams, see e.g., Figure \ref{doppler_surge4+6}, and movie doppler.mp4.    

As mentioned before, type I and type II events might be smaller versions of type III event, and all the three kinds of events could form in the same way as proposed in Figure \ref{cartoon2}. In that case each of our types I and II events may occur at an embedded flux island that is near the neutral line but is too small/weak to be detected in the HMI magnetograms. The field configuration sketched in Figure \ref{cartoon1} is for any type I or type II event in which HMI sees only a single long neutral line and no embedded flux island.
	
The eruption that drives the production of the jet/surge could be prepared and triggered by magnetic flux cancellation. In this mechanism, instead of flux emergence, flux cancellation leads to and triggers the jet/surge eruption. A twisted flux rope forms by flux cancellation \citep{van89,pane17,ster18}, which is then triggered (to erupt and drive internal and external reconnections as in Figure \ref{cartoon2}) by further flux cancellation \citep{van89,pane16qr,pane17,ster17,pane18,pane18a}.  
Recent theoretical models support this scenario \citep{wype17,wype19}. Earlier models of X-ray bright points also showed that flux cancellation can drive small-scale brightening events \citep[\eg][]{prie94}. The magnetic configuration as shown in Figure \ref{cartoon2} is similar to the configuration as found in UV bursts. There, following flux emergence the minority polarity cancels with the majority polarity of opposite sign \citep{chit_17}. Another recent reconnection modelling shows a similar situation as drawn in our Figure \ref{cartoon2}, which is found to lead to a bi-directional jet \citep{pete19_aa}.
	
In their numerical modeling \cite{shib92_flux_emergence}, and \cite{yoko95,yoko96} showed that reconnection between emerging magnetic flux and overlying magnetic field can create surges, thus advocating for magnetic reconnection as an essential process for large (flares) to small (jets and surges) scale events. The field aligned flows in surges (or apparent intensity propagation in loops) might be accelerated by the enhanced gas pressure behind the shocks driven by magnetic reconnection. The cool and hot plasma could be ejected in this process \citep{shib92,yoko96}. Because we observe both, magnetic flux emergence and cancellation, the surges might be formed in the way as proposed by \cite{shib92_flux_emergence,shib92,yoko95,yoko96}.
 
We would like to stress that most of the observed flux cancellation is plausibly a result of submergence of short loops made by convection-driven magnetic reconnection. This is what is also shown in Figures \ref{cartoon1} and \ref{cartoon2}. When we mention flux cancellation prepares and triggers an event that means a small flux-rope/minifilament is formed in the way proposed by \cite{van89} and then runaway internal reconnection under the flux rope (in the lower solar atmosphere, say in the chromosphere; of course these heights are set by the size of the closed field lobes of the jet base) unleashes the eruption that drives external reconnection that makes the jet spire \citep{moor92,moor01,ster15,pane16qr,pane17,tiw18,wype17}.

\section{Conclusions}\label{conc}
We have reported small-scale explosive energy release events observed in the core of the AR observed by Hi-C 2.1. We find three types of transient brightening events: type I dot-like, type II loop-like, and type III surge/jet-like. 
Most of the events we studied here are located at or near sharp neutral lines, and some show clear evidence of flux cancellation, often led or followed by flux emergence. Emergence-driven or converging-flow-driven flux cancellation plausibly prepares and triggers several of the three types of events we investigated.

We also mention other possible mechanisms, e.g., these events could be sudden energy release by wave dissipation, or by nanoflares in braided loops (either produced by footpoint shuffling or induced by waves). Dot-like events fit being a part of either loop-like events or surge-like events. Based on the similarities in intensity propagation and the photospheric magnetic field setting and evolution of several of types I and II events with type III events, one can expect type I and type II events to be smaller versions of type III events. However, to confirm this further detailed investigation of more cases with Hi-C-like or better instrumentation is required.

The IRIS spectra available for type III events show complex activities at their base, and upflowing cool material; as expected in a surge/jet activity these events show outflows in the initial phase and inflows in the late phase. Because the light curves (from Hi-C, IRIS, and different AIA channels) for most of type I, II and III events peak nearly simultaneously, and for none of the events show a coronal-flare trend of cooling, all three types (except for those clearly showing up in hot 94 \AA\ images -- in them the cooling time is so short that the heating is balanced by radiation) apparently have transition-region and/or chromospheric (and not coronal) temperature, but see Section \ref{diss} for caveats. 

\acknowledgments
Comments from the referee helped enhancing presentation of the manuscript. S.K.T. gratefully acknowledges support by NASA contracts NNG09FA40C (IRIS), and NNM07AA01C (Hinode). B.D.P. gratefully acknowledges support from NASA grant NNG09FA40C (IRIS). We acknowledge the High-resolution Coronal Imager (Hi-C 2.1) instrument team for making the second re-flight data available under NASA Heliophysics Technology and Instrument Development for Science (HTIDS) Low Cost Access to Space (LCAS) program (proposal HTIDS17\_2-0033).  MSFC/NASA led the mission with partners including the Smithsonian Astrophysical Observatory, the University of Central Lancashire, and Lockheed Martin Solar and Astrophysics Laboratory.  Hi-C 2.1 was launched out of the White Sands Missile Range on 2018 May 29. IRIS is a NASA small explorer mission developed and operated by LMSAL with mission operations executed at NASA Ames Research center and major contributions to downlink communications funded by ESA and the Norwegian  Space  Centre. N.K.P’s research was supported by NASA grant NNG04EA00C (SDO/AIA). R.L.M acknowledges the support from the NASA HGI program. HPW's participation was supported by NASA's Hinode program. The AIA and HMI data are courtesy of NASA/SDO and the AIA and HMI science teams. This research has made use of NASA's Astrophysics Data System and of IDL SolarSoft package. 






\newpage
\appendix
\section{Light curves for a sub-flare in the Hi-C active region}\label{app_lc_flare}
We plot AIA light curves for a small flare in the Hi-C AR during 18:39:33 -- 18:59:50 UT for a comparison with the light curves for our three types of events studied in this paper. These light curves show a systematic cooling sequence similar to observed in typical solar coronal flares.

\begin{figure*}[h]
	\centering
    \includegraphics[trim=0.8cm 2.6cm 2cm 3.4cm,clip,width=0.43\textwidth]{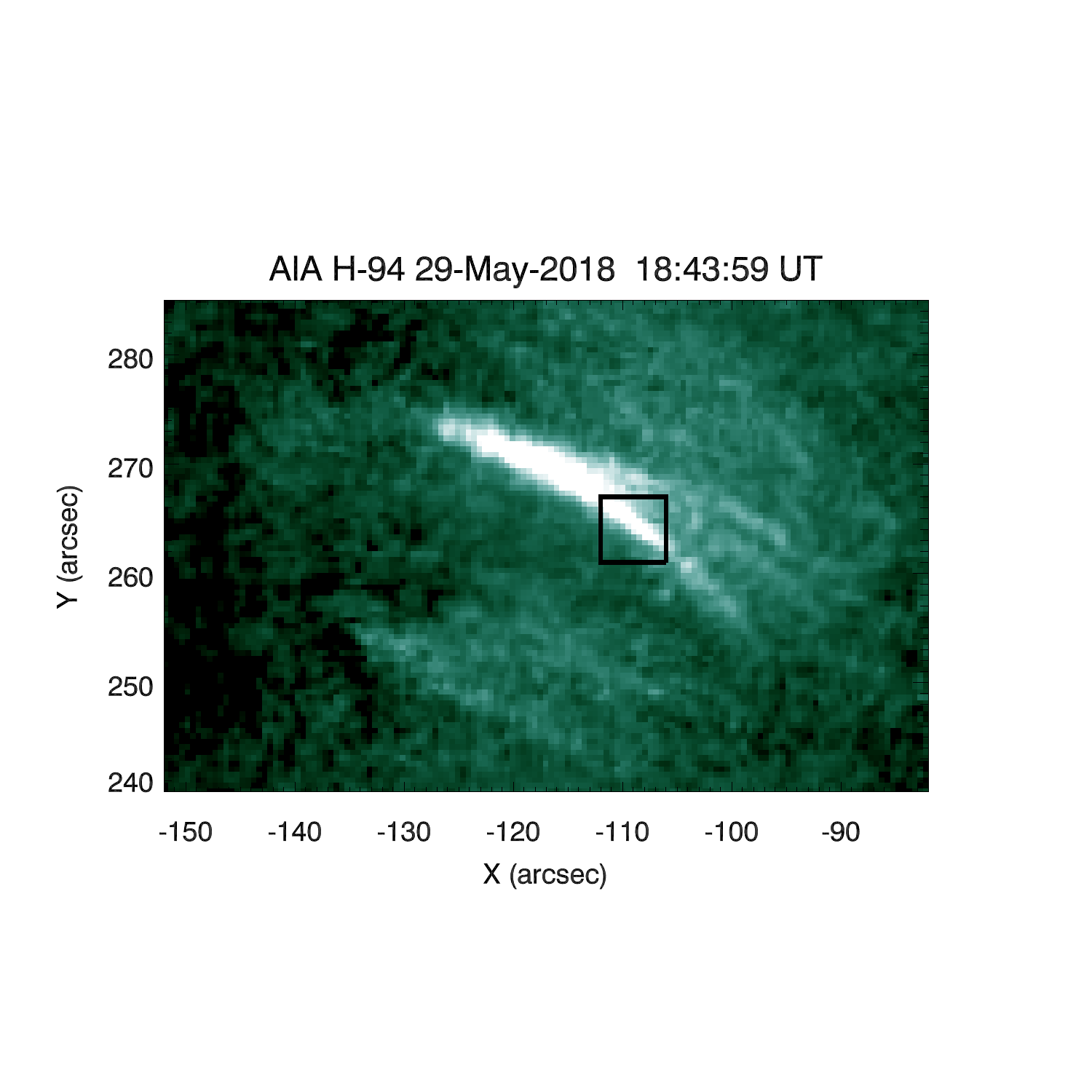}
	\includegraphics[trim=1.3cm 1.4cm 11.32cm 2.4cm,clip,width=0.56\textwidth]{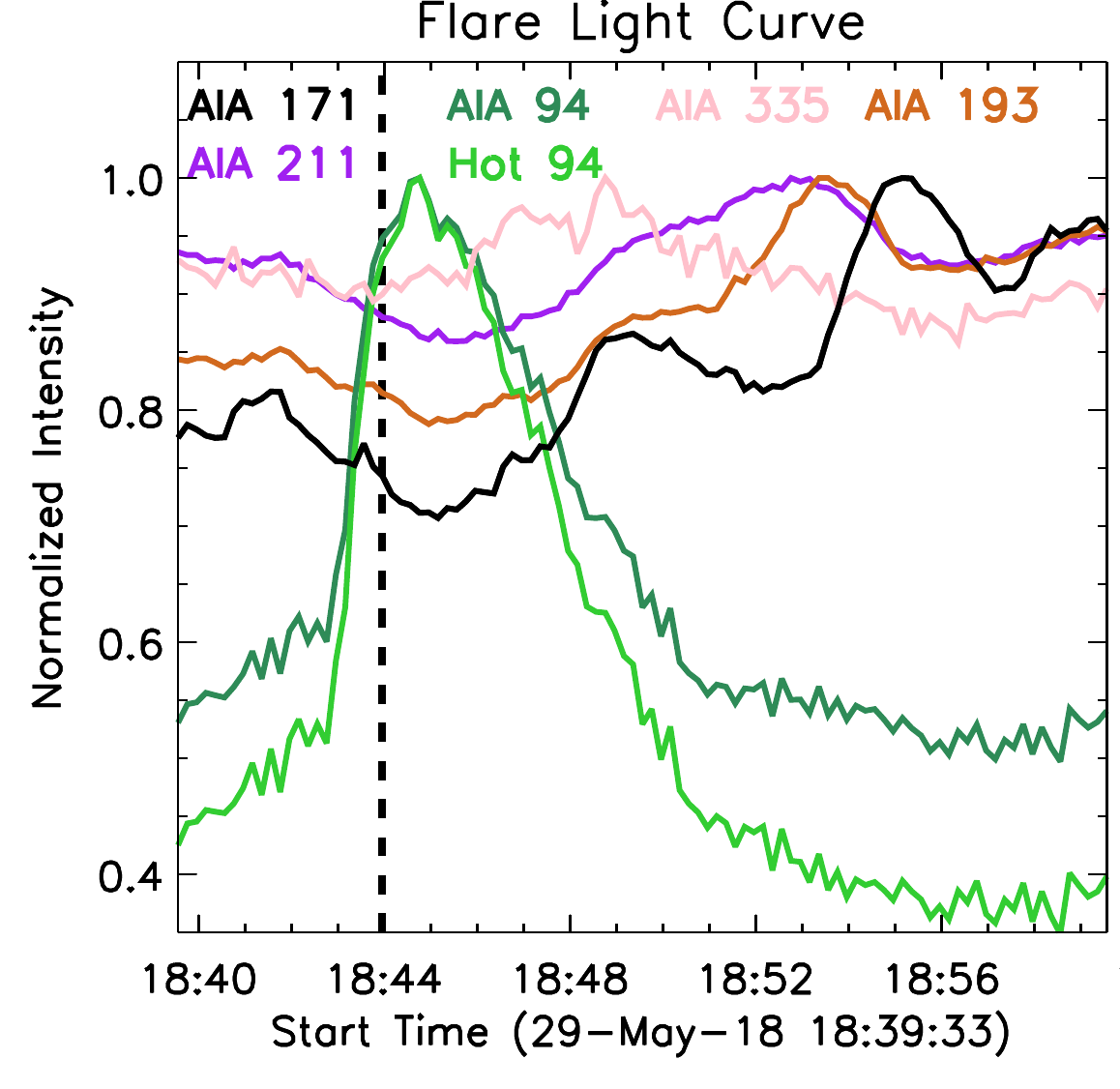}
	\caption{Light curves (right panel) from AIA intensity images for a sub-flare event peaking in AIA 94 at 18:43:59 UT. The area for the light curves is outlined by a black box in an hot 94 image of the sub-flare (on the left panel). See the flare evolution in the movie sdo\_long.mp4. A systematic cooling in this small flare can be seen in the light curves -- note the time sequence of the peaks of the successively cooler channels: hot 94/AIA 94, 335, 211, 193, and then AIA 171.
}
	\label{lc_flare}
\end{figure*}

\newpage
\section{Hi-C or AIA 171 \AA\ images at the peak time of the all brightening events listed in Table 1}\label{app_all_events}
Figure showing each of the events during its peak time as listed in Table \ref{t1}.

\begin{figure}[h]
	\centering
	\includegraphics[trim=6.9cm 0.04cm 8.8cm 0.0cm,clip,width=\columnwidth]{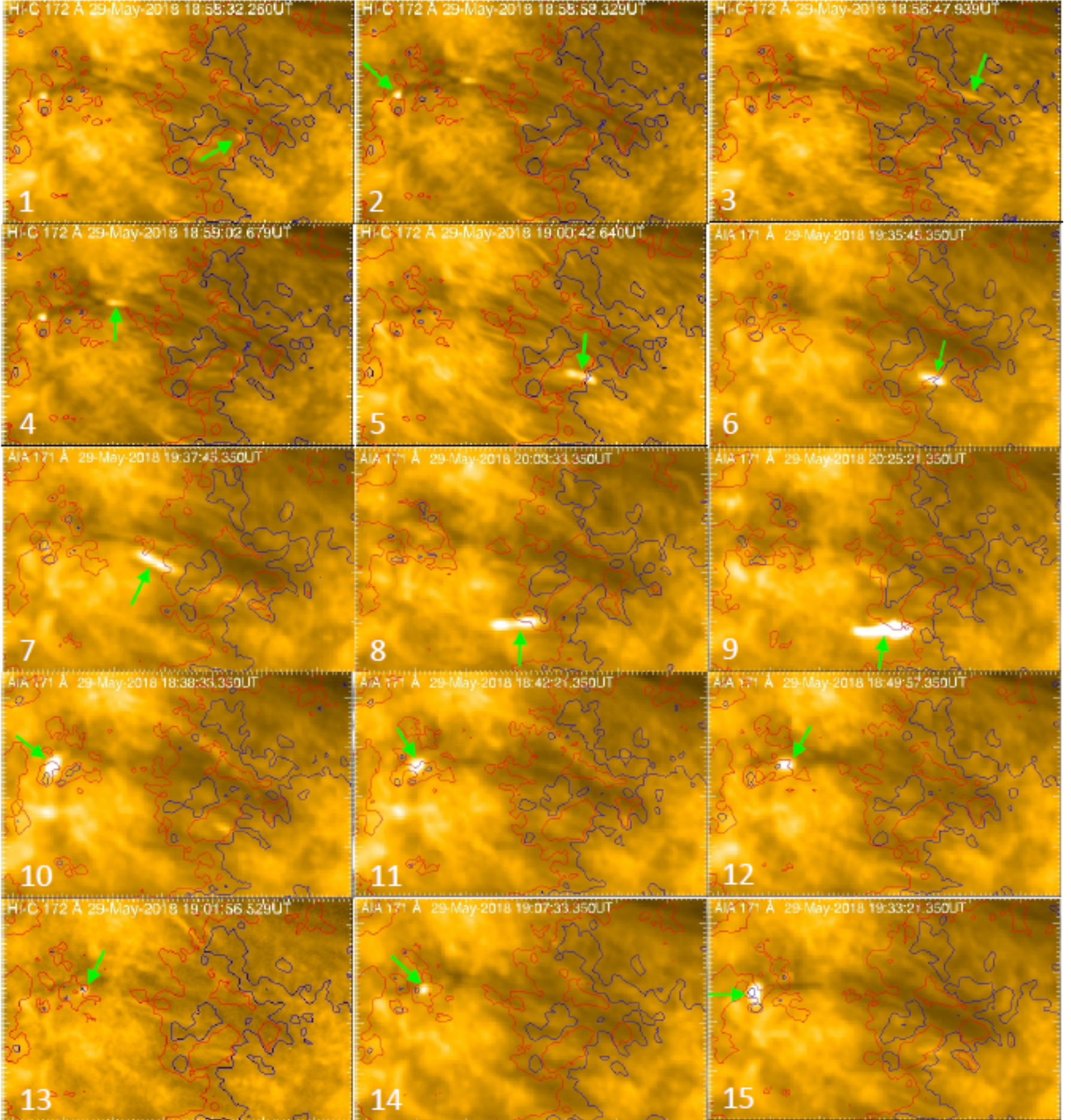}
	\caption{An image frame of each of the events (during their peak time) as listed in Table 1. Each of these are frames from either Hi-C 172 \AA\ movie ``hic\_iris\_sdo.mp4" (when available) or from AIA 171 \AA\ movie ``sdo\_long.mp4". Each event is pointed to by a green arrow.}
	\label{all_events}
\end{figure}

\newpage
\section{Examples of unidirectional and bidirectional flows}\label{app_uni_bi_flows}
\begin{figure*}[h]
	\centering
	\includegraphics[trim=1cm 3.8cm 1.3cm 1.5cm,clip,width=\textwidth]{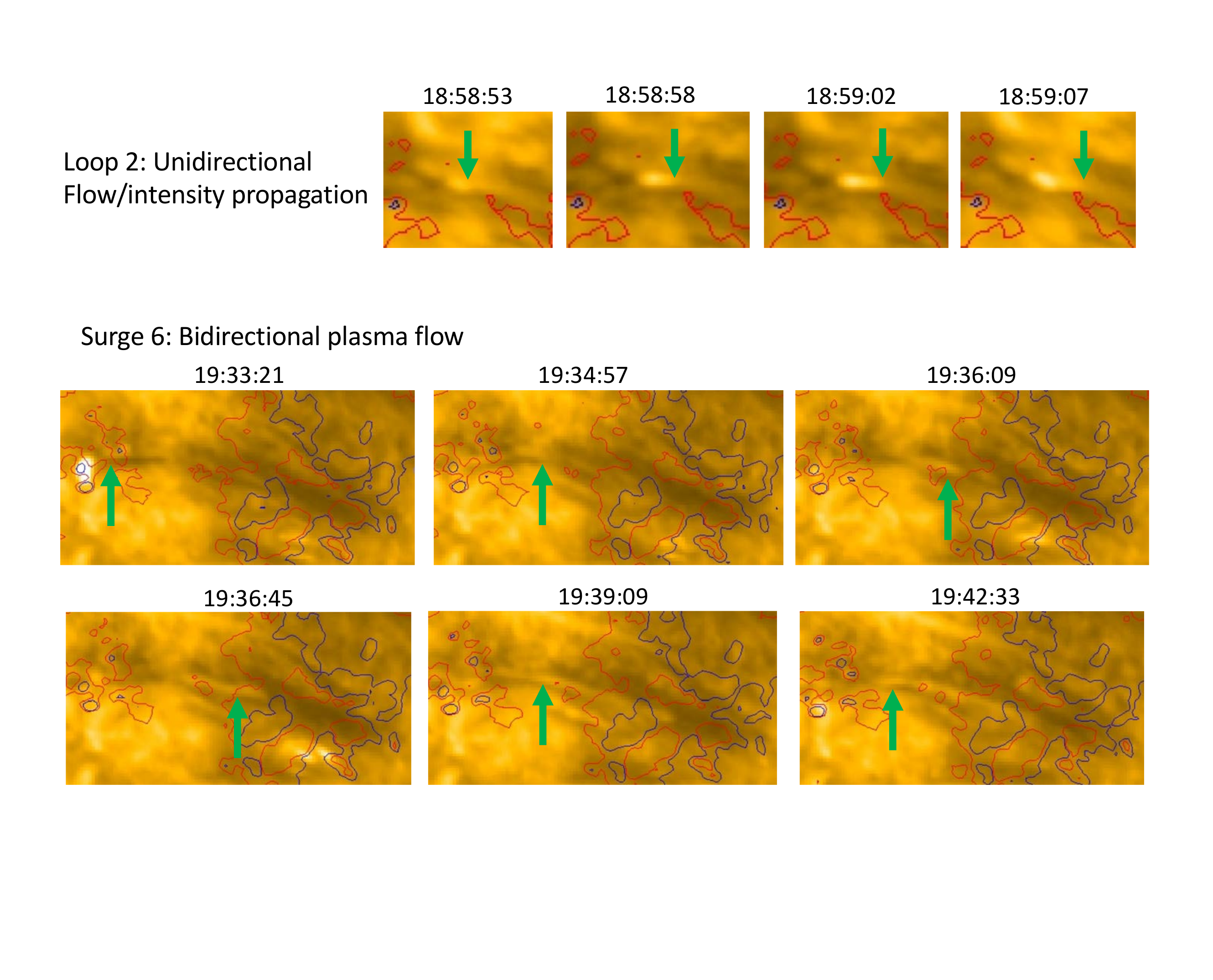}
	\caption{An example of unidirectional flow (in Loop 2) and bidirectional flow (in Surge 6). These image frames are for reference, the flows (seen as intensity propagations) are more obvious in the movies.}
	\label{lc_loop2}
\end{figure*}

\newpage
\section{Light curves of other nine events listed in Table 1}\label{app_lightcurves}

\begin{figure*}[h]
	\centering
	\includegraphics[trim=1.5cm 1.4cm 1.2cm 2.4cm,clip,width=\textwidth]{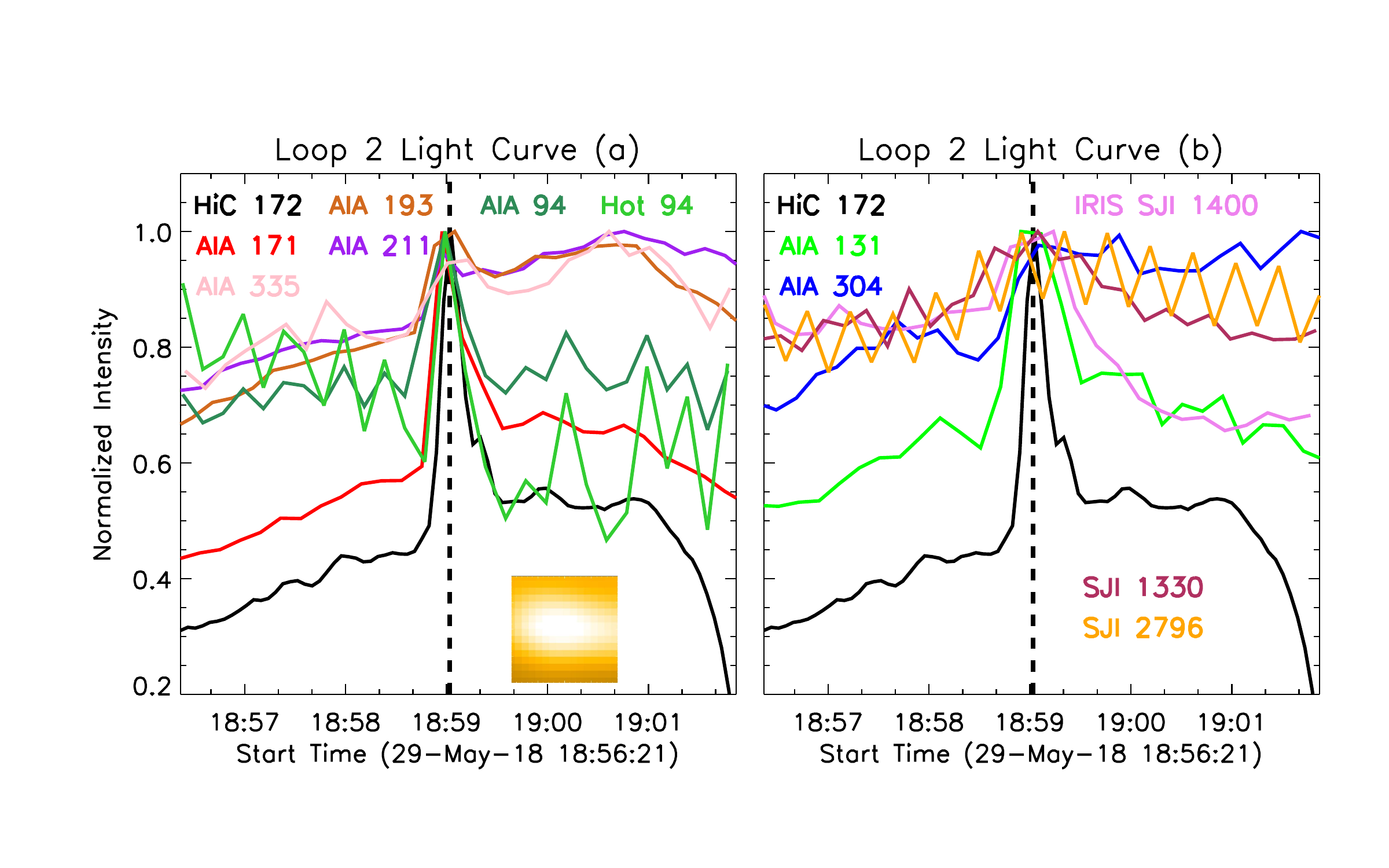}
	\caption{Light curves from AIA and IRIS intensity images over Hi-C time for Loop 2. The Hi-C area selected for making light curves is displayed as a small inset in the left panel during its peak intensity time in Hi-C.  Due to the integrated area of the SJI covering a few (dark) pixels from a dust patch some of the IRIS light curves show a repeated fluctuations. The dashed vertical line marks the peak time of the event.}
	\label{lc_loop2}
\end{figure*}

\begin{figure*}[h]
	\centering
	\includegraphics[trim=1.5cm 1.4cm 1.2cm 2.4cm,clip,width=\textwidth]{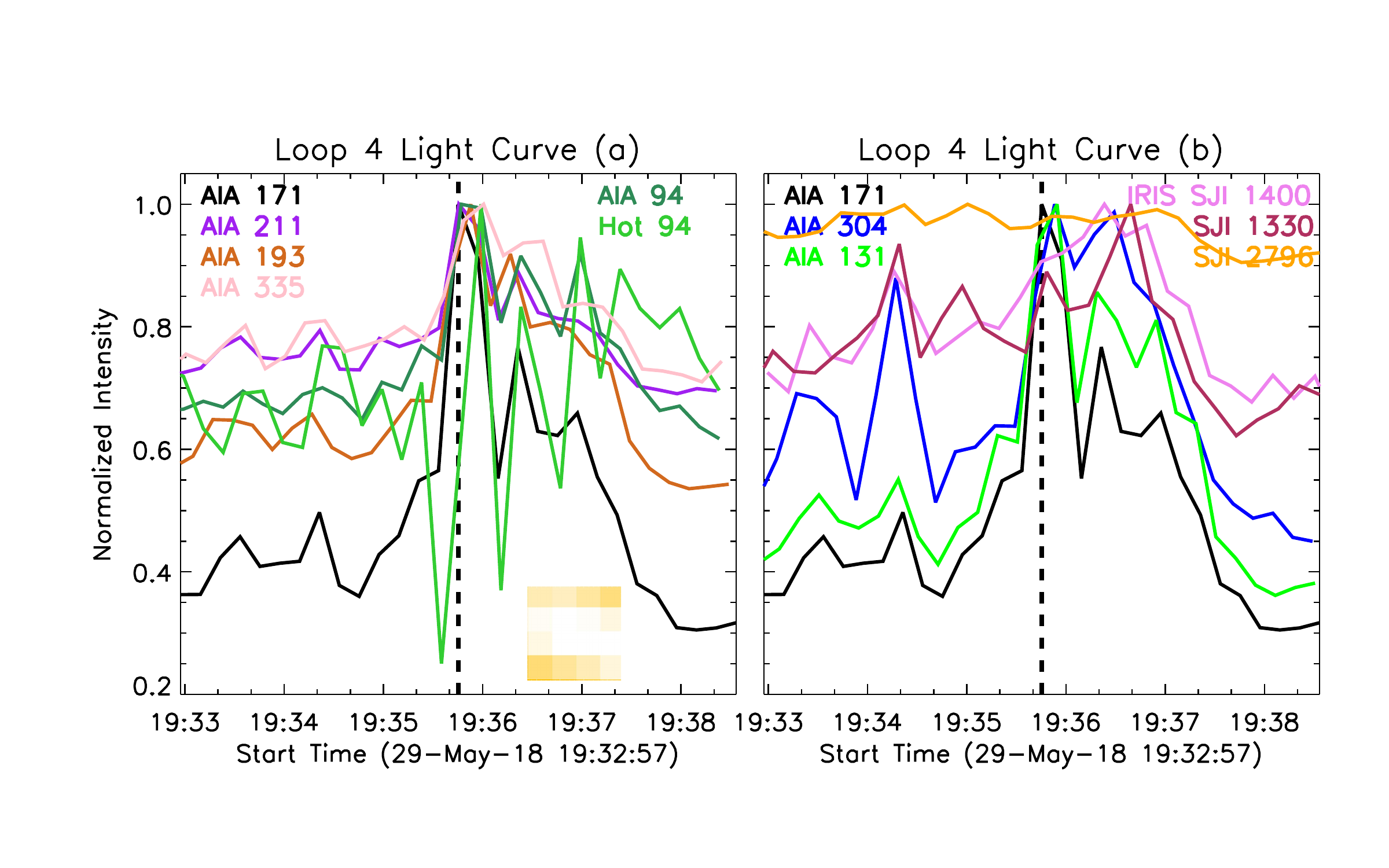}
	\includegraphics[trim=1.5cm 1.4cm 1.2cm 2cm,clip,width=\textwidth]{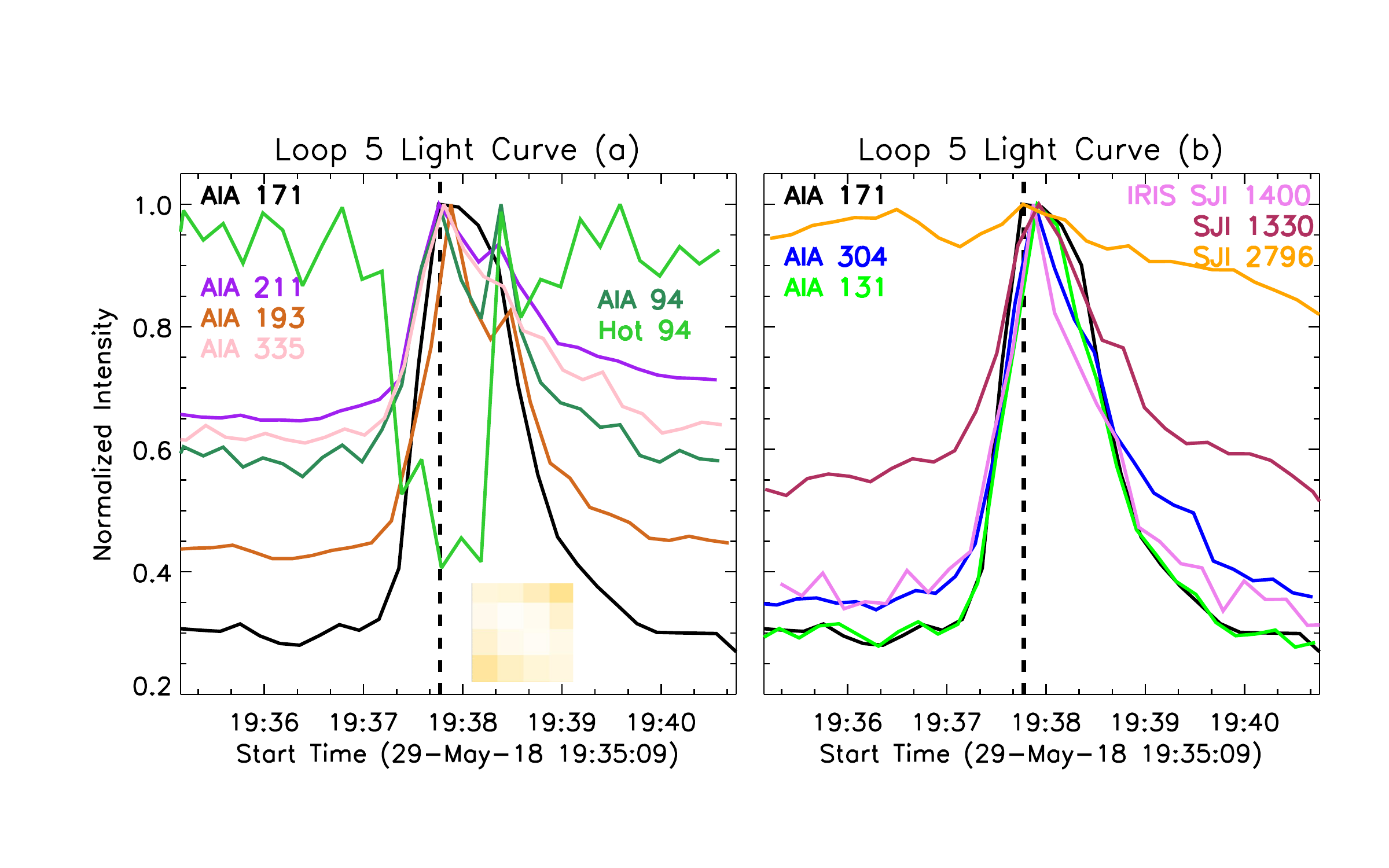}
	\caption{Light curves from AIA and IRIS intensity images for Loops 4 and 5. The AIA 171 \AA\ area selected for making light curves is displayed as a small inset in the left panel for each loop during its peak intensity time in AIA 171 \AA.  The vertical dashed lines mark times for the peak brightness of the events in AIA 171 \AA.} 
	\label{lc_loop4+5}
\end{figure*}


\begin{figure*}[h]
	\centering
	\includegraphics[trim=1.5cm 1.4cm 1.2cm 2.4cm,clip,width=\textwidth]{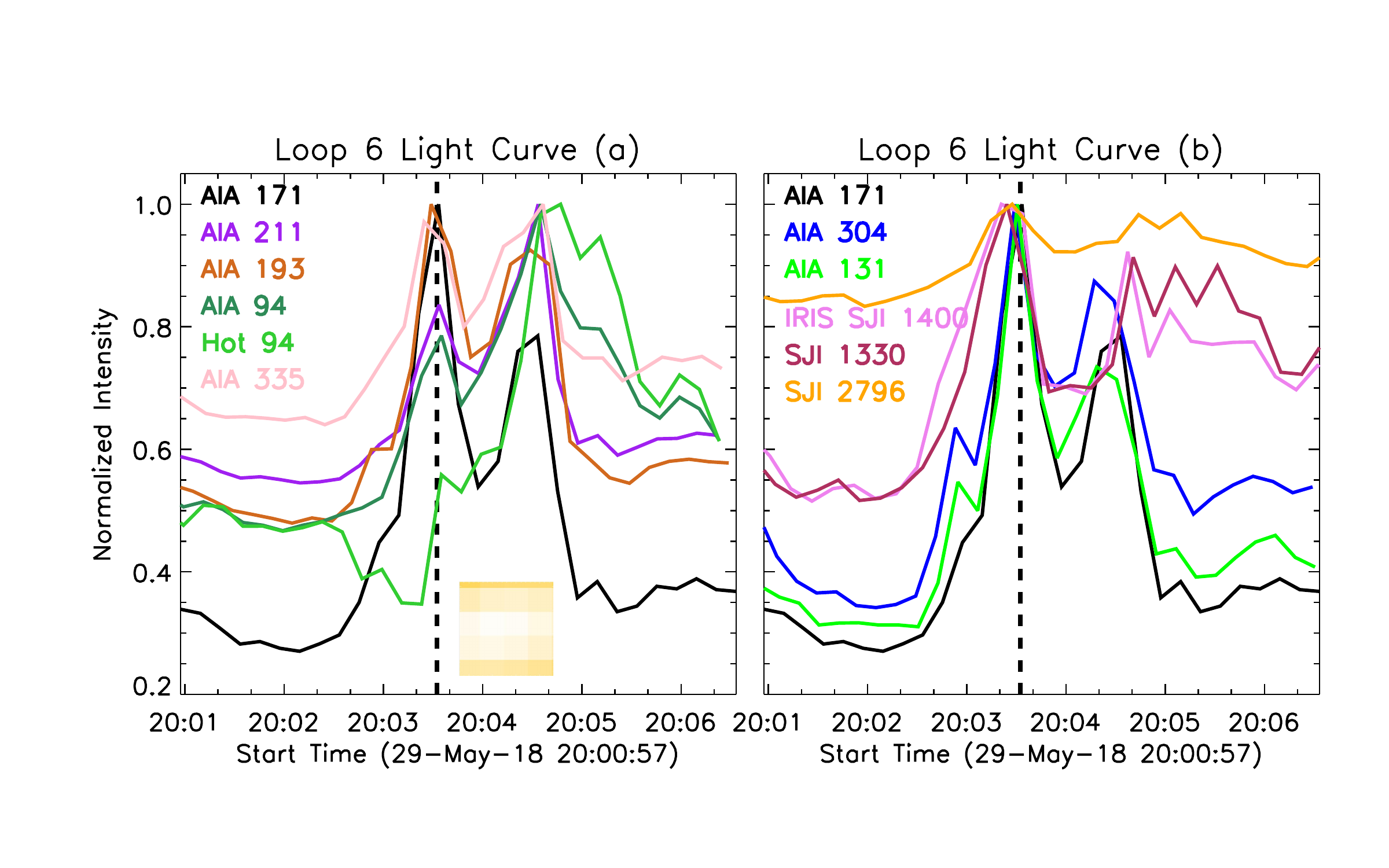}
	\includegraphics[trim=1.5cm 1.4cm 1.2cm 2cm,clip,width=\textwidth]{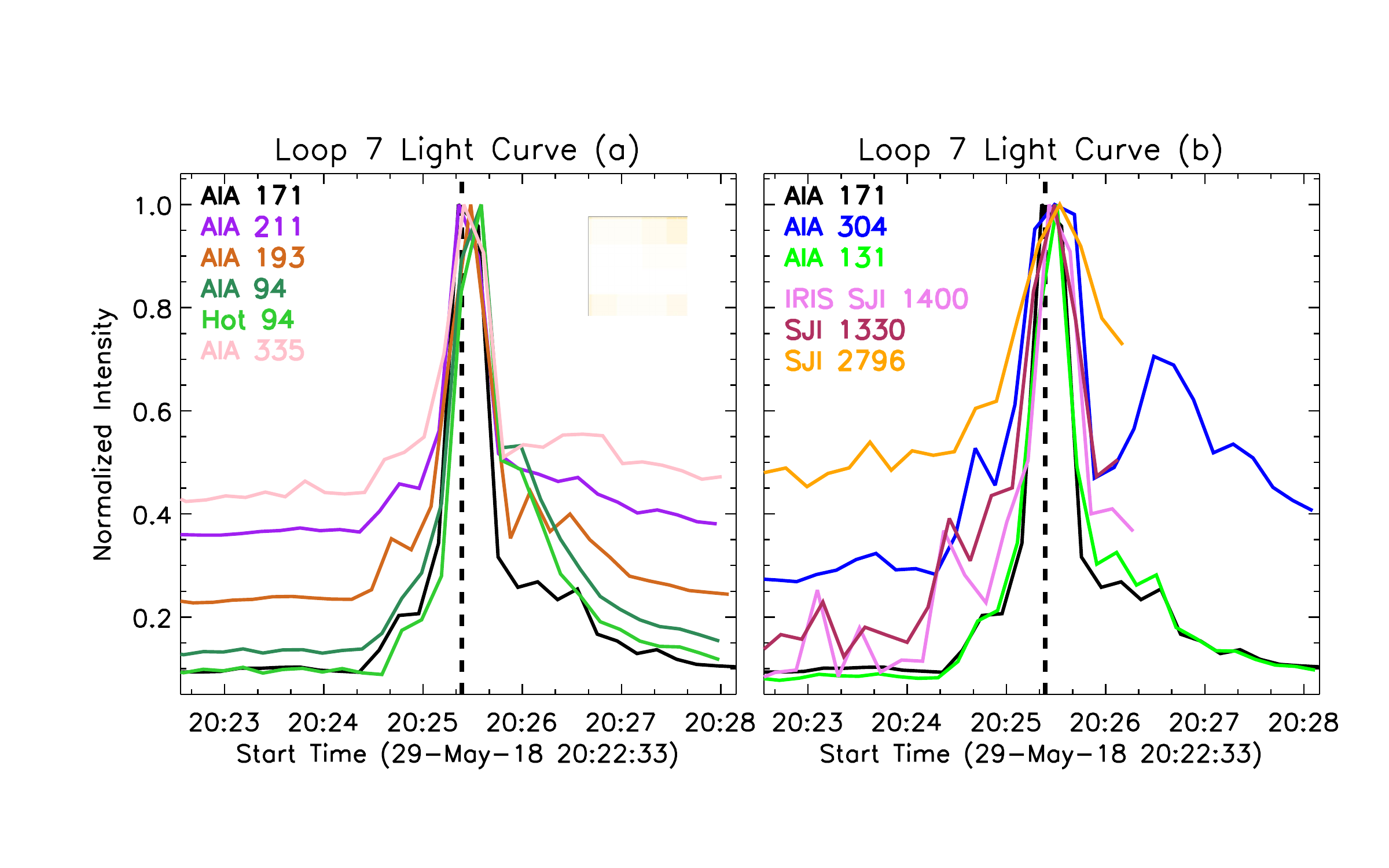}
	\caption{Light curves from AIA and IRIS intensity images for Loops 6 and 7. The AIA 171 \AA\ area selected for making light curves is displayed as a small inset in the left panel for each loop during its peak intensity time in AIA 171 \AA. The vertical dashed lines mark times for the peak brightness of the events in AIA 171 \AA. Loop 6 is a double peak event. For Loop 7, the IRIS observation time ends soon after the peak time of the event.} 
	\label{lc_loop6+7}
\end{figure*}


\begin{figure*}[h]
	\centering
	\includegraphics[trim=1.5cm 1.4cm 1.2cm 2.4cm,clip,width=\textwidth]{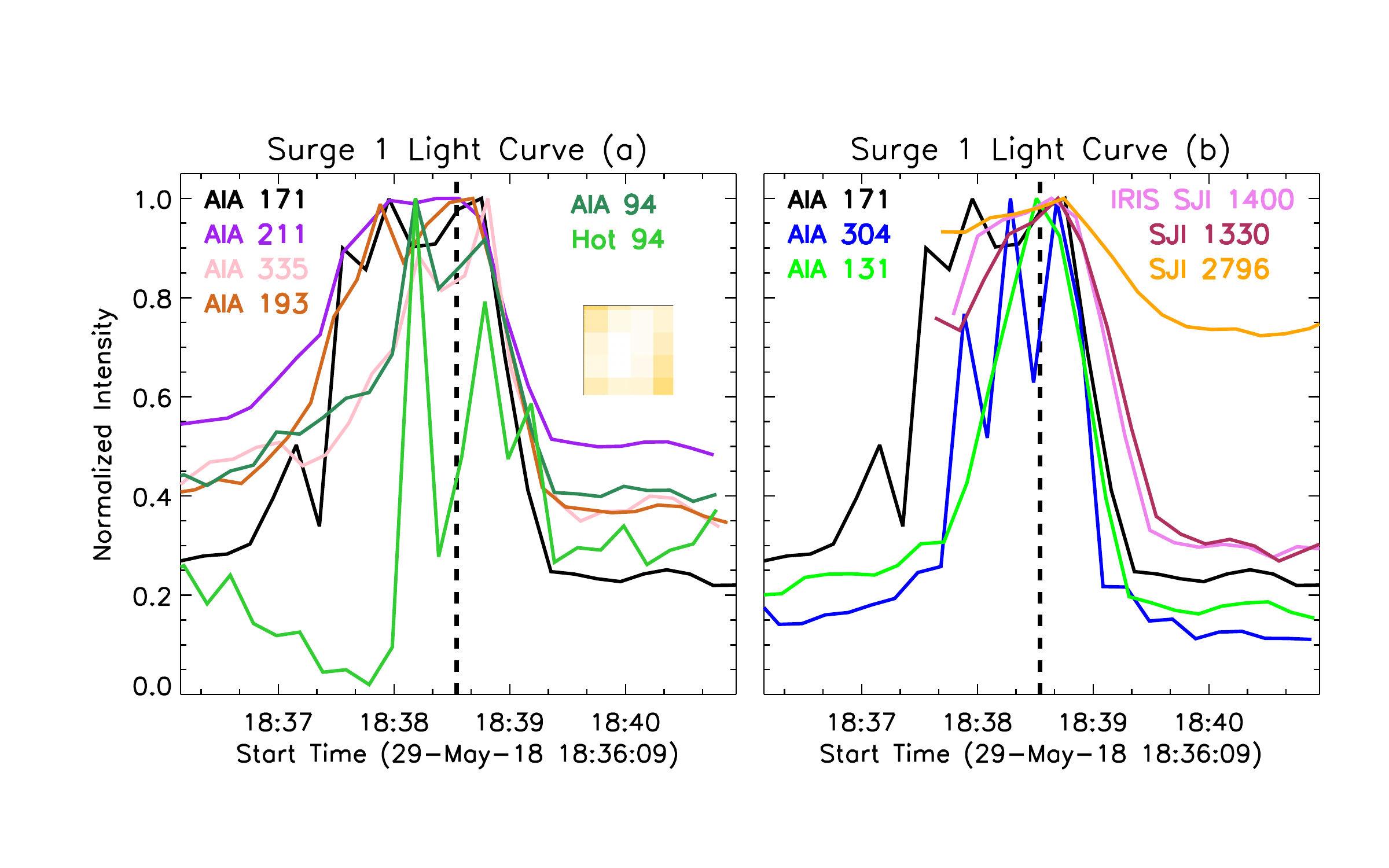}
	\includegraphics[trim=1.5cm 1.4cm 1.2cm 2cm,clip,width=\textwidth]{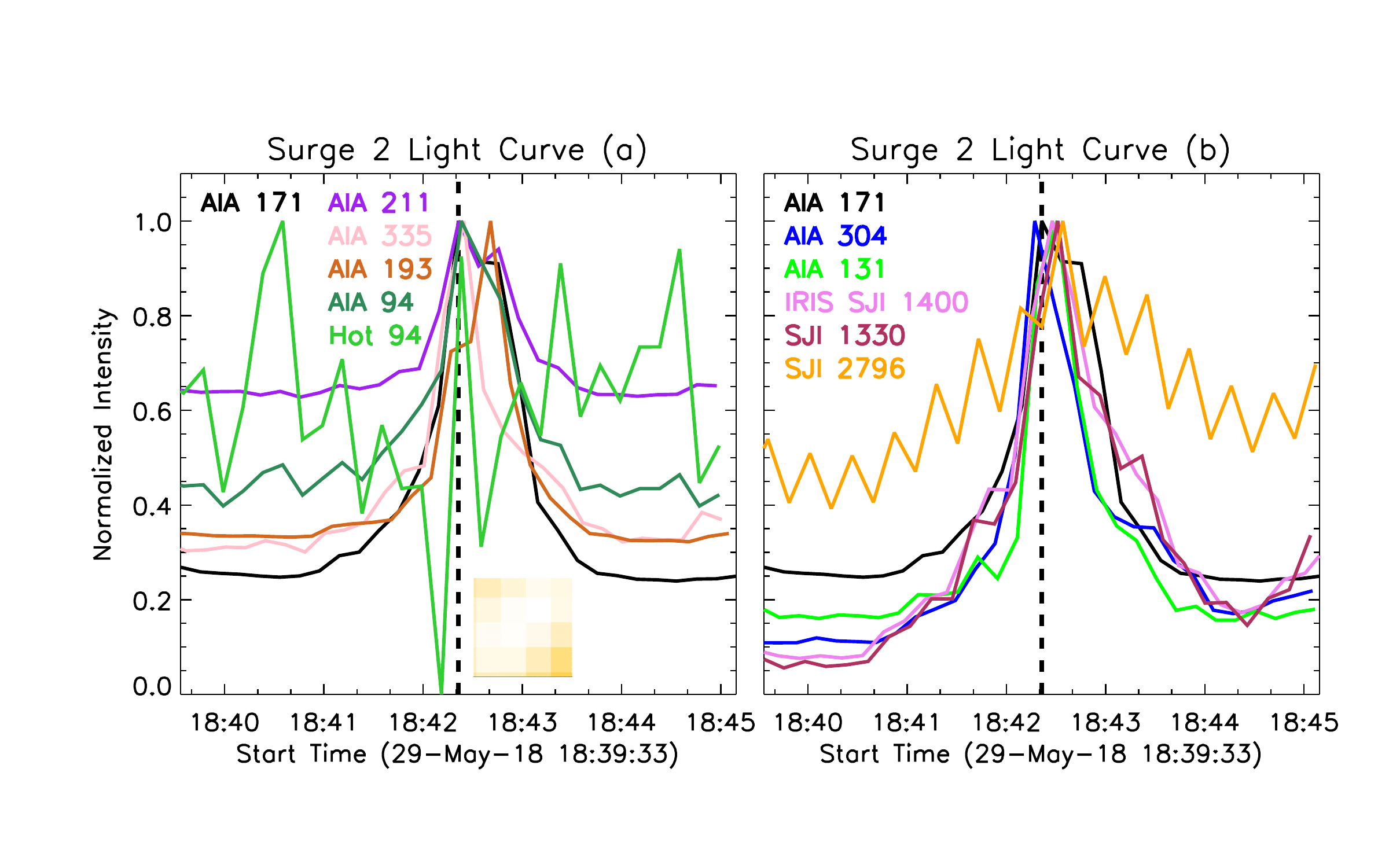}
	\caption{Light curves from AIA and IRIS intensity images for Surges 1 and 2. The AIA 171 \AA\ area selected for making light curves is displayed as a small inset in the left panel for each surge during its peak intensity time in AIA 171 \AA. The vertical dashed lines mark times for the peak brightness of the events in AIA 171 \AA. Note that for Surge 1 the IRIS coverage starts after the event has already started.}	\label{lc_surge1+2}
\end{figure*}


\begin{figure*}[h]
	\centering
	\includegraphics[trim=1.5cm 1.4cm 1.2cm 2.4cm,clip,width=\textwidth]{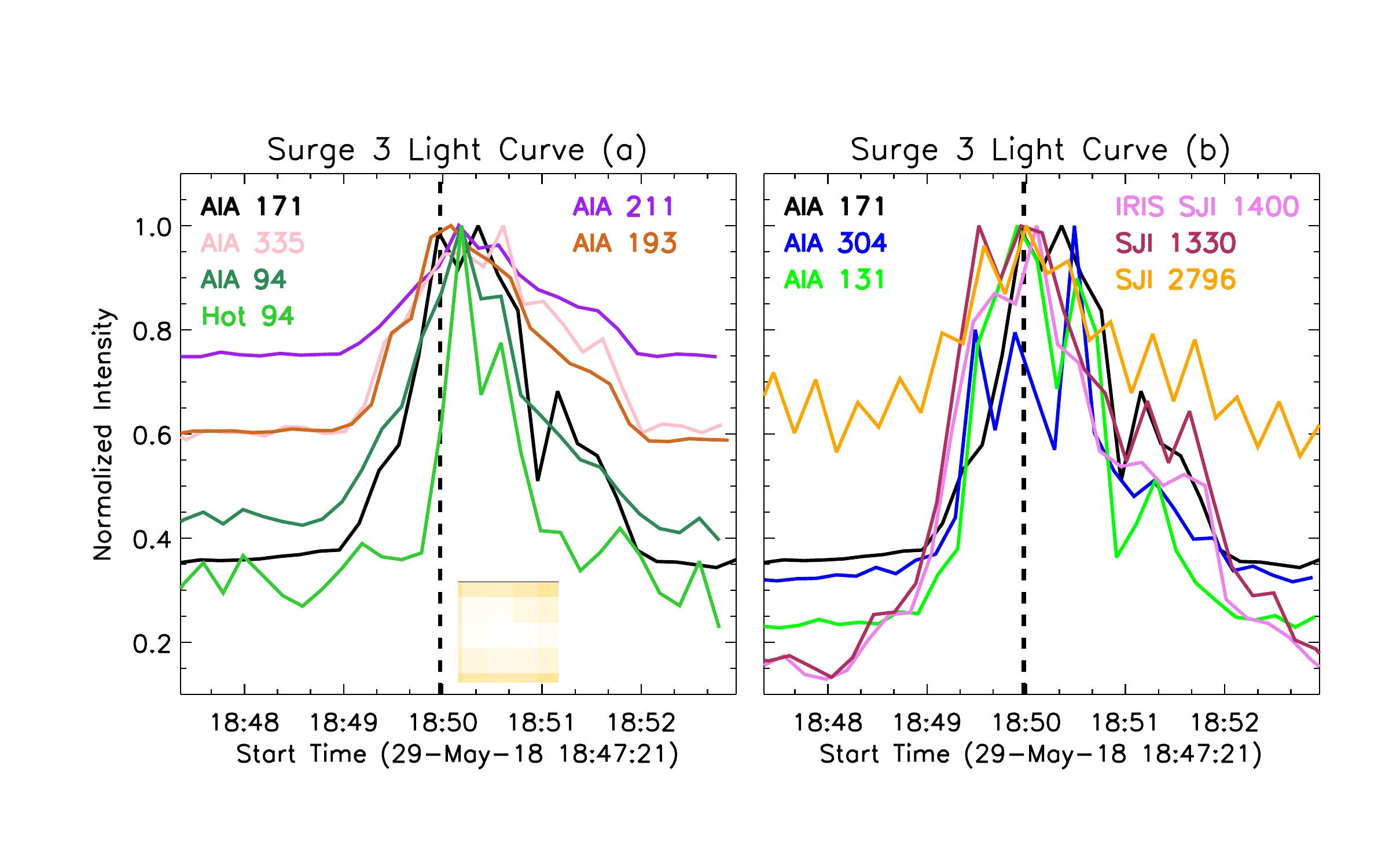}
	\includegraphics[trim=1.5cm 1.4cm 1.2cm 2cm,clip,width=\textwidth]{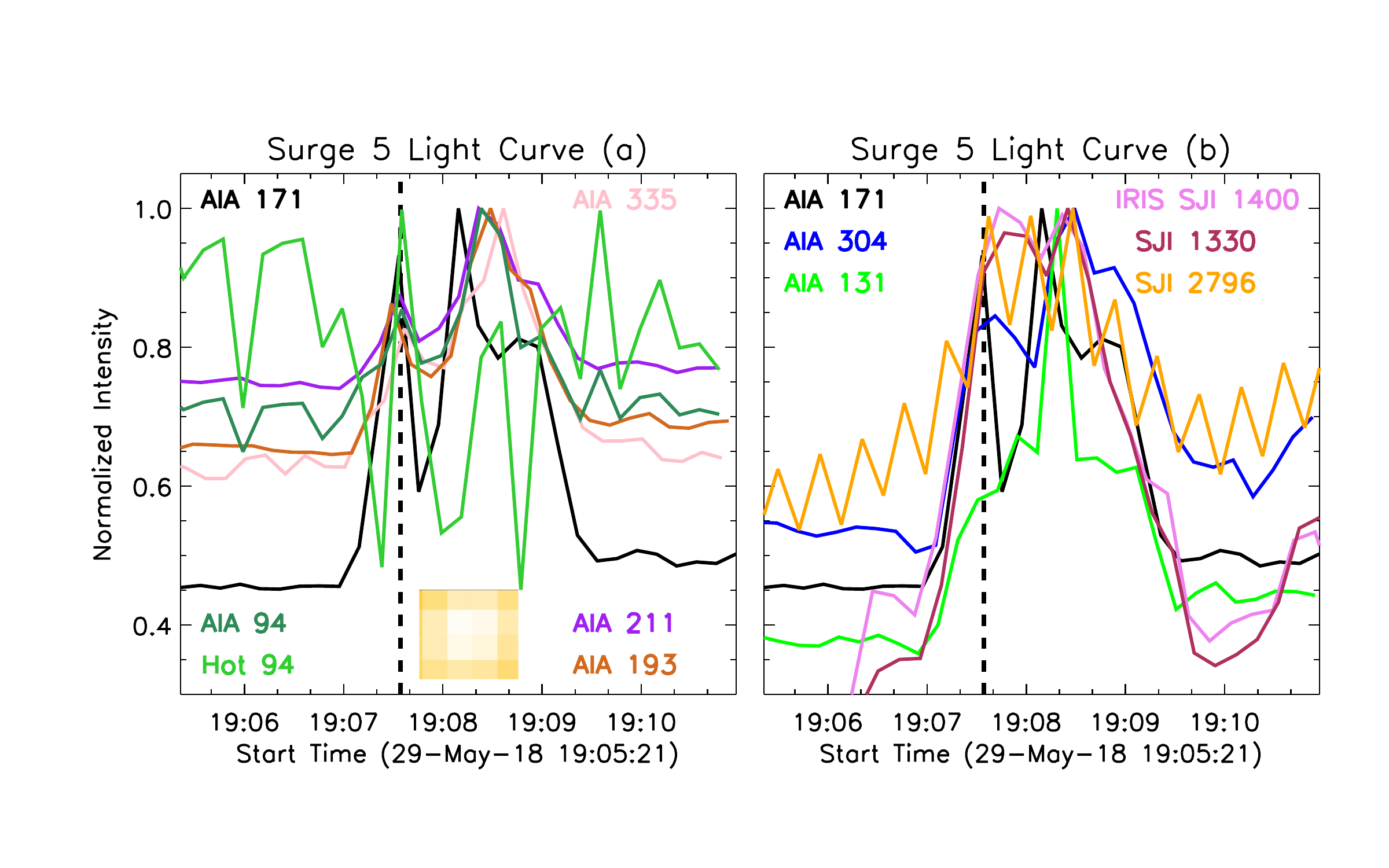}
	\caption{Light curves from AIA and IRIS intensity images for Surges 3 and 5. The AIA 171 \AA\ area selected for making light curves is displayed as a small inset in the left panel for each surge during its peak intensity time in AIA 171 \AA.  The vertical dashed lines mark times for the peak brightness of the events in AIA 171 \AA. For Surge 5, AIA 171 \AA\ (in particular) shows a double peak in the light curve. }
	\label{lc_surge3+5}
\end{figure*}


\clearpage\newpage
\section{Magnetic flux evolution in surges displaying flux emergence and cancellation}\label{app_flux_evoltion}
\begin{figure*}[h]
	\centering
	\includegraphics[trim=0.2cm 1.9cm 0.5cm 4.4cm,clip,width=0.5\textwidth]{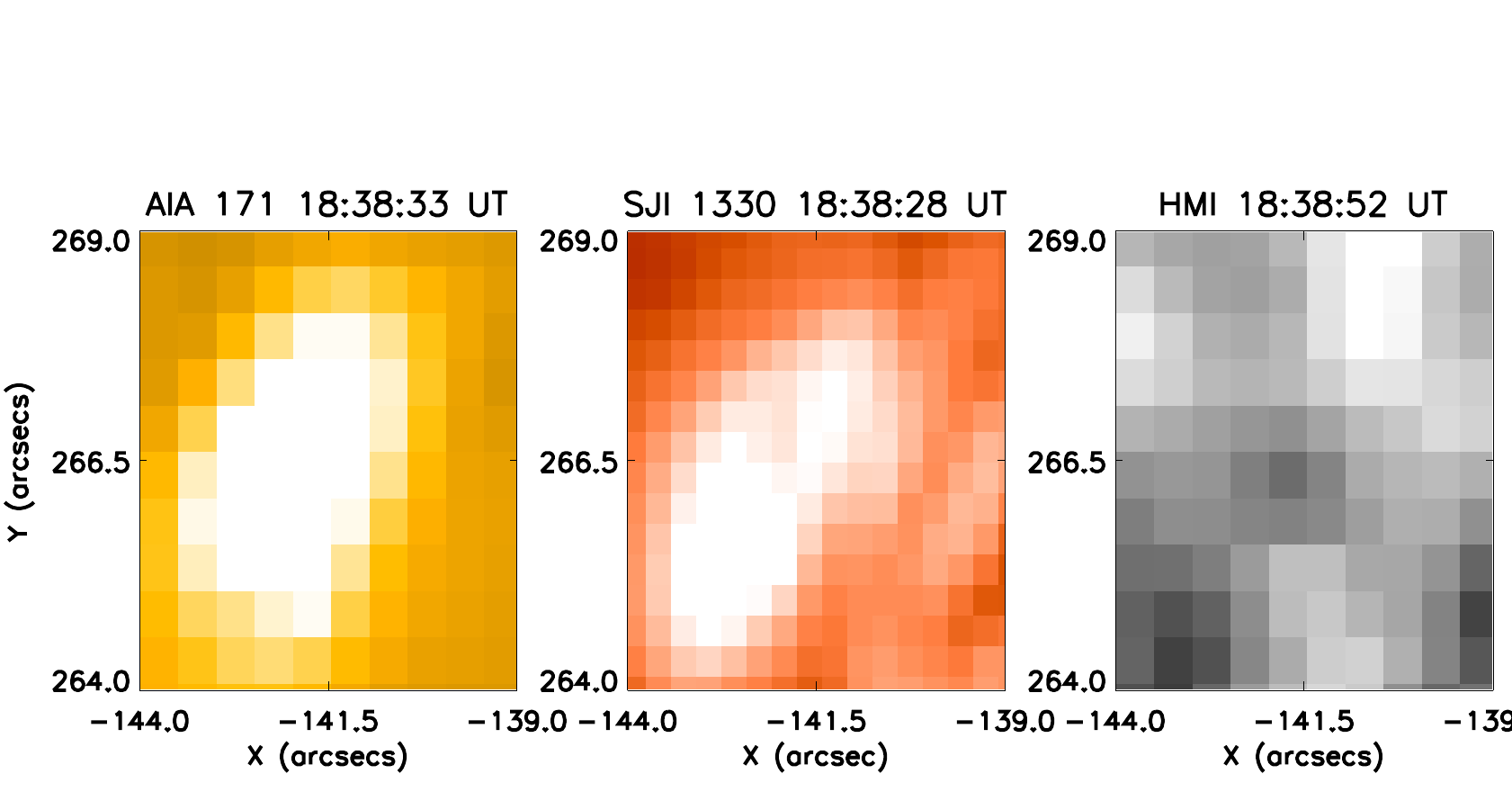}
	\put(-200,175){\Large Surge 1: Magnetic flux} \put(-200,150){\Large emergence and cancellation}
	\includegraphics[trim=0.7cm 0.4cm 0.6cm 0cm,clip,width=0.49\textwidth]{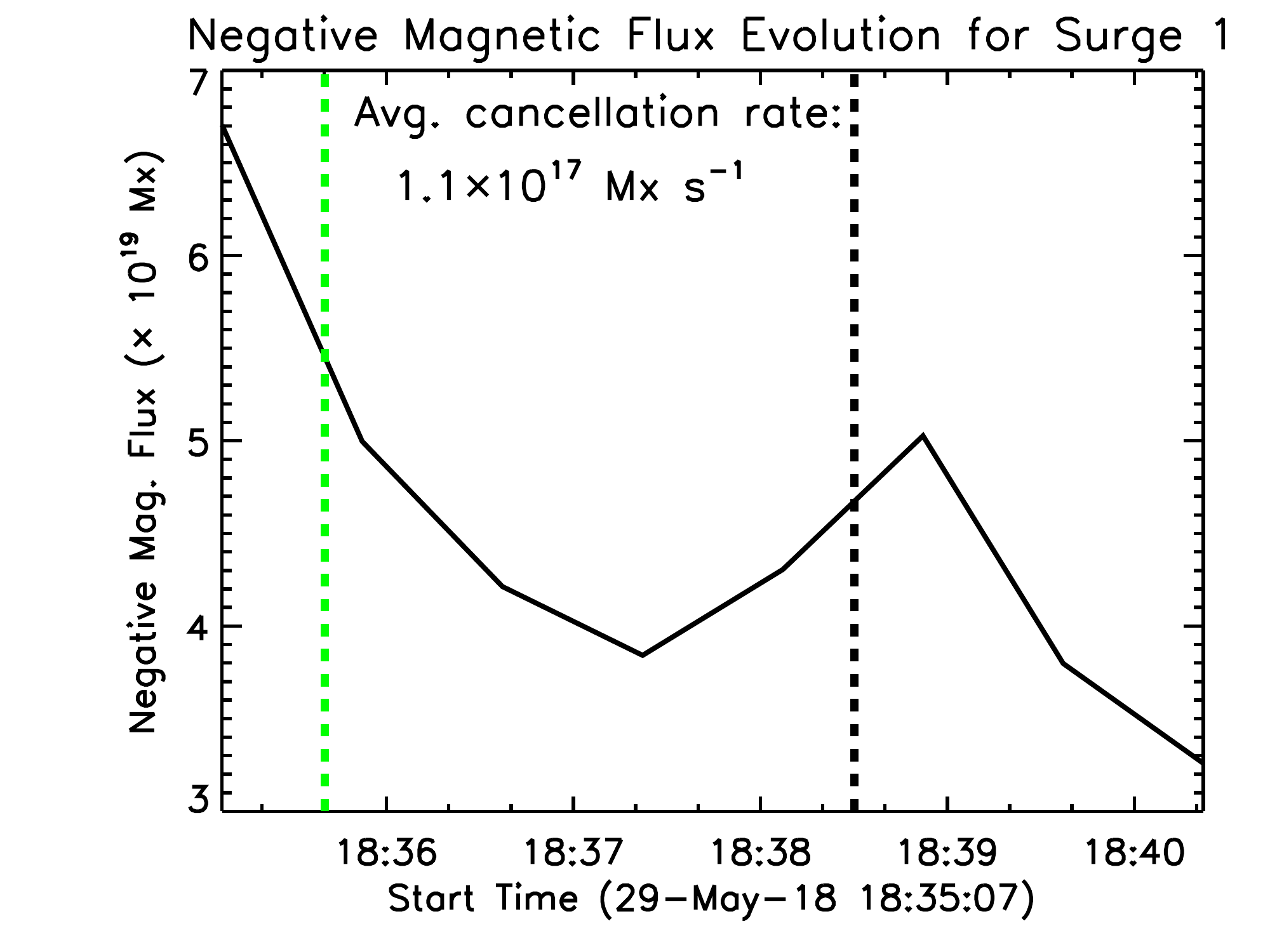}
	\includegraphics[trim=0.3cm 2cm 0.38cm 4cm,clip,width=0.5\textwidth]{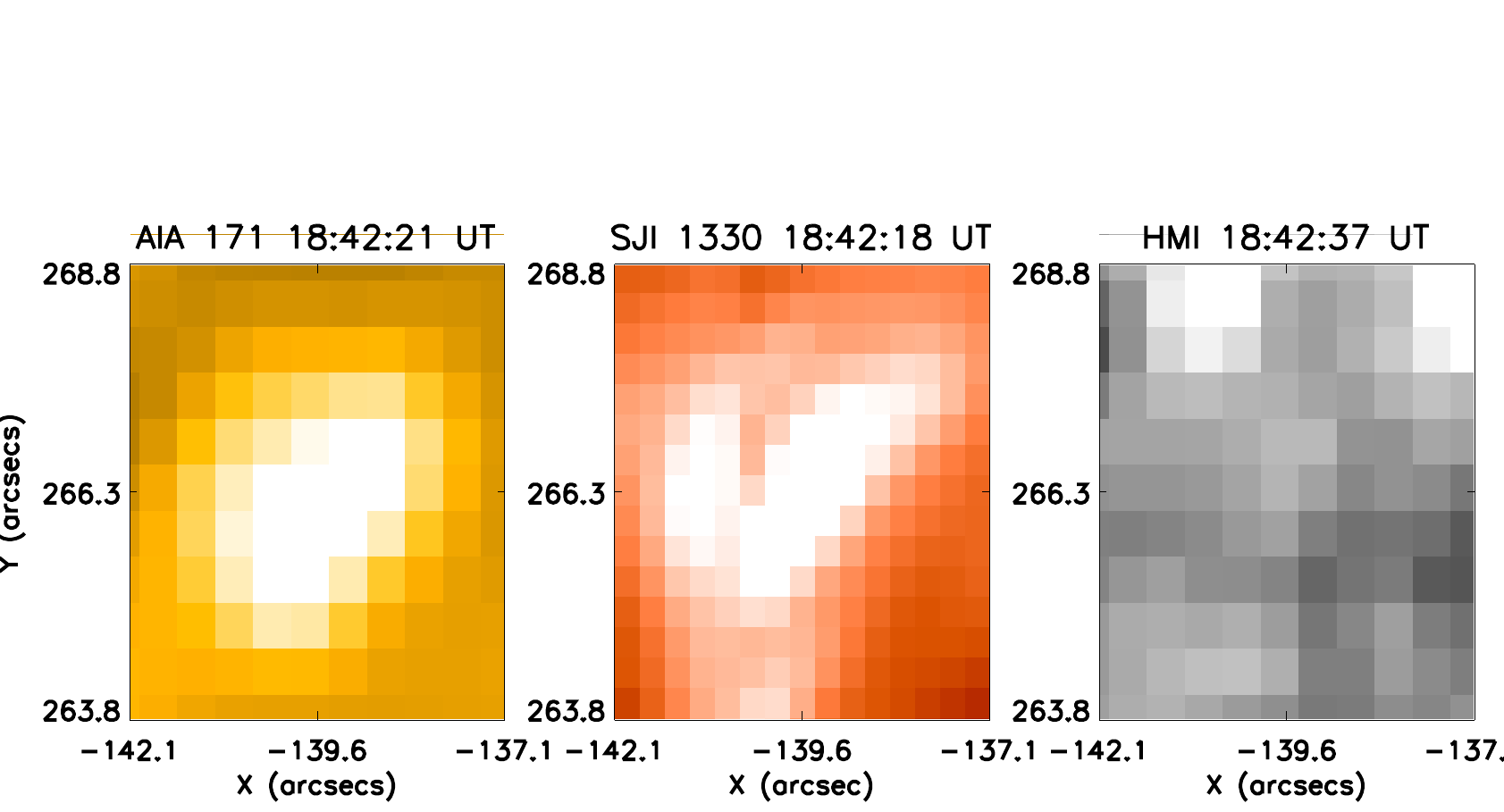}
	\put(-200,175){\Large Surge 2: Magnetic flux} \put(-200,150){\Large evolution/cancellation}
	\includegraphics[trim=1cm 0.4cm 0.6cm -1cm,clip,width=0.49\textwidth]{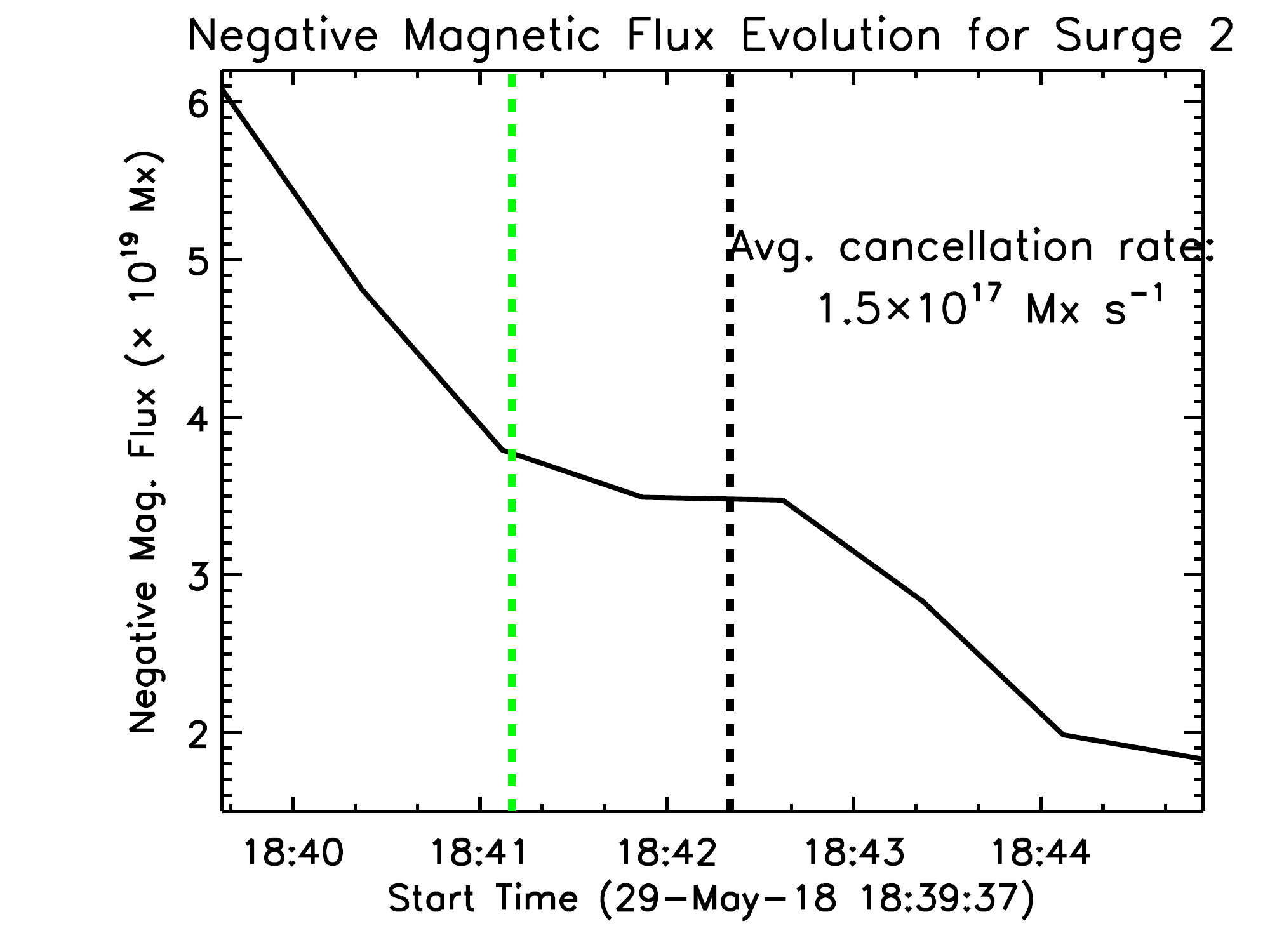}
	\caption{Magnetic flux evolution showing emergence and cancellation in Surge 1 and Surge 2. Small FOV covering the base of Surge 1 and Surge 2 (AIA 171, IRIS SJI 1330, and HMI LOS magnetogram) are shown in the upper left panels for both Surge 1 and Surge 2 -- same FOV is used to calculate flux evolution plots (negative flux for both) shown in the upper rightmost panels for each of these surges. The peak time of the event is marked by a dashed black vertical line. The vertical green dashed line marks the time when the event starts appearing in AIA 304 or SJI of \MgII\ 2796 \AA. The emergence, convergence and cancellation are also visible for each event in the movie hic\_iris\_sdo.mp4. The flux cancellation rate is mentioned on the plots. Evidently magnetic flux emergence (and convergence) - driven cancellation at the PIL triggers these events. }
	\label{fcr_surge1+2}
\end{figure*}

\begin{figure*}[h]
	\centering
	\includegraphics[trim=0.2cm 1.9cm 0.5cm 4.4cm,clip,width=0.5\textwidth]{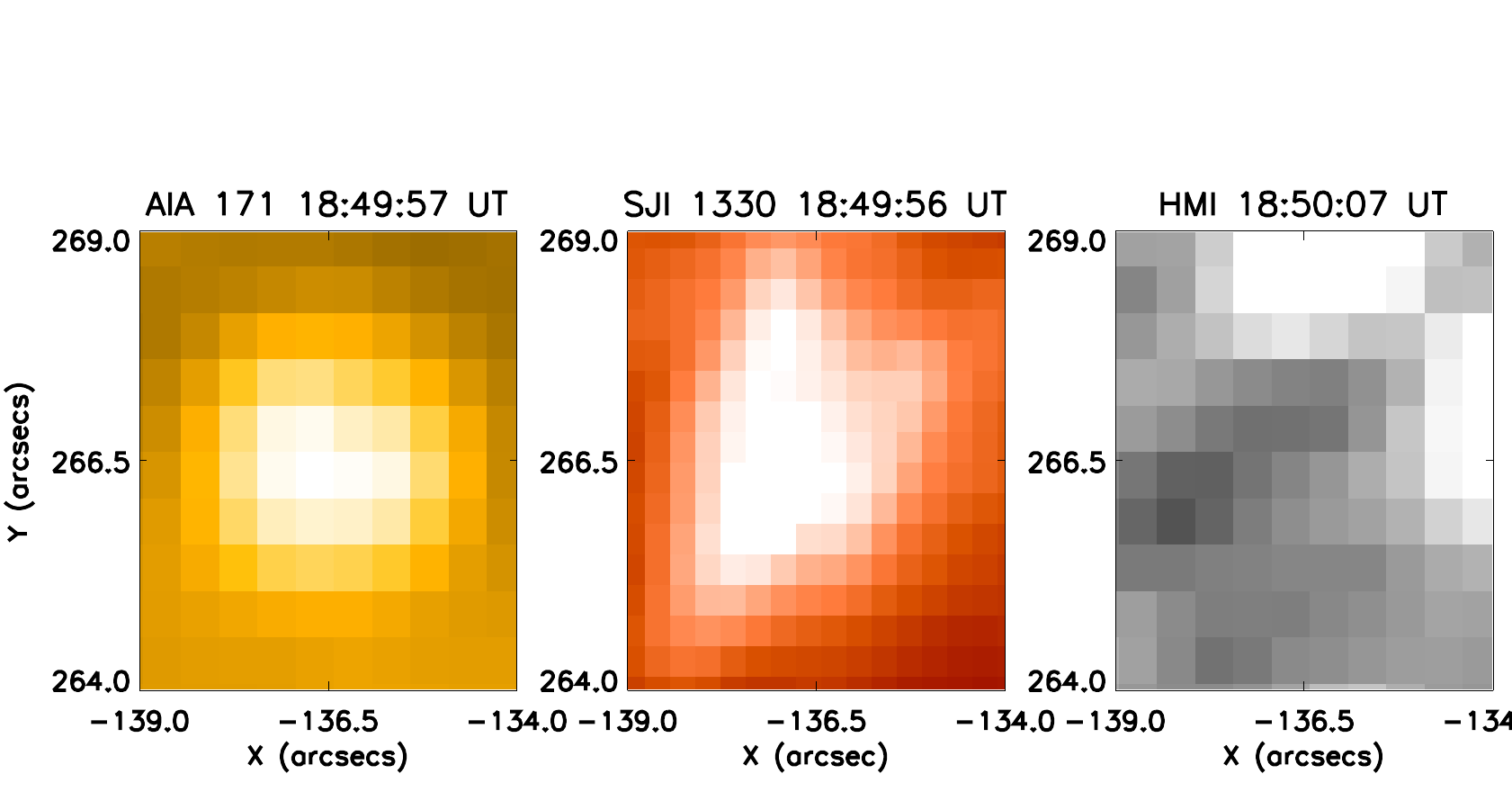}
	\put(-200,175){\Large Surge 3: Magnetic flux} \put(-200,150){\Large emergence and cancellation}
	\includegraphics[trim=0.7cm 0.4cm 0.6cm 0cm,clip,width=0.49\textwidth]{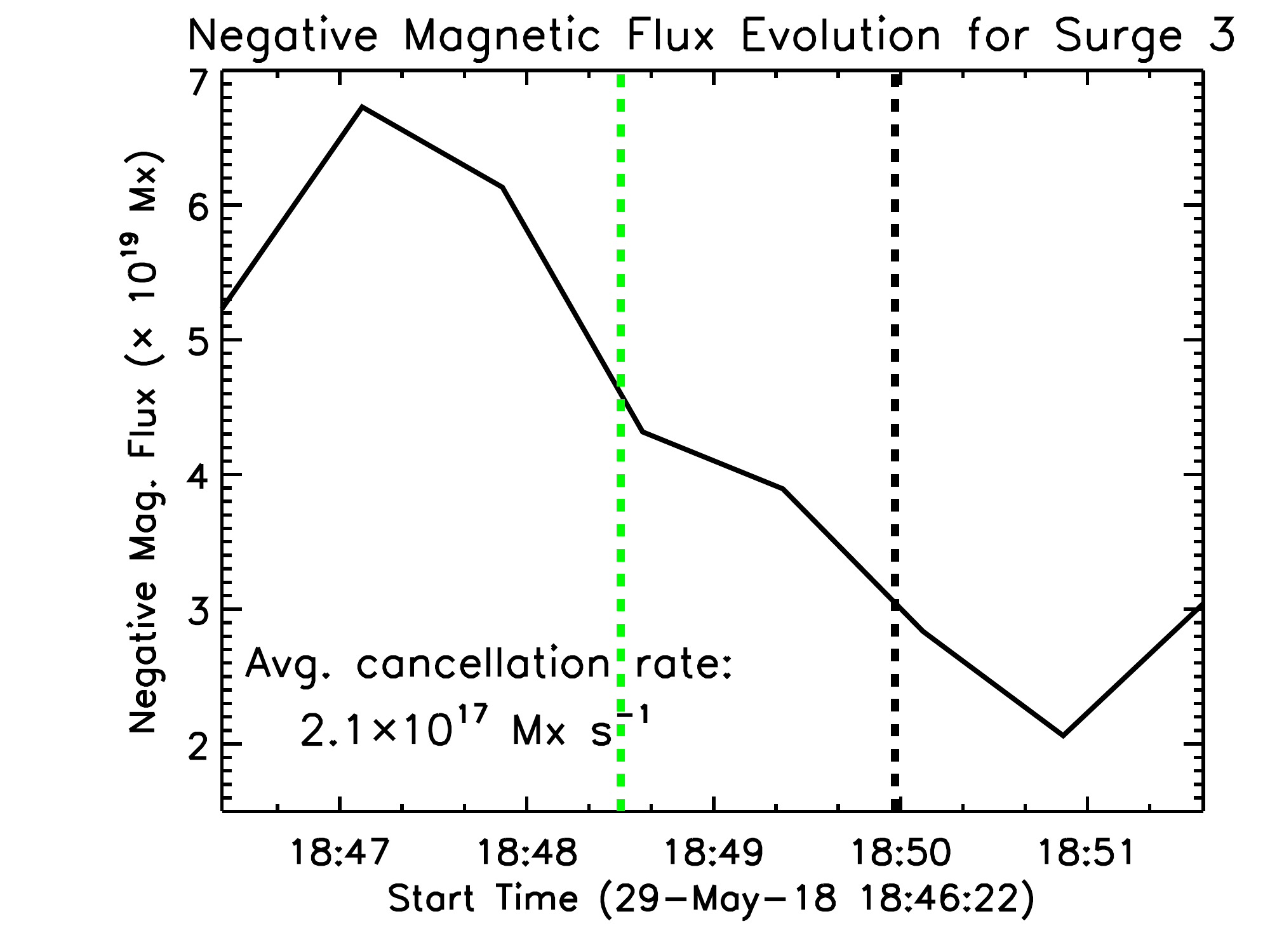}
	\includegraphics[trim=0.3cm 2cm 0.38cm 4cm,clip,width=0.5\textwidth]{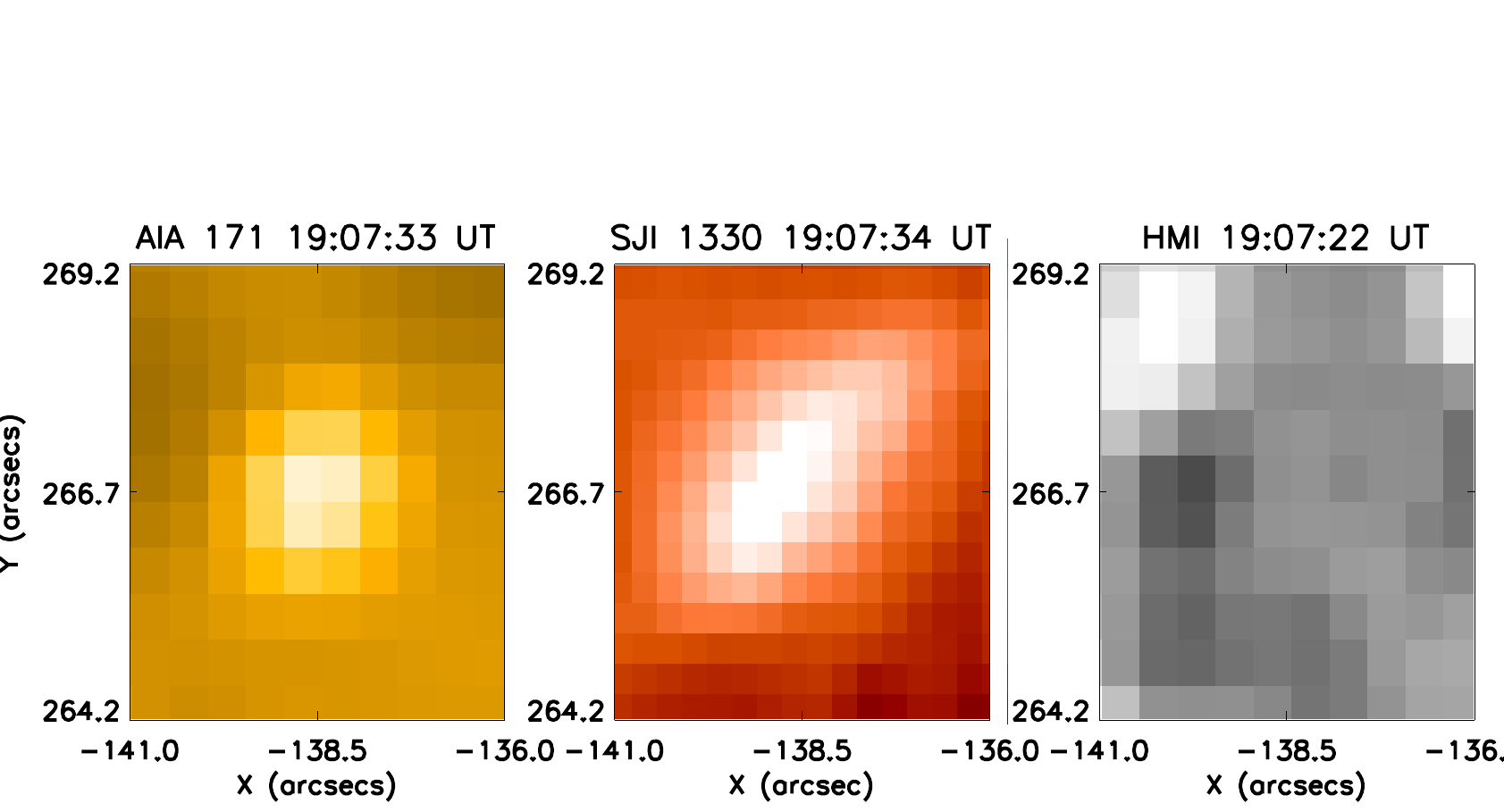}
	\put(-200,175){\Large Surge 5: Magnetic flux} \put(-200,150){\Large emergence and cancellation}
	\includegraphics[trim=1cm 0.4cm 0.6cm -1cm,clip,width=0.49\textwidth]{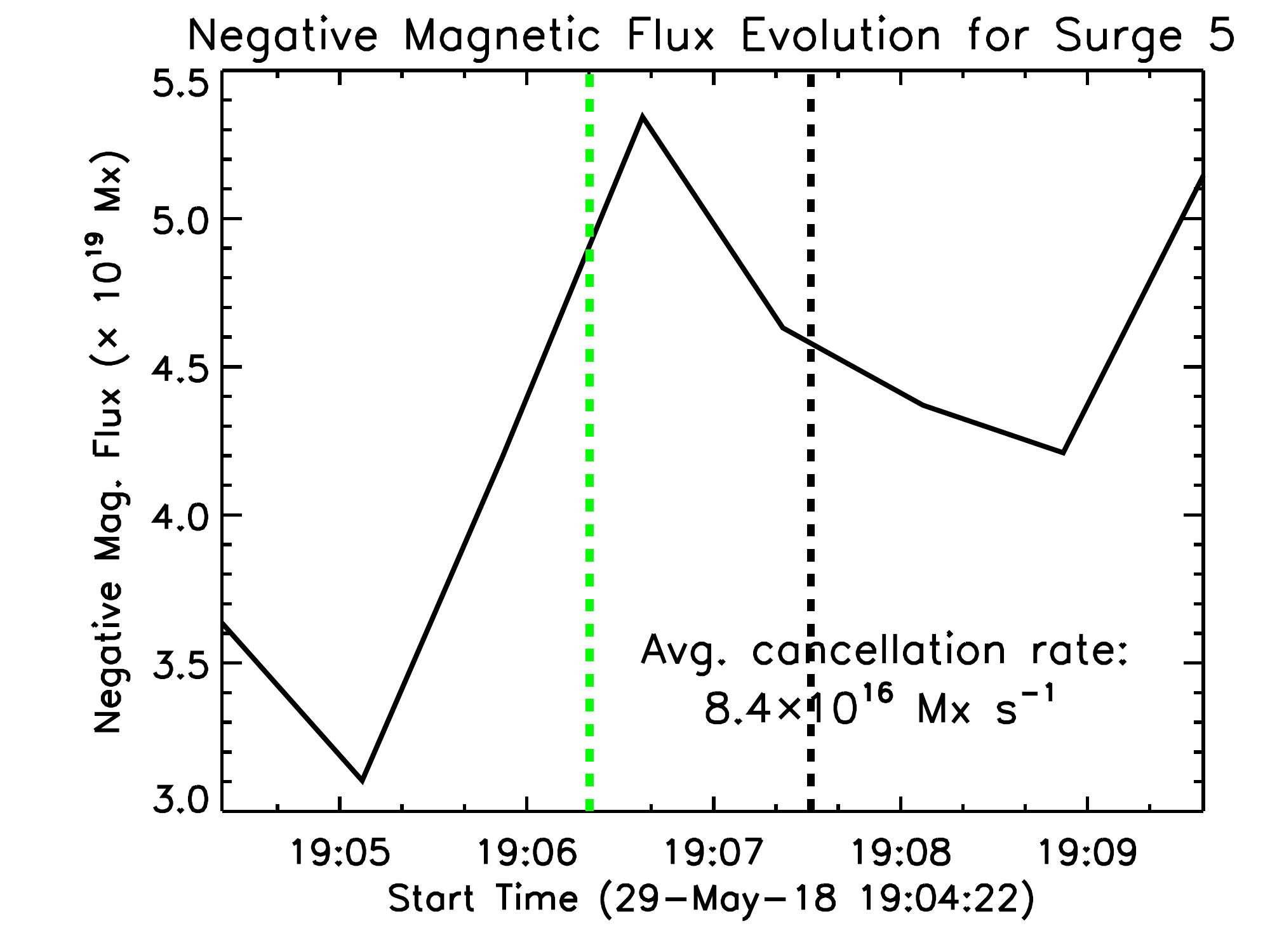}
	\caption{Magnetic flux evolution showing emergence and cancellation in Surge 3 and Surge 5. Small FOV covering the base of Surge 3 and Surge 5 (AIA 171, IRIS SJI 1330, and HMI LOS magnetogram) are shown in the upper left panels for both Surge 3 and Surge 5 -- same FOV is used to calculate flux evolution plots (negative flux for both) shown in the upper rightmost panels for each of these surges. The peak time of the event is marked by a dashed black vertical line. The vertical green dashed line marks the time when the event starts appearing in AIA 304 or SJI of \MgII\ 2796 \AA. The emergence, convergence and cancellation are also visible for each event in the movie hic\_iris\_sdo.mp4. The flux cancellation rate is mentioned on the plots. Evidently magnetic flux emergence (and convergence) - driven cancellation at the PIL triggers these events. }
	\label{fcr_surge3+5}
\end{figure*}

\end{document}